\documentclass[12pt]{article}

\usepackage{latexsym}

\usepackage{graphics}
\usepackage{graphicx}
\usepackage{epstopdf}
\usepackage{amssymb}
\newcommand{\be}{\begin{equation}}
\newcommand{\ee}{\end{equation}}
\newcommand{\ba}{\begin{eqnarray}}
\newcommand{\ea}{\end{eqnarray}}
\newcommand{\ban}{\begin{eqnarray*}}
\newcommand{\ean}{\end{eqnarray*}}

\begin{document}


\title{Myers-Perry Black Hole in an External Gravitational Field}

\author{
   Shohreh Abdolrahimi$^{1,2,3}$\footnote{E-mail: \texttt{abdolrah@ualberta.ca}},\, Jutta Kunz$^{1}$\footnote{E-mail:\texttt{jutta.kunz@uni-oldenburg.de}},\, Petya Nedkova$^{1,4}$\footnote{E-mail:\texttt{pnedkova@phys.uni-sofia.bg}}\\ \\
 {\footnotesize${}^{1}$  Institut f\"{u}r Physik, Universit\"{a}t Oldenburg}
 {\footnotesize  D-26111 Oldenburg, Germany} \\
{\footnotesize${}^{2}$ Department of Mathematics and Statistics, Memorial University, St. John's, NL A1C 5S7, Canada}\\
{\footnotesize${}^{3}$ Theoretical Physics Institute, University of Alberta, Edmonton, AB, Canada, T6G 2G7}\\
{\footnotesize ${}^{4}$ Department of Theoretical Physics,
                Faculty of Physics, Sofia University, 5 James}\\
{\footnotesize   Bourchier Boulevard, Sofia~1164, Bulgaria }}
\date{}
\maketitle

\begin{abstract}
We obtain a new exact solution of the 5D Einstein equations in vacuum describing a distorted Myers-Perry black hole with a single angular momentum. Locally, the solution is interpreted as a black hole distorted by a stationary $U(1)\times U(1)$ symmetric distribution of external matter. Technically, the solution is constructed by applying a 2-fold B\"{a}cklund transformation on a 5D distorted Minkowski spacetime as a seed. The physical quantities of the solution are calculated, and a local Smarr-like relation on the black hole horizon is derived. It possesses the same form as the Smarr-like relation for the asymptotically flat Myers-Perry black hole. It is demonstrated that in contrast to the asymptotically flat Myers-Perry black hole,  the ratio of the horizon angular momentum and the mass $J^2/ M^3$ is unbounded, and can grow arbitrarily large. We study the properties of the ergoregion and the horizon surface.  The external field does not influence the horizon topology. The horizon geometry  however is distorted, and any regular axisymmetric geometry is possible.
\end{abstract}

\section{Introduction}

Black holes are one of the most important predictions of general
relativity. They are interesting from a purely theoretical viewpoint
since they give insight into the properties of the gravitational
theory. On the other hand they represent intriguing astrophysical objects,
that are involved in 
various physical processes, including accretion disks and jets.
By now there is vast observational evidence
for their existence. Besides supermassive black holes
located in the centers of galaxies 
there are stellar mass black holes, which occur
as companions in binary systems.

Most of the theoretical investigations, concerning black holes,
consider them as isolated objects. Various exact solutions
describing such systems exist, and their properties are
comparatively well studied. Such is the Kerr-Newman family of
solutions, which represents the unique stationary charged black hole in
asymptotically flat spacetime within the classical general
relativity in four dimensions. Most astrophysical situations,
however, would suggest that the black hole is not isolated but it is
interacting with some external distribution of matter.  In general
these systems are dynamical, and due to their complexity  are subject only
to numerical or perturbative treatment. Exact solutions exist only
in idealized cases assuming stationarity, or even staticity, and a
very special form of the external matter. A plausible way to
describe a non-isolated black hole by an exact solution is to
construct a local solution, which is physically relevant only in a
close neighborhood of the black hole, but still incorporates into
itself information about the external matter fields. A major
advantage is that these solutions are valid for broad classes of
external matter, the only restrictions coming from some regularity
conditions. They are still stationary, since solution generation
techniques rely heavily on spacetime symmetries. However, they can
be considered as a possible approximation for dynamical black holes
relaxing on a time scale much shorter than the external matter, or
for equilibrium systems of black hole and matter moving in a
quasi-stationary state. Such scenarios include for example a black
hole surrounded by an accretion disk, or a galaxy with a central black
hole.

The idea was originally developed in the work of Geroch and Hartle \cite{Geroch},
where they considered general static black holes in four dimensions
in the presence of external matter fields and investigated their
properties, thermodynamical behavior and Hawking radiation. They
discussed solutions to the static Einstein equations in vacuum which
contain a regular horizon, which are free of singularities
in the domain of outer communications, and are  asymptotically
non-flat if considered as global solutions. It was suggested that
such solutions can be interesting if regarded as local solutions,
which are valid only in some neighborhood of the black hole
horizon. Provided they can be extended in some intermediate region
in spacetime to some non-vacuum solutions to the Einstein equations
which are asymptotically flat, a physically reasonable global
solution can be constructed. Then, the vacuum solution can be
interpreted as describing locally a black hole distorted by certain
external matter. To be able to apply the described argument, we
should assume moreover that the black hole and the external matter
are separated in spacetime, so that there exists a certain vacuum
neighborhood surrounding the black hole, and that the external
matter is confined to a compact region, in order for the spacetime
to be asymptotically flat.

Several local black hole solutions in the presence of external
matter were constructed explicitly, which were called after the work of Geroch and
Hartle distorted black holes. Static vacuum solutions were obtained in
\cite{Israel}-\cite{Chandrasekhar}, and rotating generalizations describing
Kerr black holes in arbitrary axisymmetric external gravitational fields
were constructed in \cite{Tomimatsu},\cite{Breton:1997} using solitonic techniques. It was observed that
distorted black holes in four dimensions can possess also a regular
horizon with toroidal topology apart from the spherical one \cite{Peters}, \cite{Xanthopoulos}, in
contrast to their non-distorted asymptotically flat counterparts.
However, in order to be able to make an asymptotically flat
non-vacuum extension of toroidal black hole solutions, there should
exist a region outside the horizon containing external matter with
negative energy density. Thus, the dominant energy condition is
violated, and distorted black holes with a toroidal horizon are
considered to be not of
astrophysical importance, although theoretically interesting.

In analogy with the vacuum solutions distorted charged black holes
were also constructed in the classical general relativity and dilaton
gravity in four dimensions. Static solutions were obtained in \cite{Fairhurst}, \cite{Yazadjiev},
while a Kerr-Newman black hole in general external gravitational field was constructed in \cite{Breton:1998}.
They are local solutions to the static Einstein-Maxwell(-dilaton) equations possessing only electric
charge. Similarly to the vacuum solutions, they should contain a
regular horizon, possess no singularities in the domain of
outer communications, and admit an extension to a solution with
external matter fields which is asymptotically flat. In the same
sense, two classes of distorted black hole solutions to the
Einstein-Klein-Gordon equations, minimally coupled to gravity, were
also obtained in \cite{Delgado}.

With the development of the string theory, the brane-world
scenarios, and the holographic ideas black hole solutions in higher
spacetime dimensions became relevant. In this relation distorted
black hole solutions to the static axisymmetric Einstein-Maxwell
equations in five dimensions were obtained, including a static
vacuum solution with a horizon with spherical topology (distorted 5D
Schwarzschild-Tangherlini black hole) \cite{Abdolrahimi:2010}, and a static electrically
charged solution with spherical horizon topology (distorted 5D
Reissner-Nordstr\"{o}m black hole) \cite{Abdolrahimi:2013}. In this framework the
restrictions on the asymptotic structure of spacetime were also naturally
relaxed, and spacetimes with one (or several) compactified spacelike
dimensions were considered. Static black holes with distorted horizon
in such spacetimes were investigated \cite{Frolov:2003},
however in this case the distortion of the horizon is not a result from
the interaction with some external matter, but follows from the compactification.

The distorted black holes are not only interesting as more realistic solutions for potential astrophysical application, but also important from a purely theoretical viewpoint. They are more general stationary and axisymmetric solutions, than the isolated black holes, and can provide deeper insights into the black hole properties. A series of works were devoted to investigating how the properties of the isolated black holes are influenced if they are distorted by external matter field, and which of them remain unaffected. It was established that the 4D static distorted black holes belong to the Petrov type D on the horizon, like their asymptotically flat counterparts, although in the rest of the spacetime they are algebraically general \cite{Papadopoulos}. It was also demonstrated that the distorted Kerr-Newman black hole satisfies on the horizon the standard Smarr relation for the non-distorted case \cite{Breton:1998}, and the same geometric inequalities between the electric charge, horizon area and angular momentum \cite{Hennig}, \cite{Ansorg}. These features were also observed in the case of the 5D distorted Reissner-Nordstr\"{o}m black hole \cite{Abdolrahimi:2013}. It was shown that for this solution, the space-time singularities are located behind the black holes inner (Cauchy) horizon, provided that the sources of the distortion satisfy the strong energy condition, and the inner horizon remains regular if the distortion fields are finite and smooth at the outer horizon. There exists a certain duality transformation between the inner and the outer horizon surfaces which links surface gravity, electrostatic potential, and space-time curvature invariants calculated at the black
hole horizons. The product of the inner and outer horizon areas depends only on the black holes electric charge and the geometric mean of the areas is the upper (lower) limit for the inner (outer) horizon area.

Within the framework of isolated horizons it was proven that a local first law of thermodynamics is valid on the distorted black holes horizon \cite{Ashtekar1}-\cite{Ashtekar3}, which possesses the same form as the first law for the corresponding asymptotically flat black holes. On the other hand, the distorted black hole can exhibit a very different horizon geometry than the isolated black holes. Even in the four-dimensional static case their horizon surfaces are axisymmetric, rather than spherically symmetric, and they can be highly elongated, or flattened \cite{Booth1}. Only in the extremal limit, the horizon of the distorted Reissner-Nordstr\"{o}m black hole is proven to be spherically symmetric \cite{Booth2}. The distorted black holes can also have different ratios between the mass and the angular momentum, than the asymptotically flat ones. It is well-known that the mass $M$ and the angular momentum $J$ of the  Kerr black hole should satisfy the inequality $\mid J \mid /M^2 \leq 1$ in order for an event horizon to exist. However, if the Kerr black hole is situated in an external matter field this ratio not only can exceed one, but become arbitrary large. This effect is also observed in a numerical solution describing a similar astrophysical situation as the distorted black holes \cite{Ansorg1},\cite{Ansorg2}. A stationary and axisymmetric configuration of a perfect fluid ring rotating around a central black hole was investigated, and it was found that the angular momentum/mass ratio of the black hole can reach $\mid J \mid/M^2> 10^4$, meaning that it is practically unbounded.

The purpose of this work is to construct a five-dimensional Myers-Perry black hole in an external gravitational field and examine its properties. The solution is obtained by applying a two-fold B\"{a}cklund transformation, which was developed by Neugebauer in order to solve the four-dimensional stationary and axisymmetric problem. The transformation is applied on a 5D vacuum Weyl solution as a seed, describing a regular region in spacetime in the presence of a static external distribution of matter. As a result, a 5D stationary and ``axisymmetric''\footnote{ This solution is actually $U(1)\times U(1)$ symmetric. A $d$-dimensional, axisymmetric space-time which admits the $SO(d-2)$ isometry group cannot be considered as an appropriate higher-dimensional generalization of the 4-dimensional Weyl form. Instead one has to consider a $d$-dimensional space-time which admits the $R^{1}\times U(1)\times U(1)$ isometry group.} vacuum solution is obtained, which is rotating only with respect to one of the symmetry axes. It possesses a regular Killing horizon with spherical topology and is asymptotically non-flat. The asymptotically flat Myers-Perry black hole is contained in it as a limiting case. Therefore, the solution is interpreted in the spirit of Geroch and Hartle as describing locally a Myers-Perry black hole in the presence of an external matter field. The solution is also a generalization of the 5D distorted Schwarzschild-Tangherlini black hole obtained in \cite{Abdolrahimi:2010}, which is recovered in the static limit. The physical properties of the distorted Myers-Perry black hole are investigated. The local mass and angular momentum on the horizon are calculated, as well as its temperature and entropy.

The paper is organized as follows. In the next section we review the 5D asymptotically flat Myers-Perry black hole and some of its distinctive features, which are relevant for our work. We also provide a representation of the 5D Myers-Perry solution with single rotation in prolate spheroidal coordinates, since we will use them in the construction of the distorted solution. In section 3, we briefly describe the B\"{a}cklund transformation, which we will apply as a solution generation technique for obtaining the distorted solution. Section 4, is devoted to the actual construction of the distorted Myers-Perry black hole. First, we construct a suitable seed solution, and then we perform a two-fold B\"{a}cklund transformation on it in prolate spheroidal coordinates. The regularity of the solution is analyzed, and appropriate restrictions on the solution parameters are imposed, so that the distorted Myers-Perry black hole is completely regular in the domain of outer communications. In section 5, some physical properties of the solution are investigated. The local mass and angular momentum of the horizon are computed, as well as the temperature and entropy, and a Smarr-like relation is derived. It is demonstrated that the ratio of the horizon angular momentum and mass is not bounded, and can grow unlimitedly. In section 6, the horizon geometry and the behaviour of the ergoregion are analyzed.

\section{ The Myers-Perry solution}

The Myers-Perry solution \cite{Myers} describes a family of black holes with spherical horizon topology in a spacetime with arbitrary dimension $D$, such that $D\geq4$. It is a stationary solution to the Einstein equations in vacuum, meaning that it possesses an asymptotically timelike Killing vector. It is also axisymmetric, in the sense that there exist $N$ spacelike Killing fields, where $N$ is the integer part of $(D-1)/2$, and they correspond to $N$ rotational axes. Therefore, the solution is characterized in general with $N$ independent angular momenta. The Myers-Perry family is the higher-dimensional generalization of the Kerr black hole, which is included as the particular case for $D=4$.

In this work, we consider the Myers-Perry black hole in five dimensional spacetime, which is represented in Boyer-Lindquist coordinates by the metric

\begin{eqnarray}
ds^2 & = & -dt^2 +\Sigma\,\left(\frac{r^2}{\Delta}\,dr^2 + d\theta^2 \right)
+(r^2 + a_1^2)\,\sin^2\theta \,d\phi^2
+(r^2 + a_2^2)\,\cos^2\theta \,d\psi^2 \nonumber \\
&  & +\,\frac{m}{\Sigma}\, \left(dt - a_1\, \sin^2\theta \,d\phi
- a_2\,\cos^2\theta \,d\psi \right)^2\,\,,
\label{metric}
\end{eqnarray}
where
\begin{eqnarray}
\Sigma=r^2+a_1^2 \,\cos^2\theta + a_2^2 \,\sin^2 \theta, \;\;\;\;\;\;
\Delta= (r^2 + a_1^2)(r^2 + a_2^2) -m \, r^2 .
\end{eqnarray}

The timelike Killing field is given by ${\partial}/{\partial t}$, while the spacelike Killing fields are ${\partial}/{\partial\phi}$ and ${\partial}/{\partial\psi}$.  The solution is characterized by three parameters - $m$, and $a_i$, $i = 1,2$. The parameter $m$ is related to the mass of the solution, while $a_1$ and $a_2$ are rotation parameters related to the angular momenta with respect to the two rotational axes $J_\phi$ and $J_\psi$. In general, an event horizon is present, located at the largest positive root $r = r_{\rm H}$ of the function

\begin{eqnarray}
\Delta(r)&= &(r^2 + a_1^2)(r^2 + a_2^2) -m \, r^2 \,\,,
\label{nullsurf}
\end{eqnarray}
which is given explicitly by
\begin{equation}
r_{\rm H}^2 = \frac{1}{2}\,\left(m - a_1^2-a_2^2 +
\sqrt{(m - a_1^2- a_2^2)^2 - 4\,a_1^2\,a_2^2} \right).
\label{hradius}
\end{equation}

The horizon exists only if the solution parameters obey the condition

\begin{equation}\label{horizonE}
a_1^2 + a_2^2 + 2 |a_1\,a_2| \,\leq m\,.
\label{extreme}
\end{equation}
If the function $\Delta(r)$ possesses another positive root $r_I < r_{\rm H}$, the solution contains an inner (Cauchy) horizon, and it can become extremal in the limit when the two horizon radii coincide. The event and the inner horizon are Killing horizons with respect to the Killing field

\begin{equation}
K =\frac{\partial}{\partial_t} + \Omega_\phi \frac{\partial}{\partial_{\phi}} + \Omega_\psi \frac{\partial}{\partial_{\psi}},
\end{equation}
where the constant coefficients $\Omega_\phi$ and $\Omega_\psi$ represent the angular velocities, with which the black hole rotates with respect to the axes of the Killing vectors $\frac{\partial}{\partial_{\phi}}$ and $\frac{\partial}{\partial_{\psi}}$. Their explicit form is given by the expressions

\begin{eqnarray}
\Omega_\phi = \frac{a_1}{(r_{\rm H}^2+a_1^2)} \ ,  \quad~~~ \Omega_\psi = \frac{a_2}{(r_{\rm H}^2+a_2^2)}.
\end{eqnarray}

In the case when the rotation parameters $a_1$ and $a_2$ vanish, we obtain the five-dimensional
Schwarzschild-Tangherlini black hole

\begin{eqnarray}
ds^2 =  -\left(1-\frac{m}{r^2}\right)\,dt^2 + \left(1-\frac{m}{r^2}\right)^{-1}\,dr^2
+ r^2\left(d\theta^2 + \sin^2\theta \,d\psi^2 + \cos^2\theta \,d\phi^2\right). \nonumber\\
\end{eqnarray}

Physically, the five-dimensional Myers-Perry solution is characterized by its ADM mass $M$ and the angular momenta with respect to the two rotational axes $J_\phi$ and $J_\psi$. They can be determined either by examining the asymptotic behavior of the metric functions $g_{tt}$,  $g_{t\phi}$, and $g_{t\psi}$, or equivalently by calculating the corresponding Komar integrals \cite{Komar}. Thus, the following quantities are obtained\footnote{We use geometrical units, i.e. the gravitational constant is set to $G=1$.}

\begin{equation}
M=\frac{3\,\pi}{8}\,m\,\,,~~~~~J_{\phi}=\frac{\pi}{4}\,m\,a_1\,\,,~~~~~
J_{\psi}=\frac{\pi}{4}\,m\,a_2\,.
\label{physical}
\end{equation}

The condition restricting the existence of the horizon $(\ref{horizonE})$ can be expressed in terms of the physical quantities as

\begin{eqnarray}\label{horizonEM}
M^3 \geq \frac{27\pi}{32}\left(J^2_\phi + J^2_\psi + 2|J_\phi\,J_\psi|\right).
\end{eqnarray}

Another form of the Myers-Perry black hole can be obtained by introducing prolate spheroidal coordinates $x$ and $y$ on the two-dimensional surfaces which are orthogonal to the Killing fields. They are particularly convenient for representing the solution in the case when one of the angular momenta, e.g. $J_\phi$, vanishes. Then, the associated rotational parameter $a_1$ vanishes as well, and the prolate spheroidal coordinates are   related to the Boyer-Lindquist coordinates $r$ and $\theta$ by the expressions
\begin{equation}
x=\frac{r^2}{2\sigma}-1, \ y=\cos 2\theta.
\end{equation}
The parameter $\sigma$ is a real number, connected to the mass parameter $m$ and the non-zero rotational parameter $a_2$ of the solution in Boyer-Lindquist coordinates as $4\sigma = m - a^2_2$. The Myers-Perry black hole with a single rotation acquires the form \cite{Mishima}

\begin{eqnarray}\label{Myers-PerryP}
ds^2&=& -\frac{x-1-\alpha^2(1-y)}{x+1+\alpha^2(1+y)}
         \left(dt + 2\sigma^{1/2}\alpha
          \frac{(1+\alpha^2)(1-y)}{x-1-\alpha^2(1-y)} d\psi\right)^2
\nonumber \\[2mm]
       &+& \sigma\frac{(x-1)(1-y)(x+1+\alpha^2(1+y))}{x-1-\alpha^2(1-y)} d\psi^2
       + \sigma(x+1)(1+y)d\phi^2
  \nonumber \\[2mm]
       &+& \frac{\sigma}{2}(x+1+\alpha^2(1+y))
         \left( \frac{dx^2}{x^2-1}+\frac{dy^2}{1-y^2}\right).
\end{eqnarray}
The prolate spheroidal coordinates $x$ and $y$ take the ranges $x\geq 1$, and $-1\leq y \leq 1$. The black hole horizon is located at $x=1$, while the axes of the spacelike Killing fields $\frac{\partial}{\partial\phi}$ and $\frac{\partial}{\partial\psi}$ are located at $y=-1$ and $y=1$ respectively. The physical infinity corresponds to the limit $x\rightarrow \infty$. The solution is characterized by the parameters $\sigma$ and $\alpha$, as the former has the meaning of a mass parameter, while the later is interpreted as a rotational parameter. It is related to the parameter set of the solution $(m, a_2)$ in Boyer-Lindquist coordinates as $\alpha^2 = a^2_2/(m-a^2_2)$. The mass $M$, the angular momentum $J$ and the angular velocity $\Omega$ of the solution are given in terms of the parameters $\{\sigma, \alpha\}$ as

\begin{eqnarray}\label{M_MP}
M=\frac{3\,\pi}{2}\,\sigma (1+\alpha^2), \quad~~~J = 2\pi\sigma^{\frac{3}{2}}\alpha (1+\alpha^2),\quad~~~  \Omega = \frac{\alpha}{2\sqrt{\sigma}(1+\alpha^2)}.
\end{eqnarray}

In the limit when the rotation parameter $\alpha$ vanishes, the solution reduces to the five-dimensional Schwarzschild-Tangherlini black hole in the prolate spheroidal coordinates

\begin{eqnarray}
ds^2&=& -\frac{x-1}{x+1}dt^2
       + \sigma(1-y)(x+1) d\psi^2
       + \sigma(x+1)(1+y)d\phi^2
  \nonumber \\[2mm]
       &+& \frac{\sigma}{2}(x+1)
         \left( \frac{dx^2}{x^2-1}+\frac{dy^2}{1-y^2}\right).
\end{eqnarray}

\section{Generation of stationary axisymmetric solutions by means of B\"{a}cklund transformations}

The construction of analytic solutions to the 5D stationary and axisymmetric Einstein equations in vacuum is comparatively well studied. It is proven that the problem is completely integrable \cite{Maison:1978}-\cite{Maison:1979a}, an associated linear problem is constructed, and a solution generation technique by means of the inverse scattering method is developed \cite{BZ1}-\cite{BZ2}. In the particular case when the solution is rotating only with respect to a single axis, i.e. one of the spacelike Killing fields is hypersurface orthogonal, a B\"{a}cklund transformation is also obtained, which can be applied to generate a new solution. In this section we will briefly describe the method of construction of solutions by using the B\"{a}cklund transformation found by Neugebauer \cite{NG2}-\cite{Neugebauer:1980}. The B\"{a}cklund transformations relate different solutions of a partial differential equation in an algebraic way. Thus, applied on an already known solution, called a seed, they lead to a new solution with minimal analytic operations. Neugebauer's transformation was originally derived in order to solve the four-dimensional stationary and axisymmetric problem. However, it can be also applied to the five-dimensional stationary and axisymmetric Einstein equations in vacuum, when one of the spacelike Killing fields is hypersurface orthogonal \cite{Mishima}-\cite{Nedkova:2010}, since the corresponding field equations closely resemble the four-dimensional case.

The general stationary and axisymmetric solution to the five-dimensional Einstein equations, which rotates only in a single plane, can be represented by the metric

\begin{eqnarray}\label{metricG}
ds^2 &=&-e^{2\chi-u}\left(dt- \omega d\psi\right)^2
       +e^{-2\chi-u}\rho^2d\psi^2
+e^{-2\chi-u}e^{2\Gamma}\left(d\rho^2+dz^2\right)
  +e^{2u}d\phi^2.
\end{eqnarray}
The asymptotically timelike Killing field is represented as ${\partial}/{\partial t}$, and the spacelike Killing fields are given by ${\partial}/{\partial\phi}$ and ${\partial}/{\partial\psi}$. The two-dimensional surfaces orthogonal to the Killing fields are parameterized by the Weyl coordinates $\rho$ and $z$, and all the metric functions depend only on them.  The Einstein equations in vacuum determining such a solution consist of a non-linear system of equations for the metric function $\chi$ and the twist potential $f$ defined as

\begin{eqnarray}\label{twist}
\partial_\rho f = - \frac{e^{4\chi}}{\rho}\partial_z\omega, \quad ~~~ \partial_z f = \frac{e^{4\chi}}{\rho}\partial_\rho\omega,
\end{eqnarray}
a Laplace equation for the metric function $u$, and a decoupled linear system for the remaining metric function $\Gamma$. It is always integrable for a particular solution $(\chi, u, f)$. In analogy to the four-dimensional case we can introduce an Ernst potential ${\cal E}$ defined as

\begin{equation}\label{Ernst}
{\cal E} = e^{2\chi} + if,
\end{equation}

\noindent
and describe the problem by means of ${\cal E}$ and its complex conjugate $\bar{{\cal E}}$ instead of the couple of functions $(\chi, f)$. The field equations acquire the form

\begin{eqnarray}
&&\left({\cal E} + \bar{{\cal E}}\right)\left(\partial^2_{\rho} {\cal E} + \rho^{-1} \partial_{\rho}{\cal E}
+ \partial^2_{z} {\cal E}\right)= 2 \left(\partial_{\rho}{\cal E}\partial_{\rho}{\cal E}
+ \partial_{z}{\cal E}\partial_{z}{\cal E} \right), \nonumber \\[2mm]
&&\partial^2_{\rho} u + \rho^{-1} \partial_{\rho}u
+ \partial^2_{z} u = 0,
\end{eqnarray}

\begin{eqnarray}\label{Gammalin}
&&\rho^{-1}\partial_{\rho}\Gamma= {1\over \left({\cal E} + \bar{{\cal E}} \right)^2}
\left[\partial_{\rho}{\cal E}\partial_{\rho}\bar{{\cal E}} - \partial_{z}{\cal E}\partial_{z}\bar{{\cal E}} \right] + \frac{3}{4}\left[\left(\partial_{\rho}u\right)^2 - \left(\partial_{z}u \right)^2\right],\nonumber \\[2mm]
 &&\rho^{-1}\partial_{z}\Gamma=  \frac{2}{\left({\cal E} + \bar{{\cal E}}\right)^2} \partial_{\rho}{\cal E} \partial_{z}\bar{{\cal E}} + \frac{3}{2}~ \partial_{\rho}u \partial_{z}u.\nonumber
\end{eqnarray}

Similar to the four-dimensional case we obtain a non-linear equation for the Ernst potential, called Ernst equation, and it represents the main difficulty for solving the system. The solutions to the Laplace equation are well-studied, and provided the Ernst potential and the metric function $u$ are known, the metric function $\Gamma$ can be obtained in a straightforward (although technically cumbersome) way.

Solutions to the Ernst equation can be constructed by using the B\"{a}cklund transformation developed by Neugebauer. The B\"{a}cklund transformation relates a new solution for the Ernst potential ${\cal E}$ to an already known solution ${\cal E}_0$, called a seed, in an algebraic way, after performing minimal analytic operations. In some cases the solution which we try to obtain is related to a particular seed solution by a sequence of $N$ subsequent B\"{a}cklund transformations, referred to as an $N$-fold B\"{a}cklund transformation. The B\"{a}cklund transformation is determined by a couple of functions ${\cal \alpha}$ and $\lambda$ which are solutions to the following system of Riccati equations \cite{NG}

\begin{eqnarray}\label{Riccati}
d\lambda &=& \rho^{-1}(\lambda -1)\left[\lambda \rho_{, \zeta}d\zeta + \rho_{, \bar\zeta}d\bar\zeta\right], \nonumber \\[2mm]
d{\cal \alpha} &=& \left({\cal E}_0 + \bar{{\cal E}}_0 \right)^{-1}\left[({\cal \alpha} - \lambda^{1/ 2})\bar{{\cal E}}_{0,\zeta}+ ({\cal \alpha}^2\lambda^{1/2}-{\cal \alpha}){\cal E}_{0,\zeta}\right]d\zeta + \nonumber \\
&&\left({\cal E}_0 + \bar{{\cal E}}_0 \right)^{-1}\left[({\cal \alpha} - \lambda^{-1/2})\bar{{\cal E}}_{0,\bar\zeta}+ ({\cal \alpha}^2\lambda^{-1/2}-{\cal \alpha}){\cal E}_{0,\bar\zeta}\right]d\bar\zeta,
\end{eqnarray}
where ${\cal E}_0$ is the Ernst potential for the seed solution, $\zeta = \rho + iz$, the bar denotes complex conjugation, and $(...),$ denotes differentiation.
The equation for $\lambda$ is solved by the function
\begin{eqnarray}\label{gamma}
\lambda &=& \frac{k - i\bar\zeta}{k + i\zeta},
\end{eqnarray}
where $k$ is a real integration constant. The equation for ${\cal \alpha}$ depends on the explicit form of the Ernst potential for the seed solution ${\cal E}_0$ and its complex conjugate $\bar{{\cal E}_0}$. An important class of seed solutions, which is convenient for the applications,  is the Weyl class of static axisymmetic solutions to the 5D Einstein equations in vacuum \cite{Emparan:2002}. They are described by the line element $(\ref{metricG})$ in the particular case when we require that the twist potential $f$ vanishes, and the Ernst potential corresponding to them is real ${\cal E}_0 = e^{2\chi_0}$. The second Riccati equation for such a seed is solved by the function
\begin{eqnarray}\label{alpha}
{\cal \alpha} &=& {\mu+ ie^{2\Phi}\over \mu - ie^{2\Phi}},
\end{eqnarray}
where $\mu$ is a real integration constant, and $\Phi$ obeys the equation
\begin{eqnarray}\label{Riccatiphi}
d\Phi = \frac{1}{2} \lambda^{1/2}~ \partial_{\zeta}(\ln{{\cal E}_0})d\zeta + \frac{1}{2} \lambda^{-1/2}~\partial_{\bar\zeta}(\ln{{\cal E}_0})d\bar\zeta.
\end{eqnarray}

A single B\"{a}cklund transformation performed on a seed solution ${\cal E}_0$ requires the integration of $(\ref{Riccati})$, which introduces a couple of integration constants $(k_1, \mu_1)$. Performing $N$ subsequent B\"{a}cklund transformations on a seed solution ${\cal E}_0$ includes solving the same Riccati equations, but each iteration introduces a new pair of integrations constants $(k_n, \mu_n)$, $n=1...N$.  The Ernst potential obtained by the application of $2N$ B\"{a}cklund transformations on a seed potential ${\cal E}_0$ is constructed in the form \cite{NG}, \cite{SK1}

\begin{eqnarray}\label{ErnstWeyl}
{\cal E}= {\cal E}_0 {\det\left({{\cal \alpha}_{p}R_{k_p} - {\cal \alpha}_{q}R_{k_q} \over k_p - k_q } -1 \right)\over
\det\left({{\cal \alpha}_{p}R_{k_p} - {\cal \alpha}_{q}R_{k_q} \over k_p - k_q } +1 \right)},
\end{eqnarray}
where $p=1,3,...,N-1$, $q=2,4,...,N$, ${\cal \alpha}_n$ is the solution to ($\ref{alpha}$) corresponding to the integration constants $(k_n, \mu_n)$, and the functions $R_{k_n}$ are given by

\begin{eqnarray}
R_{k_n}= \sqrt{\rho^2 + (z-k_n)^2}.
\end{eqnarray}

Having obtained the Ernst potential for the new solution, the metric functions $\chi$ and $\omega$ can be extracted from it by considering its definition $(\ref{Ernst})$, and integrating the relations for the twist potential $(\ref{twist})$. In the case of a 2-fold B\"{a}cklund transformation their explicit form is found to be
\begin{eqnarray}\label{chi_Back}
e^{2\chi}&=&e^{2\chi_0}\frac{W_1}{W_2} ,  \label{e^S} \\ \nonumber
\omega&=&e^{-2\chi_0}\frac{\hat{\omega}}{W_1} +  C_\omega \label{omega} ,    \\ \nonumber
\end{eqnarray}
introducing the following notations
\begin{eqnarray}\label{W_Back}
W_1&=&\left[(R_{k_1}+R_{k_2})^2-(\Delta k)^2\right] (1+ a b)^2  + \left[(R_{k_1}-R_{k_2})^2-(\Delta k)^2\right](a-b)^2  , \nonumber \\
W_2&=&\left[(R_{k_1}+R_{k_2}+\Delta k)+(R_{k_1}+R_{k_2}-\Delta k)a b \right]^2 \nonumber \\
 && +\left[(R_{k_1}-R_{k_2}-\Delta k)a - (R_{k_1}-R_{k_2}+\Delta k)b \right]^2 , \\ \nonumber  \\
{\hat \omega}&=& [(R_{k_1} + R_{k_2})^2-(\Delta k)^2](1+a b)\left[(R_{k_1}-R_{k_2} + \Delta k)b +
(R_{k_1}-R_{k_2} - \Delta k)a\right] \nonumber \\&& \,- [(R_{k_1} - R_{k_2})^2-(\Delta k)^2](b-a)
\left[(R_{k_1} + R_{k_2} + \Delta k) - (R_{k_1} + R_{k_2} - \Delta k)ab\right],  \nonumber
\end{eqnarray}
where $\Delta k = k_2-k_1$ and $C_\omega$ is a constant. The functions $a$ and $b$ are related to the functions $\Phi_1$ and $\Phi_2$, which  represent the solutions of $(\ref{Riccatiphi})$ with integration constants $k_1$ and $k_2$, respectively
\begin{eqnarray}
a = \mu_1^{-1}e^{2\Phi_1}, \quad  \bigskip b = -\mu_2 e^{-2\Phi_2}.
\end{eqnarray}

\noindent
In this way if we consider a 5D Weyl solution as a seed, which is described by the metric
\begin{eqnarray}\label{metric_seed_Back}
ds^2 &=&-e^{2\chi_0-u_0}dt^2
       +e^{-2\chi_0-u_0}\rho^2d\psi^2
+e^{-2\chi_0-u_0}e^{2\Gamma_0}\left(d\rho^2+dz^2\right)
  +e^{2u_0}d\phi^2, \nonumber
\end{eqnarray}

\noindent
we can construct a 5D stationary axisymmetric solution with single rotation in the form

\begin{eqnarray}\label{metric_BackG}
ds^2 &=&-\frac{W_1}{W_2}e^{2\chi_0-u_0}\left(dt- \omega d\psi\right)^2
       + \frac{W_2}{W_1}e^{-2\chi_0-u_0}\rho^2d\psi^2 +e^{2u_0}d\phi^2 \nonumber \\
&+& \frac{W_2}{W_1}e^{2\Gamma}e^{-2\chi_0-u_0}\left(d\rho^2+dz^2\right),
\end{eqnarray}

\noindent
by applying a 2-fold B\"{a}cklund transformation on it. The functions $W_1$, $W_2$, and $\omega$ are given by the expressions $(\ref{chi_Back})$-$(\ref{W_Back})$, and the only analytical operation involved is solving eq. $(\ref{Riccatiphi})$ for the particular Ernst potential of the seed solution. The metric function $u$ satisfies a Laplace equation in Weyl coordinates, as well as the corresponding metric function for the seed solution $u_0$. Therefore, it is convenient to preserve the metric function for the seed solution, and choose such a seed that $u_0$ coincides with the metric function $u$ for the solution we want to construct. The remaining metric function $\Gamma$ can be expressed in the form \cite{Yamazaki:1981}, \cite{Dietz:1982}

\begin{eqnarray}\label{Gamma_Back}
e^{2\Gamma}= C_1 {W_1 e^{2\gamma}\over \left(R_{k_{1}} +  R_{k_{2}}\right)^2 - (\Delta k)^2 },
\end{eqnarray}
where $C_1$ is an integration constant and $\gamma$ is a solution to the linear system

\begin{eqnarray}\label{gamma_Back}
&&\rho^{-1}\partial_{\rho}\gamma = \left(\partial_{\rho}{\tilde \chi_{0}}\right)^2
- \left(\partial_{z}{\tilde \chi_{0}}\right)^2 + \frac{3}{4}\left[\left(\partial_{\rho}{u_0}\right)^2
- \left(\partial_{z}{u_0}\right)^2\right] ,\\ \nonumber \\
&&\rho^{-1}\partial_{z}\gamma = 2 \partial_{\rho}{\tilde \chi_{0}} \partial_{z}{\tilde \chi_{0}} + \frac{3}{2} \partial_{\rho}{u_0} \partial_{z}{u_0}.\nonumber
\end{eqnarray}

The function ${\tilde\chi}_{0}$, which is involved in it, is related  to the metric function of the seed solution $\chi_0$ as

\begin{eqnarray}
{\tilde \chi}_{0} = \chi_{0} +{1 \over 2} \ln{R_{k_1}+ R_{k_2} - \Delta k\over R_{k_1} + R_{k_2} +\Delta k}.
\end{eqnarray}

\section{Construction of 5D distorted Myers-Perry black hole}

In this section, we will construct a five-dimensional distorted Myers-Perry black hole rotating only in a single plane by applying a 2-fold B\"{a}cklund transformation. The solution is an asymptotically non-flat generalization of the 5D Myers-Perry black hole presented in section 2, and can be interpreted as a local solution describing a black hole in the presence of an external distribution of matter fields. The 2-fold B\"{a}cklund transformation involves the integration of the Riccati equation $(\ref{Riccatiphi})$, and is parameterized by two couples of integration constants ($k_1$, $k_2$), and ($\mu_1$, $\mu_2$), introduced by the double integration of eqs. $(\ref{Riccati})$. The integration constants $k_1$ and $k_2$ are actually not independent. The Weyl coordinate $z$ is defined only up to a translation, therefore we can always set $k_1 = \sigma$ and $k_2 = -\sigma$ for some real positive parameter $\sigma$. This identification corresponds to a shift in the $z$ coordinate $z\rightarrow z + z_0$, where $z_0 = \frac{1}{2}(k_1 + k_2)$. Furthermore, we expect that the distorted Myers-Perry solution will be represented most conveniently in prolate spheroidal coordinates, in analogy with the distorted Kerr black hole \cite{Breton:1997}. The prolate spheroidal coordinates $x$ and $y$ are closely related to the Weyl coordinates $\rho$ and $z$. They are defined by the transformation

\begin{eqnarray}
\rho = \kappa\sqrt{(x^2-1)(1-y^2)}, \quad~~~ z=\kappa xy,
\end{eqnarray}
\noindent
where $\kappa$ is a real constant. As we already mentioned, the prolate spheroidal coordinates take the ranges $x\geq 1$, and $-1\leq y \leq 1$, and the physical infinity corresponds to the limit $x\rightarrow \infty$. We can always set $\kappa$ equal to the parameter of the B\"{a}cklund transformation $\sigma$, which will simplify the form of the constructed solution. In order to obtain a solution in prolate spheroidal coordinates by means of the 2-fold B\"{a}cklund transformation we described, we should transform the metric functions $W_1$, $W_2$ and $\omega$ in terms of $x$ and $y$, and represent the differential equations $(\ref{Riccatiphi})$ and $(\ref{gamma_Back})$, which we need to solve, in the same coordinates. Consequently, the general solution $(\ref{metric_BackG})$ acquires the form

\begin{eqnarray}\label{metric_BackP}
ds^2 &=&-\frac{W_1}{W_2}e^{2\chi_0-u_0}\left(dt- \omega d\psi\right)^2
       + \frac{W_2}{W_1}e^{-2\chi_0-u_0}\rho^2d\psi^2 +e^{2u_0}d\phi^2 \nonumber \\[2mm]
&+& C_1\frac{W_2e^{2\gamma}}{x^2-1}e^{-2\chi_0-u_0}(x^2 - y^2)\left(\frac{dx^2}{x^2-1}+\frac{dy^2}{1-y^2}\right)
  , \nonumber \\[2mm]
\omega&=&2\sigma e^{-2\chi_0}\frac{\hat{\omega}}{W_1} +  C_\omega ,
\end{eqnarray}
where the zero indices refer to the metric functions of the seed solution, and the metric functions $W_1$, $W_2$ and $\hat{\omega}$ are determined by the expressions

\begin{eqnarray}\label{W}
W_1&=&(x^2-1)(1+ab)^2-(1-y^2)(b-a)^2 , \nonumber \\[1mm]
W_2&=&[(x+1)+(x-1)ab]^2+[(1+y)a+(1-y)b]^2 \,, \nonumber \\[1mm]
\hat{\omega}&=&(x^2-1)(1+ab)[b-a-y(a+b)]  \nonumber \\[1mm]
&+& (1-y^2)(b-a)[1+ab+x(1-ab)],
\end{eqnarray}
with $C_1$ and  $C_\omega$ being real constants. Instead of representing the differential equation for $\Phi$ $(\ref{Riccatiphi})$ in term of $x$ and $y$, it is more convenient to obtain directly the equations for the related metric functions $a$ and $b$ in prolate spheroidal coordinates.  They possess the form \cite{Amanedo}
\begin{eqnarray}\label{ab}
(x-y)\partial_x a&=&
2a\left[(xy-1)\partial_x\chi_0+(1-y^2)\partial_y \chi_0\right],  \\[1mm]
(x-y)\partial_y a&=&
2a\left[-(x^2-1)\partial_x \chi_0+(xy-1)\partial_y \chi_0\right], \nonumber \\[1mm]
(x+y)\partial_x b&=&
-2b\left[(xy+1)\partial_x \chi_0+(1-y^2)\partial_y \chi_0\right] , \nonumber\\[1mm]
(x+y)\partial_y b&=&
-2b\left[-(x^2-1)\partial_x \chi_0+(xy+1)\partial_y \chi_0\right], \nonumber \\ \nonumber
\end{eqnarray}
and depend as expected on the metric function of the seed solution $\chi_0$. The equations for the remaining metric function $\gamma$ are transformed into the following system

\begin{eqnarray}\label{gamma2}
\partial_x\gamma &=& \frac{1-y^2}{(x^2-y^2)}\left[x(x^2-1)(\partial_x\chi')^2 - x(1-y^2)(\partial_y\chi')^2 - 2y(x^2-1)\partial_x\chi'\partial_y\chi'\right]  \nonumber \\[1mm]
&+& \frac{3(1-y^2)}{4(x^2-y^2)}\left[x(x^2-1)(\partial_x u_0)^2 - x(1-y^2)(\partial_y u_0)^2 - 2y(x^2-1)\partial_x u_0\partial_y u_0\right],\nonumber \\[2mm]
\partial_y\gamma &=& \frac{x^2-1}{(x^2-y^2)}\left[y(x^2-1)(\partial_x\chi')^2 - y(1-y^2)(\partial_y\chi')^2 + 2x(1-y^2)\partial_x\chi'\partial_y\chi'\right] \nonumber \\[1mm]
&+& \frac{3(x^2-1)}{4(x^2-y^2)}\left[y(x^2-1)(\partial_x u_0)^2 - y(1-y^2)(\partial_y u_0)^2 + 2x(1-y^2)\partial_x u_0\partial_y u_0\right],
\end{eqnarray}

\noindent
where $\chi' = \frac{1}{2}\ln\left(\frac{x-1}{x+1}\right) + \chi_0$.

\subsection{Seed solution}
A non-trivial step in the construction of solutions by means of B\"{a}cklund transformations is choosing the seed solution. As we mentioned, a convenient class of seed solutions is the static Weyl class, since they are comparatively simple and minimize the technical difficulties. A general 5D Weyl solution in prolate spheroidal coordinates possesses the form

\begin{eqnarray}\label{WeylPS}
ds^2 &=&-e^{2\chi_0-u_0}dt^2 + e^{-2\chi_0-u_0}\rho^2d\psi^2 +e^{2u_0}d\phi^2 \nonumber \\[2mm]
&+& e^{2\gamma}e^{-2\chi_0-u_0}(x^2 - y^2)\left(\frac{dx^2}{x^2-1}+\frac{dy^2}{1-y^2}\right),
\end{eqnarray}
where all the metric functions depend only on $x$ and $y$, and the metric functions $\chi_0$ and $u_0$ obey a Laplace equation in 3D flat space. The most general solution to the Laplace equation, which is regular on the symmetry axes, can be presented in the form \cite{Quevedo:1989}, \cite{Quevedo:1990}

\begin{eqnarray}\label{Laplace}
\chi_0 = \sum^{\infty}_{n=0}\frac{c_n}{R^{n+1}}P_n\left(\frac{xy}{R}\right) + \sum^{\infty}_{n=0}b_nR^nP_n\left(\frac{xy}{R}\right),
\end{eqnarray}
and equivalently for $u_0$. The parameters $c_n$ and $b_n$ are real constants, $n$ is a natural number, $P_n$ are the Legendre polynomials, and the function $R$ is defined by

\begin{equation}
R = \sqrt{x^2 +  y^2 - 1}.
\end{equation}
The first part of the sum in the solution of the Laplace equation corresponds to an asymptotically flat solution of the static Einstein equations in vacuum. It describes a deformed mass source, and the parameters $c_n$ are related to the mass multipole moments of the solution. The second part of the sum corresponds to an asymptotically non-flat solution, which is interpreted as a local solution in an external gravitational field, and the constants $b_n$ are related to the characteristics of the external field. The 2-fold B\"{a}cklund transformation preserves the asymptotic structure of the seed solution. Therefore, if we want to obtain a solution describing a black hole in an external gravitational field, we should choose a seed with non-zero constants $b_n$. A suitable seed solution is a regular Weyl solution describing a static external distribution of matter, which contains no horizons, and in the limit when the external gravitational fields vanish reduces to the 5D Minkowski spacetime. The 5D Minkowski spacetime can be represented in prolate spheroidal coordinates by the metric

\begin{eqnarray}\label{Minkovski}
ds^2 &=& -dt^2 + e^{-2W_0}d\psi^2 + e^{-2U_0}d\phi^2 + \frac{\sigma}{2}(x-y)\left(\frac{dx^2}{x^2-1}+\frac{dy^2}{1-y^2}\right), \nonumber \\[2mm]
e^{-2W_0} &=& \sigma(x-1)(1-y), \quad ~~~ e^{-2U_0} = \sigma(x+1)(1+y).
\end{eqnarray}
The functions $U_0$ and $W_0$ are solutions to the Laplace equation in 3D flat space, and in Weyl coordinates their sum satisfies $U_0 + W_0 = \ln\rho$, as required for a 5D Weyl solution \cite{Emparan:2002}. If we replace the functions $U_0$ and $W_0$ by asymptotically non-flat solutions of the Laplace equation, and calculate the remaining metric function $\gamma$, we can obtain an asymptotically non-flat Weyl solution which contains no horizons\footnote{ The metric function $\gamma$ is a solution to ($\ref{gamma2}$) with $\chi'=\chi_0$, where $\chi_0$ and $u_0$ are the metric functions appearing in the line element ($\ref{WeylPS}$).}. It can be interpreted as describing locally a vacuum region in spacetime which is influenced by the presence of some static and axisymmetric  distribution of matter situated in its exterior. Therefore, it can be called distorted Minkowski spacetime in analogy with the distorted black hole solution. For general static and axisymmetric external gravitational fields the metric of the distorted Minkowski spacetime acquires the form

\begin{eqnarray}\label{Minkovski_dist}
ds^2 &=& -e^{2(\widehat{U} + \widehat{W})} dt^2 + e^{-2W}d\psi^2 + e^{-2U}d\phi^2   \nonumber \\[2mm]
&+&  \frac{\sigma}{2}(x-y)e^{2V}\left(\frac{dx^2}{x^2-1}+\frac{dy^2}{1-y^2}\right),
\end{eqnarray}

\noindent
where the functions $U = U_0 + \widehat{U}$ and $W=W_0 + \widehat{W}$ include a contribution from the non-distorted Minkowski metric ($\ref{Minkovski}$) denoted with zero index, and terms $\widehat{U}$ and $\widehat{W}$ characterizing the external sources. The most general form of $\widehat{U}$ and $\widehat{W}$ is given by the expressions

\begin{eqnarray}\label{UW_dist}
\widehat{U} &=& \sum^{\infty}_{n=0}a_nR^n P_n\left(\frac{xy}{R}\right), \quad ~~~
\widehat{W} = \sum^{\infty}_{n=0}b_nR^n P_n\left(\frac{xy}{R}\right), \\[1mm]
\end{eqnarray}
according to the solution of the Laplace equation $(\ref{Laplace})$.  The metric function $V$ is obtained  in the form

\begin{eqnarray}
V &=& \sum^{\infty}_{n,k=1}\frac{nk}{n+k}\left(a_na_k + a_nb_k + b_nb_k\right)R^{n+k}\left(P_n P_k - P_{n-1}P_{k-1}\right) \nonumber \\
&+& \frac{1}{2}\sum^{\infty}_{n=1}(a_n - b_n)\sum^{n-1}_{k=0}(-1)^{n-k+1}(x+y)R^k P_k\left(\frac{xy}{R}\right) \nonumber \\
&-&\frac{1}{2}\sum^{\infty}_{n=0}(a_n + b_n)R^n P_n\left(\frac{xy}{R}\right).
\end{eqnarray}

For general values of the parameters $a_n$ and $b_n$ characterizing the external gravitational field the solution is not completely regular, since it contains conical singularities. They can be removed by requiring that the solution is Lorenzian (elementary flat) in the vicinity of the axes of the spacelike Killing fields $\partial/\partial\phi$ and $\partial/\partial\psi$ located at $y=-1$ and $y=1$, respectively. The condition ensuring elementary flatness is given by the relation \cite{Mars}, \cite{SK1}

\begin{eqnarray}\label{conicalG0}
\frac{1}{4X}g^{\mu\nu}\partial_\mu X\partial_\nu X \longrightarrow 1,
\end{eqnarray}
which should be satisfied by the norm $X$ of each of the spacelike Killing fields in the neighbourhood of their axes. It leads to the following constraints

\begin{eqnarray}
&&\exp\left[-\sum^{\infty}_{n=0}(-1)^n(a_n - b_n)\right] = 1, \quad~~~~~ y = 1, \nonumber \\[2mm]
&&\exp\left[\sum^{\infty}_{n=0}(-1)^n(a_n - b_n)\right] = 1, \quad~~~~~~ y=-1.
\end{eqnarray}

Therefore, the distorted Minkowski solution is free of conical singularities only if the parameters characterizing the external gravitational field satisfy the relation $\sum^{\infty}_{n=0}(-1)^n(a_n - b_n) = 0$.

\subsection{Distorted 5D Myers-Perry black hole}

We will apply the 2-fold B\"{a}cklund transformation  on the distorted Minskowski spacetime ($\ref{Minkovski_dist}$) as a background. The transformation can be performed with respect to either of the spacelike Killing fields, leading to a solution rotating around its axis. Since we aim to construct a Myers-Perry black hole, which is axially symmetric, both approaches will result in equivalent solutions. Here we choose to perform the transformation with respect to the Killing field $\partial/\partial\psi$. Then, comparing the metric of the distorted Minkowski spacetime with the general expression ($\ref{WeylP}$) we conclude that the functions $\chi_0$ and  $u_0$ are given   by

\begin{eqnarray}
2\chi_0 =  -U_0 + 2\widehat{W} + \widehat{U}, \quad~~~ u_0 = -U_0 - \widehat{U}.
\end{eqnarray}

Taking advantage of these expressions we solve equations ($\ref{ab}$) and obtain the functions $a$ and $b$ in a similar way as in \cite{Breton:1997}

\begin{eqnarray}
a &=& \alpha\,\sqrt{\frac{x+1}{y+1}}\exp\left[\sum^{\infty}_{n=1}(a_n + 2b_n)\sum^{n-1}_{k=0}(x-y)R^kP_k\left(\frac{xy}{R}\right)\right], \\[2mm]
b &=& \beta\,\frac{\sqrt{(x+1)(y+1)}}{x+y}\exp\left[\sum^{\infty}_{n=1}(a_n + 2b_n)(x+y)\sum^{n-1}_{k=0}(-1)^{n-k}R^kP_k\left(\frac{xy}{R}\right)\right], \nonumber
\end{eqnarray}

\noindent
where $\alpha$ and $\beta$ are real constants\footnote{ The parameters $\alpha$ and $\beta$ are the equivalent of the parameters of the B\"{a}cklund transformation $\mu_1$ and $\mu_2$, which we introduced before. They are related as $\alpha = \mu_1^{-1}$ and $\beta = -\mu_2$. }. The equations for the metric function $\gamma$ can be also solved in resemblance to \cite{Breton:1997}, leading  to the following result

\begin{eqnarray}\label{gamma_dist}
\gamma &=& \gamma_0 + \hat{\gamma}, \nonumber \\
\gamma_0 &=& \frac{1}{2}\ln(x^2-1) - \frac{1}{2}\ln(x^2-y^2) +  \frac{1}{2}\ln(y+1), \nonumber \\
\hat{\gamma} &=& \sum^{\infty}_{n,k=1}\frac{nk}{n+k}\left(a_na_k + a_nb_k + b_nb_k\right)R^{n+k}\left(P_n P_k - P_{n-1}P_{k-1}\right) \nonumber \\
&+& \frac{1}{2}\sum^{\infty}_{n=1}(2a_n+b_n)\sum^{n-1}_{k=0}(-1)^{n-k+1}(x+y)R^k P_k\left(\frac{xy}{R}\right) \nonumber \\
&-& \frac{1}{2}\sum^{\infty}_{n=1}(a_n+2b_n)\sum^{n-1}_{k=0}(x-y)R^k P_k\left(\frac{xy}{R}\right) \nonumber \\
&-&\frac{1}{2}\sum^{\infty}_{n=0}(a_n - b_n)R^n P_n\left(\frac{xy}{R}\right),
\end{eqnarray}

\noindent
where all the Legendre polynomials have the same argument $xy/R$.

These functions determine completely a new stationary and axisymmetric solution to the 5D Einstein equations in vacuum with metric given by ($\ref{metric}$). The constructed solution possesses an event horizon with $S^3$ topology located at $x=1$, $-1\leq y\leq 1$, which rotates with respect to the axis  of the Killing field $\partial/\partial\psi$. The axes of the spacelike Killing fields $\partial/\partial\phi$ and $\partial/\partial\psi$ correspond to $y=-1$ and $y=1$, respectively. The solution is not regular outside the black hole horizon for general values of the transformation parameters, since the metric function $W_1$ gets singular at $x=1$, $y=-1$. This pathological feature can be avoided if we set the parameter $\beta$ equal to zero. Then,  the function $b$ vanishes and the explicit form of the solution simplifies considerably. It can be presented as

\begin{eqnarray}
ds^2&=& -\frac{x-1-\hat{a}^2(1-y)}{x+1+\hat{a}^2(1+y)}e^{2(\widehat{U} + \widehat{W})}
         \left(dt -\omega d\psi\right)^2
       +\frac{x+1+\hat{a}^2(1+y)}{x-1-\hat{a}^2(1-y)}e^{-2W} d\psi^2
\nonumber \\[2mm]
    &+& e^{-2U}d\phi^2 + C_1\left[x+1+\hat{a}^2(1+y)\right]e^{2(\hat{\gamma} - \widehat{W})}
         \left( \frac{dx^2}{x^2-1}+\frac{dy^2}{1-y^2}\right), \label{metric_dist} \\
\omega &=& -2\sqrt{\sigma}\frac{(x-y)\hat{a}e^{-2\widehat{W}-\widehat{U}}}{(x-1)-(1-y)\hat{a}^2} + C_\omega, \label{omega_dist} \\[2mm]
\hat{a} &=& \alpha \exp{\left[\sum^{\infty}_{n=1}(a_n+2b_n)\sum^{n-1}_{k=0}(x-y)R^k P_k\left(\frac{xy}{R}\right)\right]},
\end{eqnarray}
where the metric functions $U$ and $W$ correspond to the seed solution ($\ref{Minkovski_dist}$), and $\hat{\gamma}$ is given by the expression ($\ref{gamma_dist}$). The solution contains two integration constants $C_1$ and $C_\omega$, which should be chosen appropriately in order to avoid pathological behavior. The value of $C_\omega$ should be determined by the requirement that the solution is regular on the axis of rotation, i.e. it should be satisfied that $\omega = 0$ on the axis $y=1$. Examining the behavior of the function $(\ref{omega_dist})$ at $y=1$ we obtain

\begin{eqnarray}\label{C_omega}
C_\omega = 2\sqrt{\sigma}\alpha\exp{\left[-\sum^\infty_{n=0}(a_n + 2b_n)\right]}.
\end{eqnarray}

The value of the constant $C_1$ is connected with the requirement that the solution is Lorentzian (elementary flat) in the vicinity of the axes of the spacelike Killing fields located at $y=-1$ and $y=1$. Otherwise, the solution will contain conical singularities. The absence of conical singularities is ensured if the norm $X$ of each of the spacelike Killing fields $\partial/\partial\phi$ and $\partial/\partial\psi$ satisfies the condition \cite{Mars}, \cite{SK1}

\begin{eqnarray}\label{conicalG}
\frac{1}{4X}g^{\mu\nu}\partial_\mu X\partial_\nu X \longrightarrow 1,
\end{eqnarray}
in the vicinity of the corresponding rotational axis. It is equivalent to requiring that the orbits of the Killing fields are $2\pi$-periodic in the neighbourhood of their axes. Evaluating $(\ref{conicalG})$  at $y=1$ leads to the constraint

\begin{eqnarray}
\frac{2C_1}{\sigma}\exp{\left[\sum^\infty_{n=0}(b_{2n} - a_{2n}) + 3\sum^\infty_{n=0}(a_{2n+1} + b_{2n+1}) \right]} = 1,
\end{eqnarray}
while at $y=-1$ we obtain the relation
\begin{eqnarray}
\frac{2C_1}{\sigma}\exp{\left[-\sum^\infty_{n=0}(b_{2n} - a_{2n}) -3\sum^\infty_{n=0}(a_{2n+1} + b_{2n+1}) \right]} = 1.
\end{eqnarray}
Both conditions are compatible only if the parameters $a_n$ and $b_n$ characterizing the external gravitational field satisfy
\begin{eqnarray}\label{conical}
\sum^\infty_{n=0}(b_{2n} - a_{2n}) + 3\sum^\infty_{n=0}(a_{2n+1} + b_{2n+1})= 0.
\end{eqnarray}
Then, the constant $C_1$ is determined to be $C_1 = \frac{\sigma}{2}$.

Assigning the constants $C_\omega$ and $C_1$ the values we obtained, leads to a black hole solution, which contains a non-singular horizon, and is completely regular in the domain of outer communications. If we take the limit when  $a_n = 0$, $b_n = 0$ for every $n$ the metric functions describing the external sources vanish. Then, the solution reduces to the asymptotically flat Myers-Perry black hole with single rotation, which is represented in prolate spheroidal coordinates by $(\ref{Myers-PerryP})$. Another interesting limit is if we set the rotation parameter $\alpha = 0$. In this case the solution becomes static, and we recover the 5D distorted Schwarschild-Tangherlini black hole, which is obtained in \cite{Abdolrahimi:2010}. The coordinates $(\eta,\theta)$, in which the solution is represented, are related to the prolate spheroidal coordinates $(x,y)$ as $x = \eta$, $y = \cos\theta$, and the solution parameter $r_0$ is connected with $\sigma$ as $\sigma = r_0^2/4$. Furthermore, the parameters $a_n$ and $b_n$ characterizing the external gravitational field are interchanged in the two solutions $a_n\longleftrightarrow b_n$.

\paragraph{}The 5D stationary and axisymmetric solutions are conveniently described by their interval structure \cite{Hollands:2007}. It specifies the location of the horizons and the axes of the spacelike Killing fields in the factor space of the spacetime with respect to the isometry group. If we introduce the Weyl coordinates $\rho$ and $z$ on the two-dimensional surfaces orthogonal to the Killing fields, the factor space is represented by the upper $(\rho, z)$ half-plane, and its boundary coincides with the $z$-axis. The fixed point sets of the spacelike Killing fields and the horizons correspond to intervals on the $z$-axis. In addition, a direction vector consisting of integer numbers is associated with each interval. It specifies the coefficients in the linear combination of Killing fields which vanishes on it.

\begin{figure}[htbp]
\vbox{ \hfil \scalebox{0.65}{\includegraphics{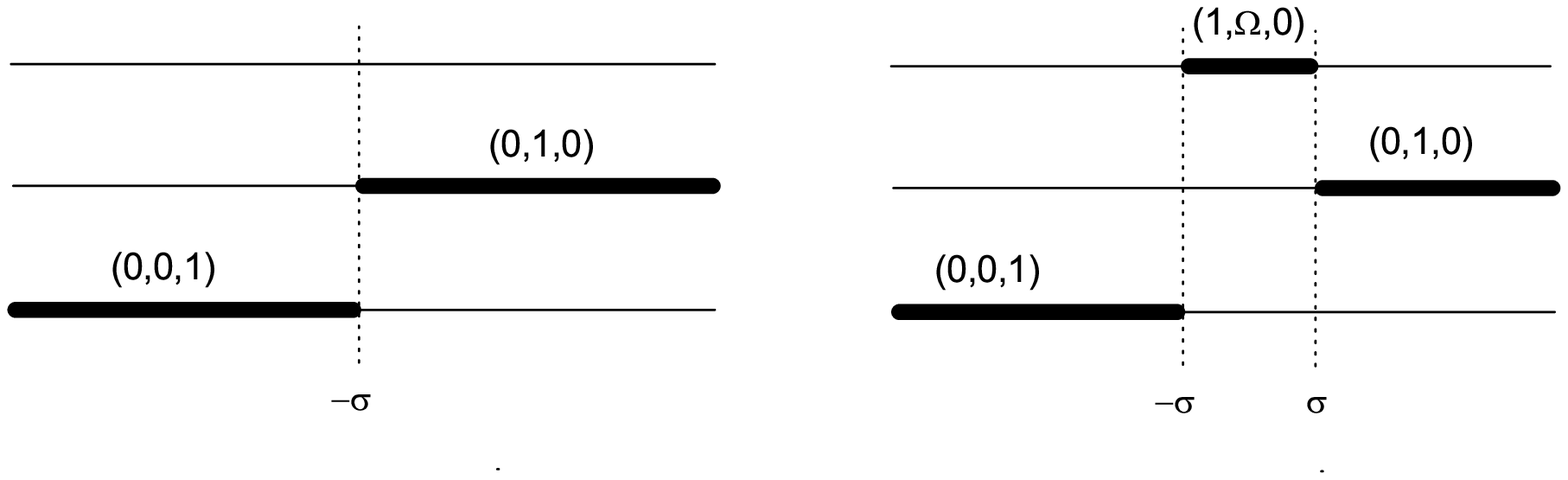}}%
\bigskip%
\caption{Rod structure of the seed solution (left) and the distorted Myers-Perry black hole with single rotation (right).}
\label{rodstr_MP}}
\end{figure}

The interval structure of the 5D distorted Minkowski spacetime and the distorted Myers-Perry black hole are presented on fig. $\ref{rodstr_MP}$. The distorted solutions are interpreted as local solutions describing a certain vacuum neighbourhood which is influenced by external matter sources. Therefore, the interval structure corresponding to them should also be interpreted as characterizing only the region where the distorted solution is valid, and not the whole spacetime. The direction vectors corresponding to each interval are specified above it, as the directions are given with respect to a basis of Killing vectors $\{\partial/\partial t, \partial/\partial \psi, \partial/ \partial\phi \}$.

On fig. $\ref{rodstr_MP}$ it is demonstrated that the axes of the spacelike Killing fields $\partial/\partial\phi$ and $\partial/\partial\psi$ for the distorted Myers-Perry black hole are located in Weyl coordinates at $z\leq -\sigma$, and $z \geq \sigma $, respectively. The interval $-\sigma\leq z \leq \sigma$ corresponds to a Killing horizon for the Killing field $V = \partial/\partial t + \Omega \partial/\partial\psi$, i.e. $V$ becomes null on the 3D hypersurface at constant $t$ located at $-\sigma\leq z \leq \sigma$, $\rho=0$. The constant coefficient in the linear combination of  Killing fields $\Omega$ represents the angular velocity of the horizon rotating with respect to the axis of the Killing field $\partial/\partial\psi$. It is equal to the value of the metric function $\omega^{-1}$ on the hypersurface $-\sigma \leq z \leq \sigma$, $\rho=0$, or equivalently at $x=1$, $-1\leq y \leq 1$ in prolate spheroidal coordinates. Taking into account the relations $(\ref{omega_dist})$ and $(\ref{C_omega})$, the angular velocity is calculated to be

\begin{eqnarray}\label{Omega_dist}
\Omega = \frac{\alpha}{2\sqrt{\sigma}(1+\alpha^2)}\exp{\left[\sum^\infty_{n=0}(a_n + 2b_n)\right]}.
\end{eqnarray}

In the limit when the parameters $a_n$ and $b_n$ vanish, the expression reduces to the angular velocity of the asymptotically flat Myers-Perry black hole ($\ref{M_MP}$). Therefore, the exponential factor can be interpreted as describing the influence of the external matter fields on the horizon rotation. The interval structure contains also information about the topology of the horizon hypersurface. If we consider the directions of the intervals adjacent to the horizon interval, we can determine whether the horizon is topologically a sphere, a ring ($S^1\times S^2$), or a general lens space $L(p,q)$ \cite{Hollands:2007}. In our case the horizon possesses $S^3$ topology, as the non-distorted Myers-Perry black hole. In fact the interval structure for the distorted Myers-Perry black hole on fig. $\ref{rodstr_MP}$ is equivalent in the vicinity of the horizon to the interval structure of the non-distorted Myers-Perry solution $(\ref{Myers-PerryP})$, differing only in the value of the angular velocity $\Omega$. Consequently, the horizon location also coincides with that in the isolated Myers-Perry case.

\section{Mass, angular momentum and Smarr-like relations}

The distorted Myers-Perry solution is stationary and axisymmetric, therefore we can define locally mass and angular momentum of a particular region of the spacetime by Komar integrals \cite{Komar}. We denote by $\xi$ the 1-form dual to the Killing field $\partial/\partial t$, and by $\zeta$ the 1-form dual to the Killing field $\partial/\partial\psi$. Then, the Komar mass $M_H$ and angular momentum $J_H$ associated with the black hole horizon are defined as
\begin{eqnarray}\label{Komar}
M_{H} &=&  - {3\over 32\pi} \int_{H} \star d\xi , \nonumber \\
J_H &=&  {1\over 16\pi} \int_{H} \star d \zeta,
\end{eqnarray}
where the integration is performed over the horizon cross-section. Calculating the Komar integrals we obtain the expressions
\begin{eqnarray}
M_H &=&\frac{3\,\pi}{2}\,\sigma (1+\alpha^2),  \label{M_dist} \\
J_H &=& 2\pi\sigma^{\frac{3}{2}}\alpha (1+\alpha^2)\exp{\left[-\sum^\infty_{n=0}(a_n + 2b_n)\right]}, \label{J_dist}
\end{eqnarray}
which can be interpreted as the intrinsic mass and the angular momentum of the black hole. Comparing with the corresponding expressions for the asymptotically flat Myers-Perry black hole $(\ref{M_MP})$, we see that the mass possesses the same form, while the angular momentum differs by the exponential factor.  We should also note that the mass expression in the asymptotically flat case refers both to the Komar mass at the horizon, and the $ADM$ mass of the spacetime, since in that case they coincide. The mass of the distorted black hole $(\ref{M_dist})$ is only the local Komar mass on the horizon. Since the solution is not asymptotically flat, and an appropriate extension is required in order to construct an asymptotically flat solution, the $ADM$ mass is not defined. Assuming that the solution is extended outside the horizon to some asymptotically flat solution containing matter fields, the $ADM$ mass, which will correspond to it, will contain also terms characterizing the matter fields. The $ADM$ mass is equivalent to the Komar integral $(\ref{Komar})$, however evaluated on a $3D$-sphere at infinity $S_\infty$, instead of the horizon. Using the Stokes' theorem, the integral can be represented as a sum of the Komar integral on the horizon, representing the local mass of the black hole, and a bulk integral of the Ricci 1-form\footnote{We assume that the solution contains no other horizons or bolts for the spacelike Killing fields.}

\begin{eqnarray}\label{M_ADM}
M_{ADM} =  - {3\over 32\pi} \int_{S_\infty} \star d\xi = - {3\over 32\pi} \int_{H} \star d\xi - {3\over 16\pi} \int_{\hat{M}} \star R(\xi).
\end{eqnarray}

The Ricci 1-form $R(\xi)$ is connected to the stress-energy tensor of the matter fields through the field equations. Therefore, the bulk integral will introduce corrections to the horizon mass $M_H$, which depend on the external matter fields, in the expression for the total $ADM$ mass of the spacetime. A similar argument applies also for the angular momentum on the horizon, which will not coincide with the total angular momentum of the spacetime if external matter is present.

\paragraph{}The horizon mass and angular momentum satisfy a local Smarr-like relation defined on the black hole horizon,  which is equivalent to the Smarr relation for the non-distorted Myers-Perry black hole. We can define a surface gravity on the horizon by the standard relation

\begin{eqnarray}
\kappa^2_H = -\frac{1}{4\lambda} g^{\mu\nu}\partial_{\mu}\lambda
\partial_{\nu}\lambda ,
\end{eqnarray}

\noindent
where $\lambda=g(V,V)$ is the norm of the Killing field $V = \partial/\partial t + \Omega\partial/\partial\psi$, which becomes null on the horizon. Performing the calculations we obtain the expression

\begin{eqnarray}
\kappa_H = \frac{1}{2\sqrt{\sigma}(1+\alpha^2)}\exp{\left[\frac{3}{2}\sum^\infty_{n=0}(a_{2n} + b_{2n}) + \frac{1}{2}\sum^\infty_{n=0}(b_{2n+1} - a_{2n+1})\right]}.
\end{eqnarray}
The surface gravity is defined only up to a constant scale factor, since we don't have a normalization of the Killing field $V$ at infinity. We have fixed the freedom by assuming that the distorted solution can be extended to an asymptotically flat one. We can calculate also the horizon area using the restriction of the metric on the horizon $g_H$ defined by $x=1$ and $t=const.$

\begin{eqnarray}
{\cal A}_H &=& \int_H \sqrt{\det{g_H}}d\phi\, d\psi\, dy \nonumber \\[2mm]
&=& 16\pi^2\sigma^{\frac{3}{2}}(1 + \alpha^2)\exp{\left[-\frac{3}{2}\sum^\infty_{n=0}(a_{2n} + b_{2n}) - \frac{1}{2}\sum^\infty_{n=0}(b_{2n+1} - a_{2n+1})\right]}.
\end{eqnarray}
In the limit when no external matter is present, i.e. $a_n=b_n=0$, the surface gravity and the horizon area reduce to the corresponding quantities for the non-distorted Myers-Perry black hole. The surface gravity is related to the temperature associated with the horizon  $T=\kappa_H/2\pi$, and the horizon area is proportional to the entropy of the black hole $S = {\cal A}_H/4$.

Taking advantage of the expression for the angular velocity $(\ref{Omega_dist})$, we see that a local Smarr-like relation is satisfied for the physical quantities defined on the black hole horizon

\begin{eqnarray}
M_H = \frac{3}{16\pi}\kappa_H {\cal A}_H + \frac{3}{2}\Omega J_H.
\end{eqnarray}

It coincides exactly with the local Smarr-like relation for the non-distorted Myers-Perry black hole, which in this case is also a global relation for the $ADM$ mass and angular momentum, since they are equivalent to the corresponding quantities on the black hole horizon.

Finally, we will discuss the possible values that the horizon mass $M_H$ and angular momentum $J_H$ can attain. In the case of the asymptotically flat Myers-Perry black hole their values are restricted, since the following relation should be satisfied

\begin{eqnarray}\label{MJ}
\frac{27\pi}{32}\frac{J^2_H}{M^3_H}< 1,
\end{eqnarray}
in order for the event horizon to exist (see  $(\ref{horizonEM})$). In the representation of the solution in prolate spheroidal coordinates the parameters $\sigma$ and $\alpha$ are introduced in such a way, that $(\ref{MJ})$ is automatically satisfied,

\begin{eqnarray}
\frac{27\pi}{32}\frac{J^2_H}{M^3_H} = \frac{\alpha^2}{1+\alpha^2}<1,
\end{eqnarray}
i.e. the metric $(\ref{Myers-PerryP})$ always describes a black hole. If we examine the relation $J^2_H/M^3_H$ for the distorted Myers-Perry black hole we obtain the expression
\begin{eqnarray}
\frac{27\pi}{32}\frac{J^2_H}{M^3_H} = \frac{\alpha^2}{1+\alpha^2}\exp{\left[-2\sum^\infty_{n=0}(a_{n} + 2b_{n})\right]}.
\end{eqnarray}

The parameters $a_n$ and $b_n$ characterize the external matter field, and their values are not connected with the existence of a horizon. Therefore, for some matter distributions they can certainly possess such values that the ratio $\frac{27\pi}{32}\frac{J^2_H}{M^3_H}$ exceeds one. Moreover, since it is not bounded, it is allowed to grow unlimitedly.

A similar effect was noticed in the investigation of a numerical solution representing the Kerr black hole surrounded by a perfect fluid ring \cite{Ansorg1}, \cite{Ansorg2}. The corresponding ratio of the angular momentum and the mass for an isolated Kerr black hole should satisfy the bound $\mid J \mid / M^2 \leq 1$. However, when a matter ring is present, it is demonstrated that the ratio can grow arbitrarily large. The effect can be easily foreseen from the analytical expressions for the horizon mass and the angular momentum of the Kerr black hole situated in an arbitrary stationary and axisymmetric external gravitational field, which were obtained in \cite{Breton:1997}. The horizon mass and angular momentum ratio is given by

\begin{eqnarray}
\frac{\mid J \mid}{M^2} = \frac{2\alpha}{1 + \alpha^2}\exp{\left[-2\sum^\infty_{n=1}a_{2n}\right]}, \quad~~~ 0<\alpha<1,
\end{eqnarray}
where $\alpha$ is a rotation parameter and the parameters $a_n$ characterize the external field. Again, for certain values of the parameters $a_n$ characterizing the external field the ratio can grow unlimitedly.

\subsubsection*{Particular case}

We consider a particular case of the general distorted Myers-Perry solution $(\ref{metric_dist})$, in which the parameters $a_n$ and $b_n$ characterizing the external matter field obey the relation $a_n + 2 b_n = 0$ for every $n$. Then, the metric function ${\hat a}$ reduces to the rotation parameter $\alpha$ and the metric acquires the form

\begin{eqnarray}
ds^2&=& -\frac{x-1-\alpha^2(1-y)}{x+1+\alpha^2(1+y)}e^{-2\widehat{W}}
         \left(dt -\omega d\psi\right)^2
       +\frac{x+1+\alpha^2(1+y)}{x-1-\alpha^2(1-y)}e^{-2W} d\psi^2
\nonumber \\[2mm]
    &+& e^{-2U}d\phi^2 + \frac{\sigma}{2}\left[x+1+\alpha^2(1+y)\right]e^{2(\hat{\gamma} - \widehat{W})}
         \left( \frac{dx^2}{x^2-1}+\frac{dy^2}{1-y^2}\right), \label{metric_dist2} \\
\omega &=& -2\sqrt{\sigma}\alpha\frac{(x-y)}{(x-1)-(1-y)\alpha^2} + 2\sqrt{\sigma}\alpha ,  \\[2mm]
\end{eqnarray}
where $\hat{\gamma}$ is given by the expression ($\ref{gamma_dist}$) with $a_n =- 2 b_n$. The influence of the external matter fields appears only in the exponents containing the functions $\widehat{U}$, $\widehat{W}$ and $\hat{\gamma}$, while the remaining part of the line element coincides with the metric of the non-distorted solution. In particular, the metric function $\omega$ is equal in this case to the corresponding metric function for the asymptotically flat Myers-Perry black hole. Consequently,  the angular velocity of the horizon is also equal in the two cases, and the ergoregion behaves as in the asymptotically flat case (see section 6). It is confined to a compact region in spacetime encompassing the horizon, and the ergosurface intersects the axes of the spacelike Killing fields at the points $\{x=1, y=1\}$, and $\{x = 1 + 2\alpha^2, y = -1\}$. Since the metric function $\omega$ is already asymptotically flat, it is in general possible to extend the distorted black hole solution to an asymptotically flat solution involving matter fields, which do not possess intrinsic angular velocity.

The condition for absence of conical singularities takes the form
\begin{equation}
\sum^\infty_{n=0} (-1)^n b_{n} = 0,
\end{equation}
and examining the expressions for the horizon area, surface gravity and angular momentum, we see that they coincide with the corresponding quantities for the asymptotically flat Myers-Perry black hole. Thus, the physical characteristics of the solution remain unaffected by the external distribution of matter.

\section{Properties of the horizon and the ergoregion}

The 3-dimensional surface of the horizon is defined by $t=const.$ and $x=1$. The corresponding metric on the horizon surface is given by the expression
\begin{eqnarray}\label{metric_H}
ds^2_{H} &=& \frac{4\sigma(1+\alpha^2)^2}{2 + {\hat a}^2(1+y)}(1-y)e^{-2{\widehat W}}d\psi^2 + 2\sigma(1+y)e^{-2{\widehat U}}d\phi^2   \nonumber \\[2mm]
&+& \frac{\sigma}{2}\left[2 + {\hat a}^2(1+y)\right]e^{2({\hat\gamma}-{\widehat W})}\frac{dy^2}{1-y^2},
\end{eqnarray}
where the metric functions ${\hat a}$, ${\widehat W}$, ${\widehat U}$ and ${\hat\gamma}$ possess the following form
\begin{eqnarray}
&&{\widehat U} = {\sum^\infty_{n=0}a_n\,y^n}, \quad~~~{\widehat W} = {\sum^\infty_{n=0}b_n\,y^n},  \nonumber \\[2mm]
&&{\hat a} = \alpha\exp{\left[-\sum^\infty_{n=0}(a_n + 2b_n)(y^n-1)\right]},  \nonumber \\[2mm]
&&{\hat\gamma} = {\sum^\infty_{n=0}(a_n + 2b_n)y^n -\frac{3}{2}\sum^\infty_{n=0}(a_{2n} + b_{2n}) - \frac{1}{2}\sum^\infty_{n=0}(b_{2n+1} - a_{2n+1})} \nonumber.
\end{eqnarray}
The structure of the metric on the horizon cross-section is more transparent if we introduce the angular coordinate $0\leq\theta\leq {\pi}$, such that $y=\cos \theta$. Then, it acquires the form:
\begin{eqnarray}\label{HorizonM}
ds^2_{H} &=& \frac{4\sigma(1+\alpha^2)^2}{1 + {\hat a}^2(\theta)\cos^2\left(\frac{\theta}{2}\right)}\sin^2\left(\frac{\theta}{2}\right) e^{-2{\widehat W}(\theta)}d\psi^2 + 4\sigma \cos^2\left(\frac{\theta}{2}\right) e^{-2{\widehat U}({\theta})}d\phi^2 \nonumber \\[2mm]
&+& \sigma\left[1 + {\hat a}^2({\theta})\cos^2\left(\frac{\theta}{2}\right)\right]e^{2({\hat\gamma}(\theta)-{\widehat W}(\theta))} d\theta^2,
\end{eqnarray}
\noindent
which can be written also as
\begin{eqnarray}\label{horizon_H}
ds^2_{H} &=& 4\sigma\bigg[F(\lambda)e^{2({\hat\gamma}(\lambda)-{\widehat W}(\lambda))}d\lambda^2 + \frac{(1+\alpha^2)^2e^{-2{\widehat W}(\lambda)}}{F(\lambda)}\sin^2\lambda d\psi^2 \nonumber \\
&+& e^{-2{\widehat U}({\lambda})}\cos^2\lambda d\phi^2\bigg],
\end{eqnarray}
\noindent
where $\lambda=\theta/2$, and $F(\lambda) = 1 + \hat{a}^2(\lambda)\cos^2\lambda$. In the limit when the rotation parameter $\alpha$, and the external matter parameters $a_n$ and $b_n$ vanish, we obtain the metric on the 3-sphere with radius $R = 2\sqrt{\sigma}$
\begin{eqnarray}
ds^2_{H} = 4\sigma\left[d\lambda^2 + \sin^2\lambda d\psi^2 + \cos^2\lambda d\phi^2\right],\label{Horizonnorotation}
\end{eqnarray}
\noindent
in the Hopf coordinates $\{0\leq\lambda\leq \pi/2, 0<\psi<2\pi, 0<\phi<2\pi\}$. Comparing the metric on the 3-sphere with ($\ref{horizon_H}$) we can see that the horizon geometry is deformed from the spherical, and all the analytic axisymmetric horizon geometries are possible depending on the type of the external matter fields. In the absence of external matter the horizon is also deformed
\begin{eqnarray}
ds^2_{H} = 4\sigma\left[F(\lambda)d\lambda^2 + \frac{(1+\alpha^2)^2}{F(\lambda)}\sin^2\lambda d\psi^2 + \cos^2\lambda d\phi^2\right].
\end{eqnarray}
\noindent
However, the deformation is more restricted since it is determined only by the function $F(\lambda) = 1 + \alpha^2\cos^2\lambda$, where $\alpha$ is the rotation parameter.

\begin{figure}[htp]
\setlength{\tabcolsep}{ 0 pt }{\scriptsize\tt
		\begin{tabular}{ ccc }
            \includegraphics[width=4.1 cm]{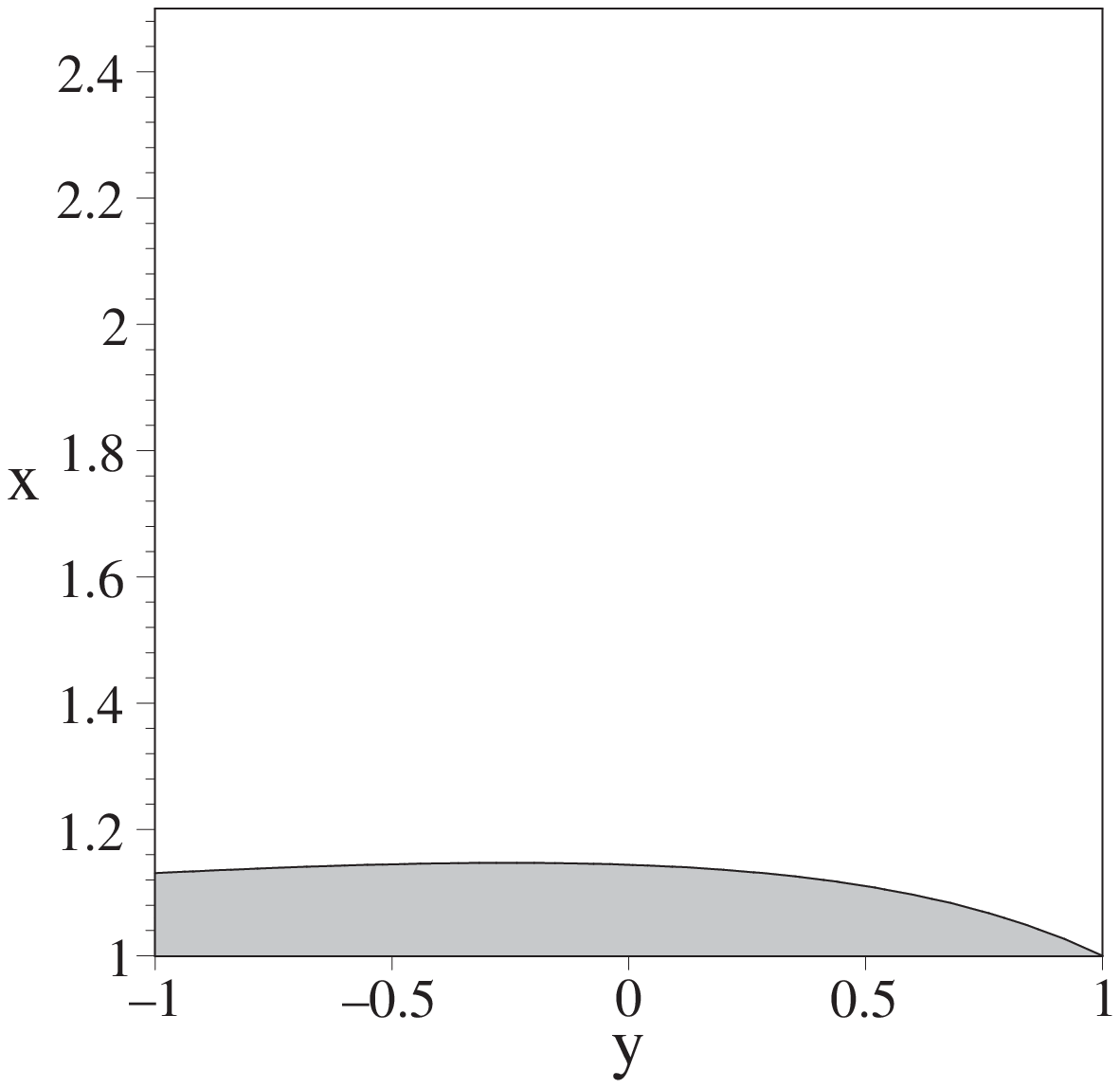} &
            \includegraphics[width=4.1 cm]{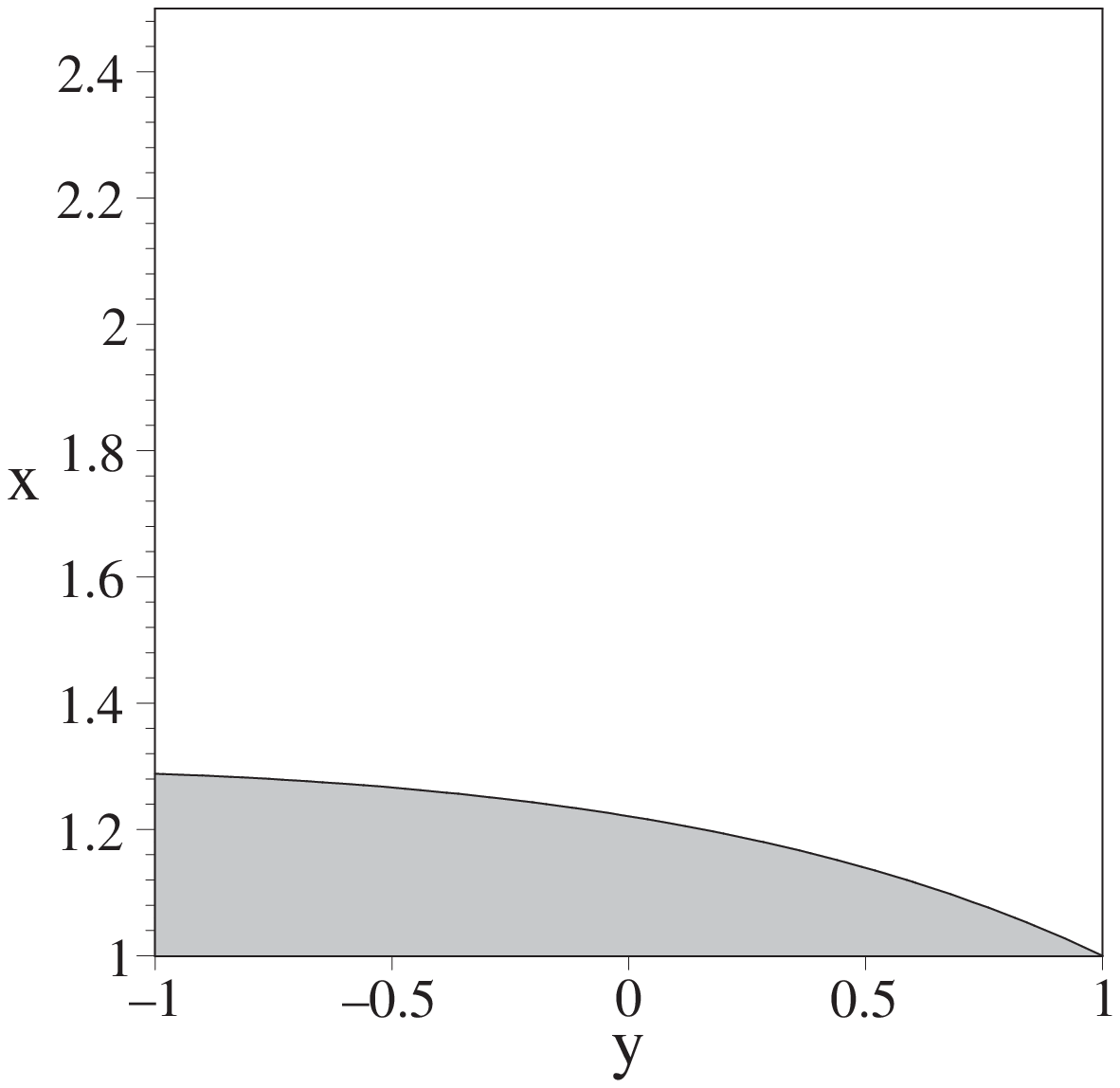}&
            \includegraphics[width=4.1 cm]{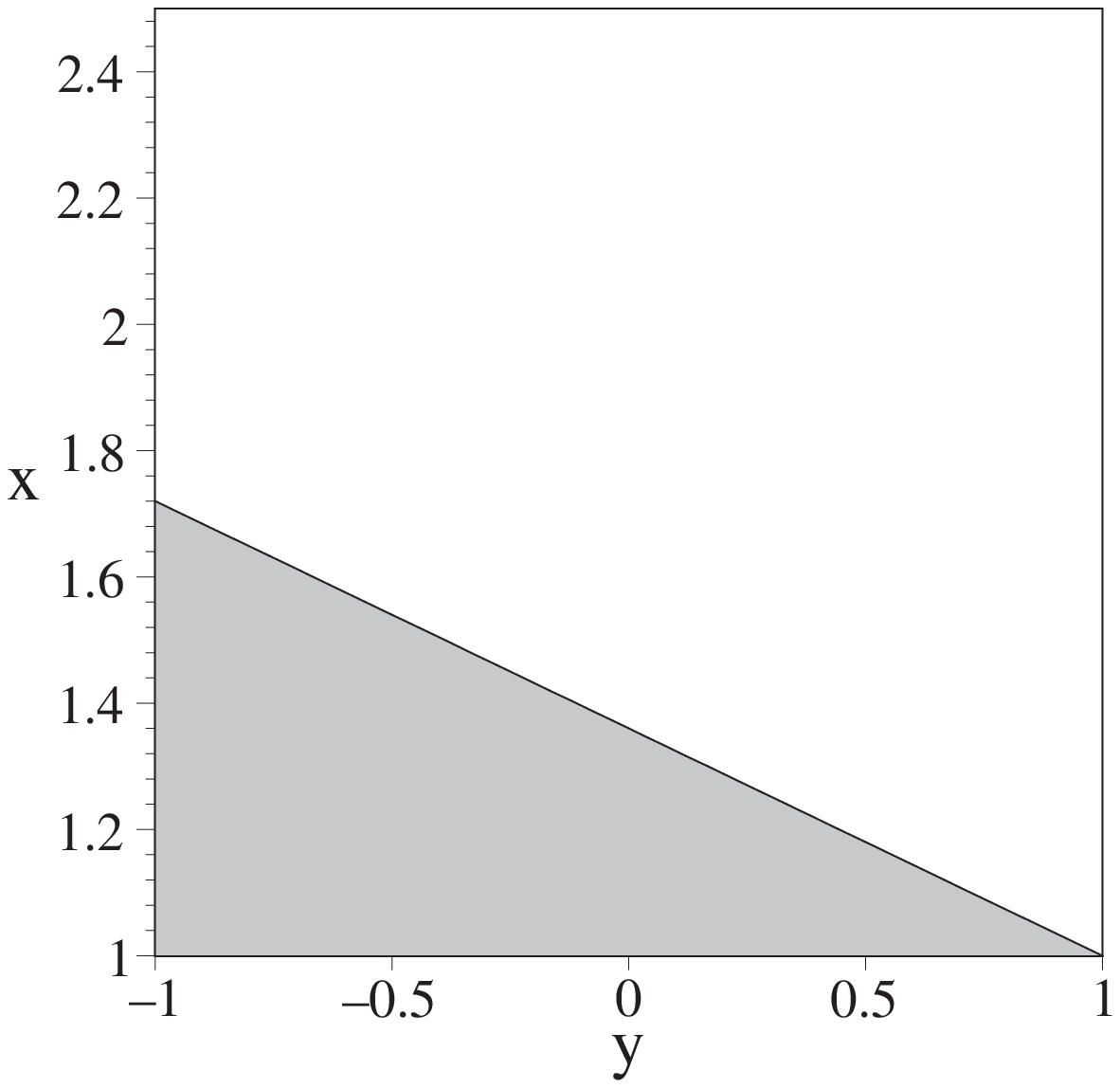}  \\
            $a)\,\,\,$ $\alpha = 0.6$, $b_1 = - 0.2$  &
			$\alpha = 0.6$, $b_1 = - 0.1$ &
            $\alpha = 0.6$, $b_1 = 0$\\[2mm]
            \includegraphics[width=4.1cm]{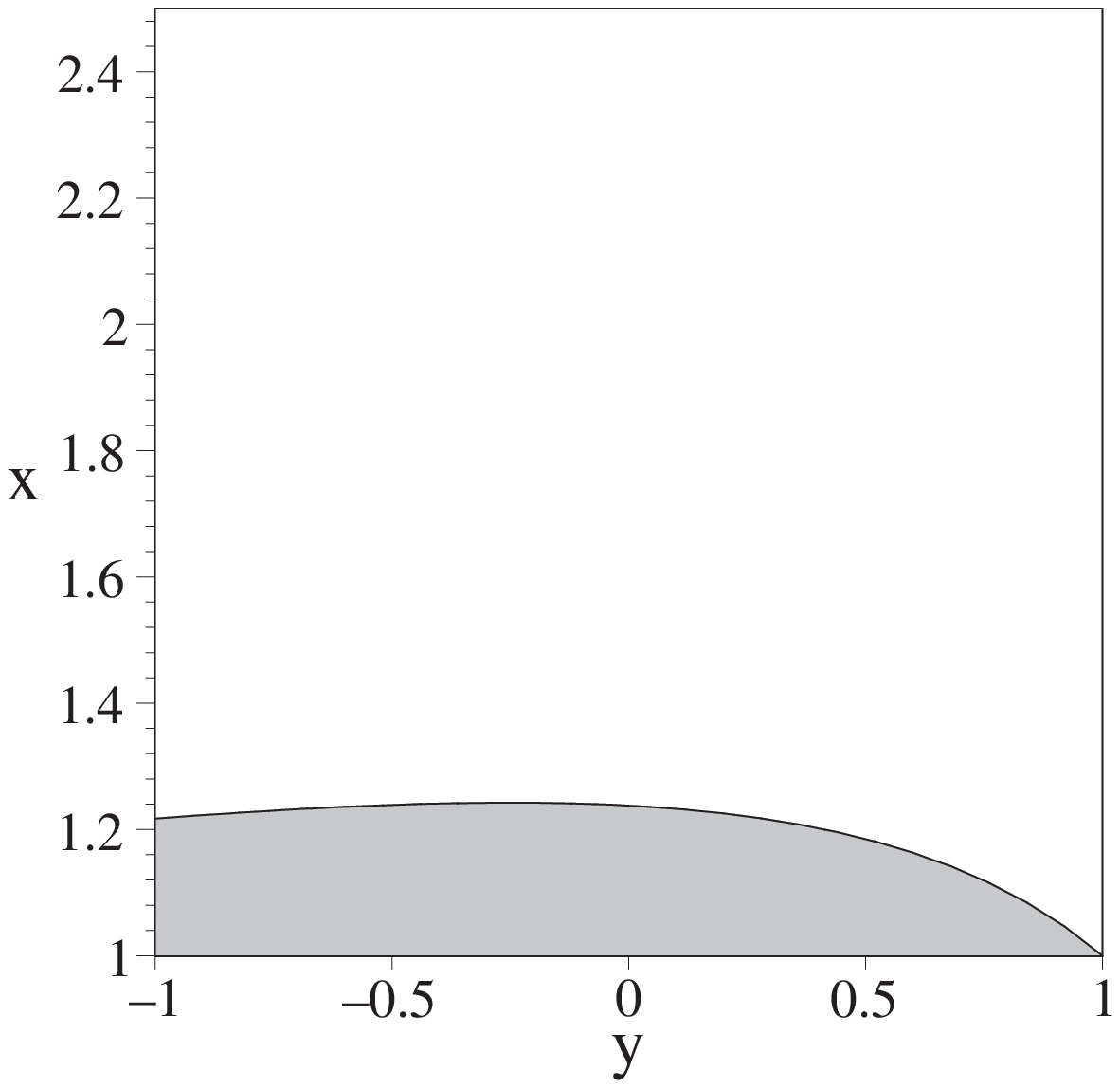} &
            \includegraphics[width=4.1cm]{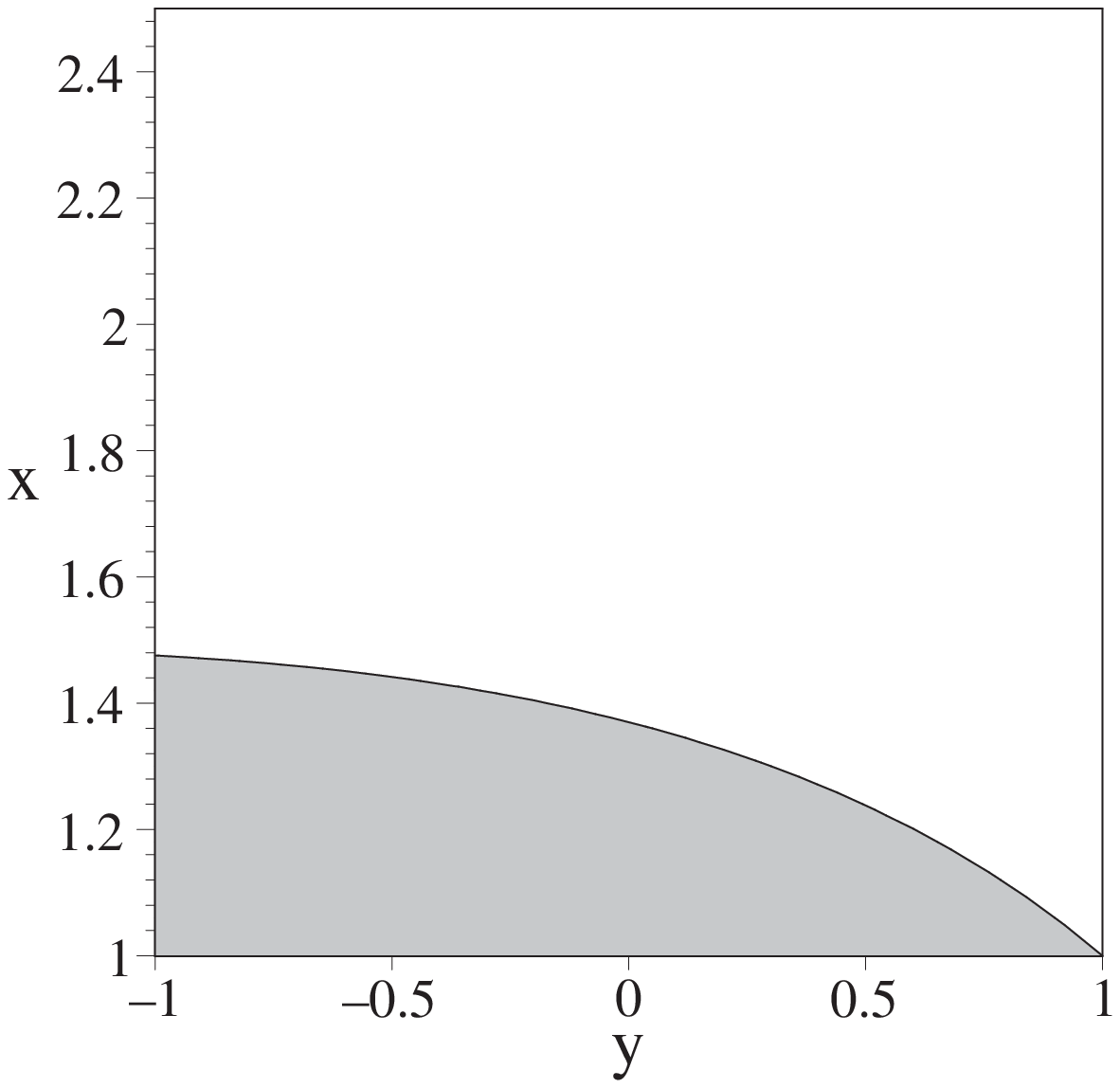} &
            \includegraphics[width=4.1cm]{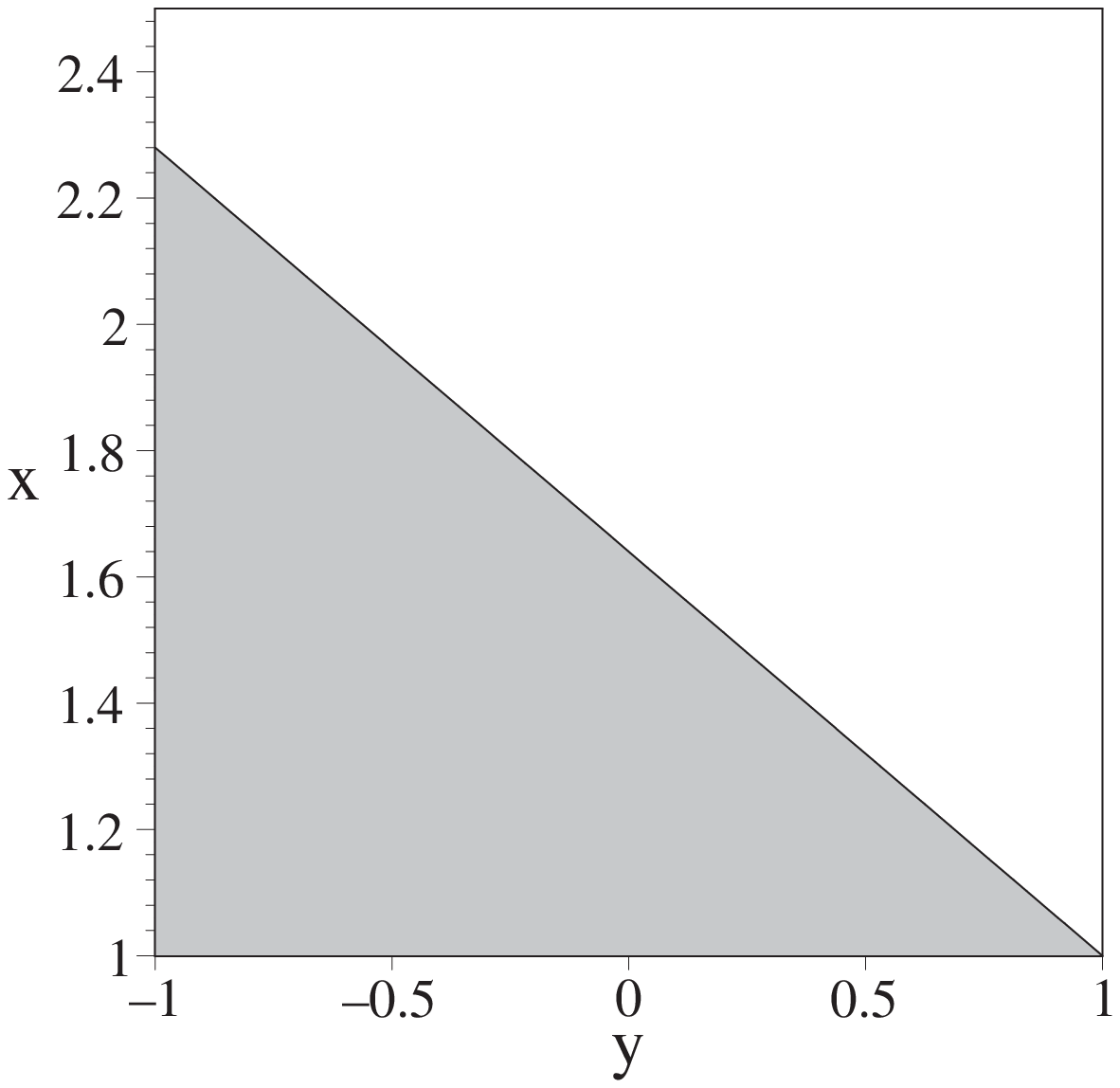}\\
			$b)\,\,\,$ $\alpha = 0.8$, $b_1 = - 0.2$  &
			$\alpha = 0.8$, $b_1 = - 0.1$ &
            $\alpha = 0.8$, $b_1 = 0$ \\
\end{tabular}}
\caption{\footnotesize{Behavior of the ergoregion (grey area) for negative distortion parameter $b_1$ and different values of the rotation parameter $\alpha$. The ergoregion for the non-distorted Myers-Perry black hole ($b_1 = 0$) is also presented.}}
		\label{Erg_neg}
\end{figure}

\paragraph{}The horizon is encompassed by an ergoregion defined as the region where the Killing field $\partial/\partial t$ is spacelike
\begin{eqnarray}
g_{tt}= -\frac{x-1-\hat{a}^2(1-y)}{x+1+\hat{a}^2(1+y)}e^{2(\widehat{U} + \widehat{W})}>0.\label{ergo1}
\end{eqnarray}
Since the denominator is always positive, the ergoregion is determined by the sign of the function
\begin{eqnarray}\label{ergoregion}
&& G = x-1-(1-y){\hat a}^2, \\
&&\hat{a} = \alpha \exp{\left[\sum^{\infty}_{n=1}(a_n+2b_n)\sum^{n-1}_{k=0}(x-y)R^k P_k\left(\frac{xy}{R}\right)\right]}, \nonumber
\end{eqnarray}
and the boundary of the ergoregion $g_{tt}=0$ defines the ergosurface. In the case of the non-distorted Myers-Perry black hole the ergosurface is always a compact 3D hypersurface which touches  the horizon at the point $\{x=1,y=1\}$, and intersects the symmetry axis $ y = -1$ at $x = 1 + \alpha^2$. Increasing the rotation parameter $\alpha$ the ergoregion extends further from the horizon (see fig. $\ref{Erg_neg}$). In the case of the Myers-Perry black hole in an external gravitational field the ergoregion has a more complicated behavior. We can see that there always exists a small neighbourhood of the horizon $x\rightarrow 1$, $x>1$, where ($\ref{ergoregion}$) is positive. Since the ergoregion should be axially symmetric, we can estimate its qualitative behaviour by investigating the points in which the ergosurface intersects the axes of the spacelike Killing fields. As for the non-distorted Myers-Perry black hole the ergosurface always touches the horizon $x=1$ at the axis $y=1$.  The intersection point with the axis of the Killing field $\partial/\partial\phi$ depends on the values of the parameters $a_n$, $b_n$ and $\alpha$. The restriction of the function ${\hat a}$ on the axis $y=-1$ is given by the expression
\begin{eqnarray}
\hat{a} = \alpha \exp{\left[\sum^{\infty}_{n=0}(a_n+2b_n)(1 - (-1)^nx^n)\right]},  \nonumber
\end{eqnarray}
and examining eq. $(\ref{ergoregion})$ we can see that there exist values of the parameters $a_n$ and $b_n$, for which it is never satisfied. This behavior means that the ergoregion extends to infinity, if we consider the distorted Myers-Perry black hole as a global solution. Considering it as a local solution, which is valid only in some neighbourhood of the horizon, it means that in some cases the ergoregion is not compact in the region of validity.

We demonstrate some of the possible configurations of the ergoregion by investigating some simple cases of distortion. The ergoregion is independent of the solution parameters $\sigma$, $a_0$ and $b_0$.  Therefore, the most simple non-trivial type of distortion is described by setting $b_1\neq 0$, and requiring the other external matter parameters to vanish. We will refer to this case as dipole distortion. Setting $a_1\neq 0$ instead, and requiring the other distortion parameters to vanish leads to the same qualitative behavior of the ergoregion, and the two cases are related by the shift $a_1\longleftrightarrow 2b_1$. Similarly, all the arguments for the case $b_1\neq 0$ refer also for distortions of the type $a_1\neq 0$, $b_1\neq 0$, and the other distortion parameters vanishing.

Examining the behavior of $(\ref{ergoregion})$ in the case of dipole distortion $b_1\neq0$ we obtain that for negative values of $b_1$ the ergoregion is always a compact region encompassing the horizon, similar to the non-distorted solution. If the distortion parameter $b_1$ is positive, however, the ergoregion always contains a non-compact part which extends to infinity if we consider our solution as a global one. Furthermore, the ergoregion can be either simply connected, or include two parts separated in spacetime, one of which is a compact region in the vicinity of the horizon.

In the case of the dipole distortion the number of the separated parts of the ergoregion is determined by the number of the crossing points of the ergosurface with the axis $y=-1$, or equivalently by the number of the real roots of the equation

\begin{eqnarray}\label{Erg_int}
G = x-1-(1-y){\hat a}^2 = 0,
\end{eqnarray}
evaluated at  $y=-1$. For negative values of the external matter parameter $b_1$, eq. ($\ref{Erg_int}$) has always a single real root, and consequently a single intersection point of the ergosurface with the $y=-1$ axis. This corresponds to a compact ergoregion encompassing the horizon for all values of $b_1<0$, and the rotation parameter $\alpha$. If we keep $\alpha$ fixed and vary $b_1$, we obtain that the ergoregion gets smaller as $\mid b_1 \mid$ grows.  Keeping $b_1$ fixed and varying $\alpha$, we observe that the ergoregion gets larger with the increase of the rotation parameter, and extends further from the horizon. The behavior of the ergoregion for negative $b_1$ is presented on fig. $\ref{Erg_neg}$.  For comparison we have also illustrated  the ergoregion for the non-distorted Myers-Perry black hole for the same values of the rotation parameter $\alpha$.

If we consider positive values of the distortion parameter $b_1>0$, eq. ($\ref{Erg_int}$) can possess at most two real roots. Provided we keep the rotation parameter $\alpha$ constant and vary $b_1$, two roots occur for small $b_1$, and there exists a critical value $b_{1_{crit}}$, depending on the particular value of $\alpha$, when eq. ($\ref{Erg_int}$) possesses only a single real root. For $b_1 > b_{1_{crit}}$ no real roots exist. Consequently, for $b_1 < b_{1_{crit}}$ we observe two separated ergoregions, one of which extends to infinity. When $b_1$ approaches $b_{1_{crit}}$ the two regions get closer to each other, and for $b_1 = b_{1_{crit}}$ they touch at the common crossing point of their boundaries with the axis $y=-1$. When  $b_1$ exceeds $b_{1_{crit}}$ the two parts of the ergoregion merge into a single non-compact ergoregion which extends to infinity. This is in agreement with the fact that for $b_1 > b_{1_{crit}}$ the ergosurface doesn't intersect the $y=-1$ axis.

\begin{figure}[htp]
		\setlength{\tabcolsep}{ 0 pt }{\scriptsize\tt
		\begin{tabular}{ cccc }
            \includegraphics[width=4.1 cm]{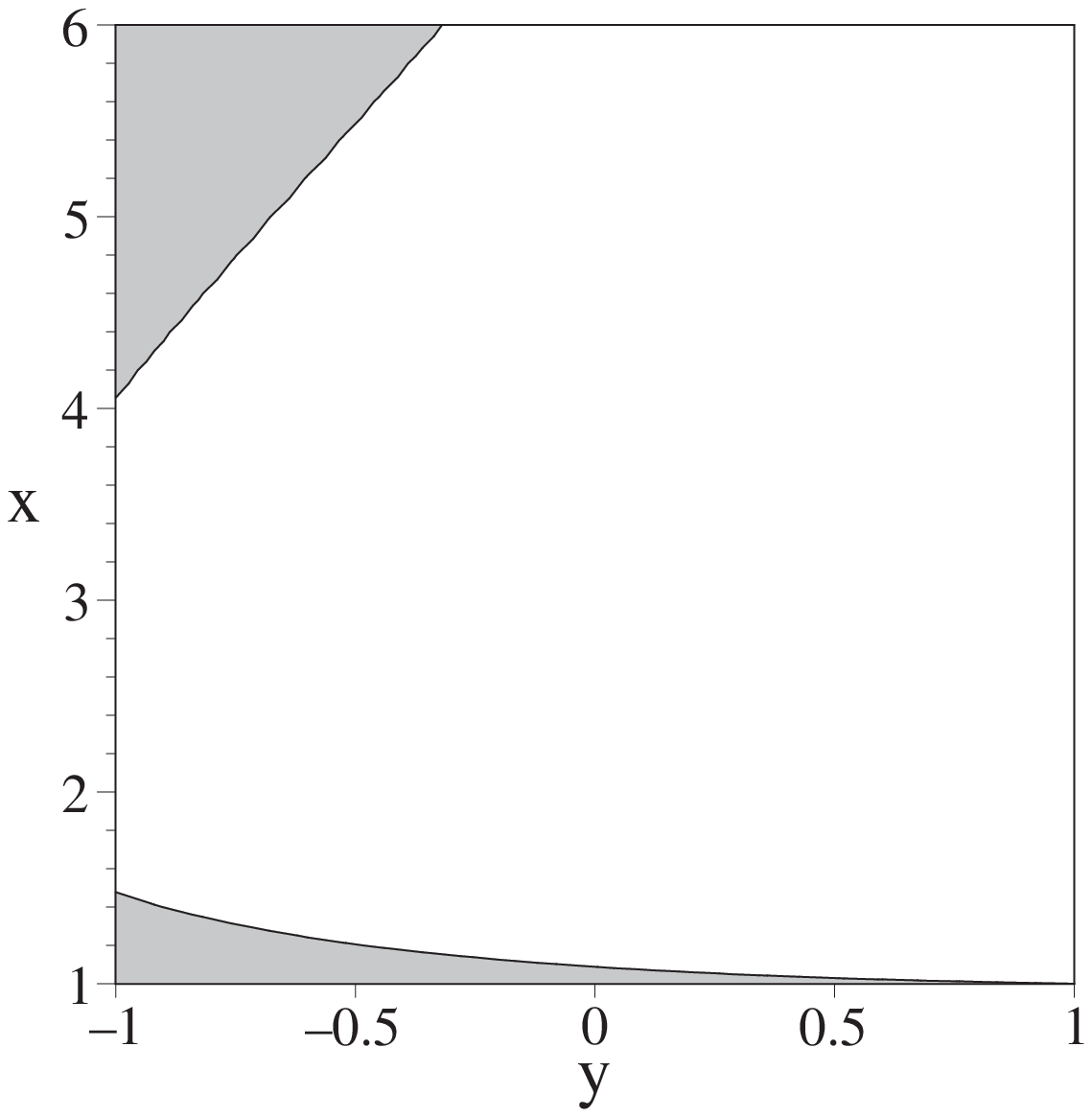} &
            \includegraphics[width=4.1 cm]{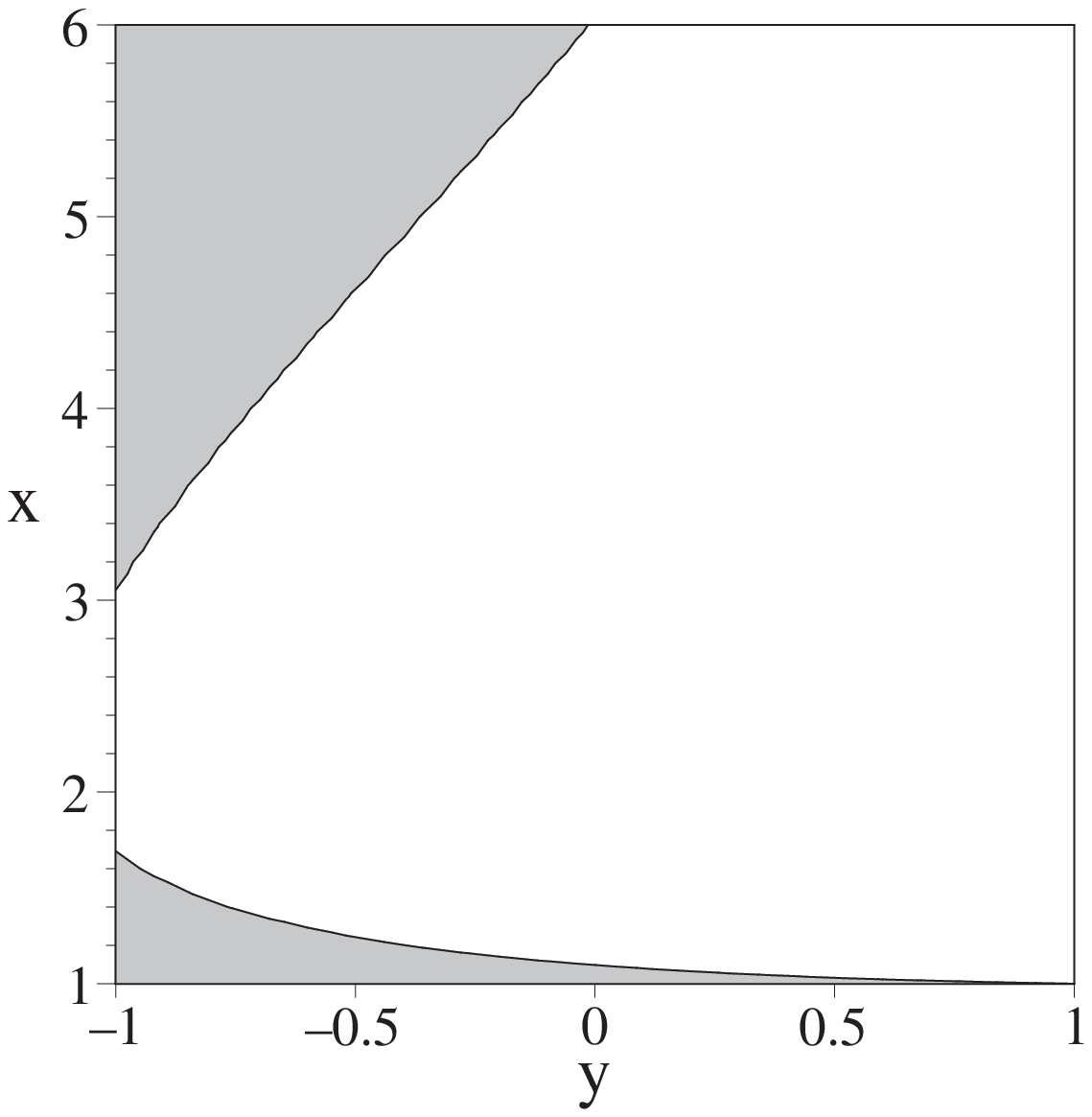} &
            \includegraphics[width=4.1 cm]{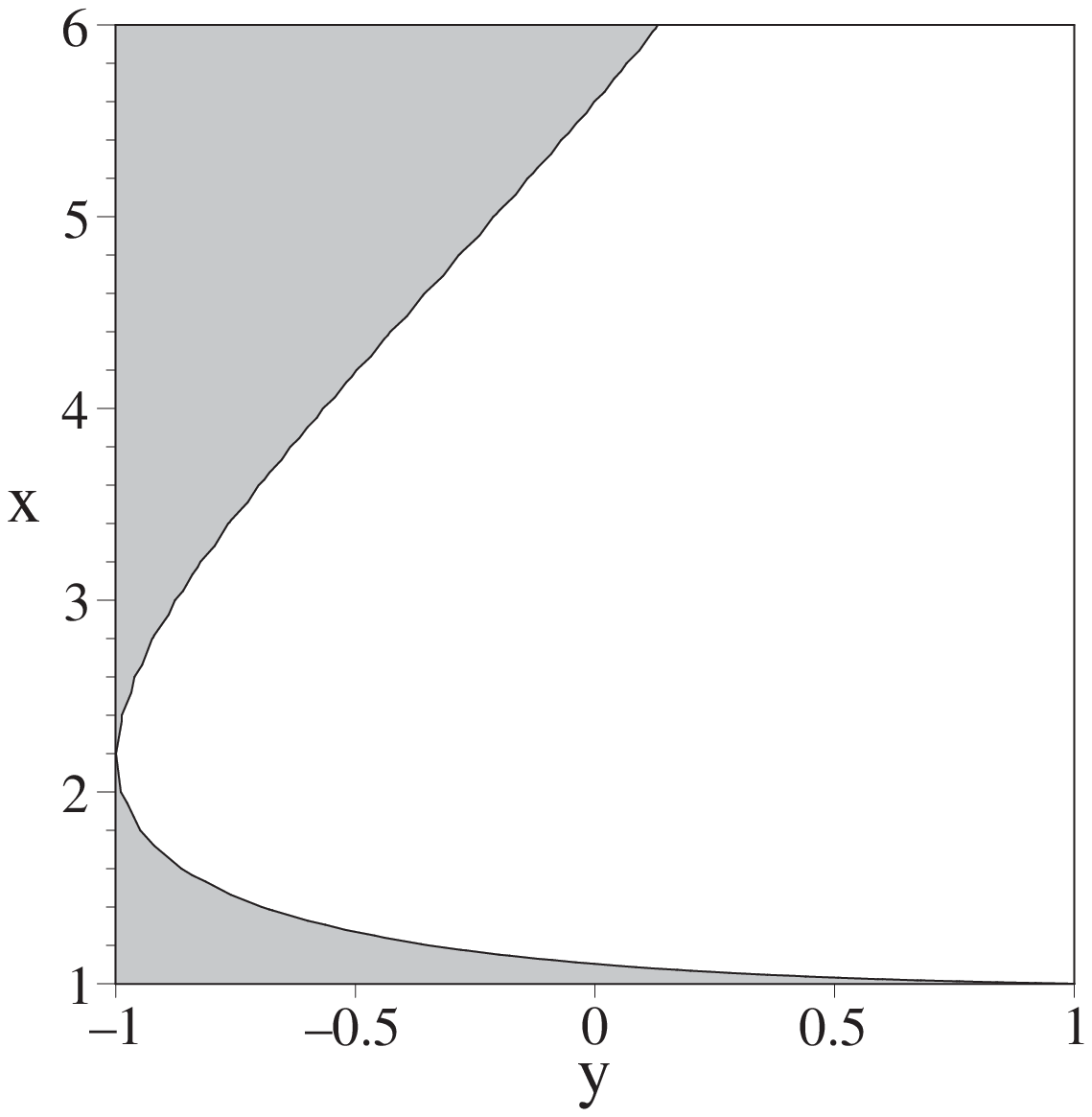} &
			\includegraphics[width=4.1 cm]{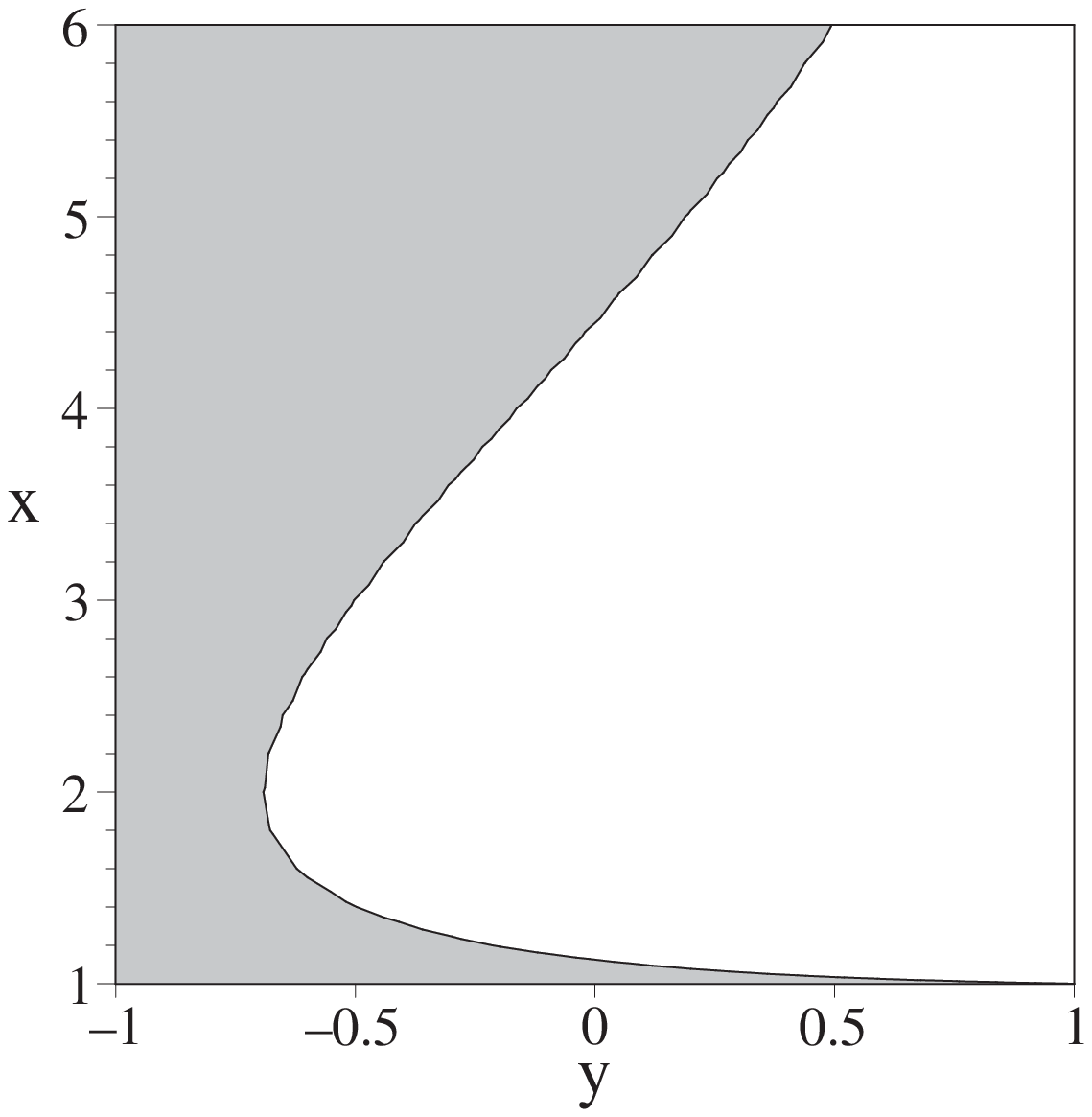} \\
            $a)\,\,\,$ $\alpha = 0.2$, $b_1 = 0.18$\  &
			$\alpha = 0.2$, $b_1 = 0.2$\  &
            $\alpha = 0.2$, $b_1 = 0.2116$\  &
			$\alpha = 0.2$, $b_1 = 0.25$ \\[2mm]
            \includegraphics[width=4.1cm]{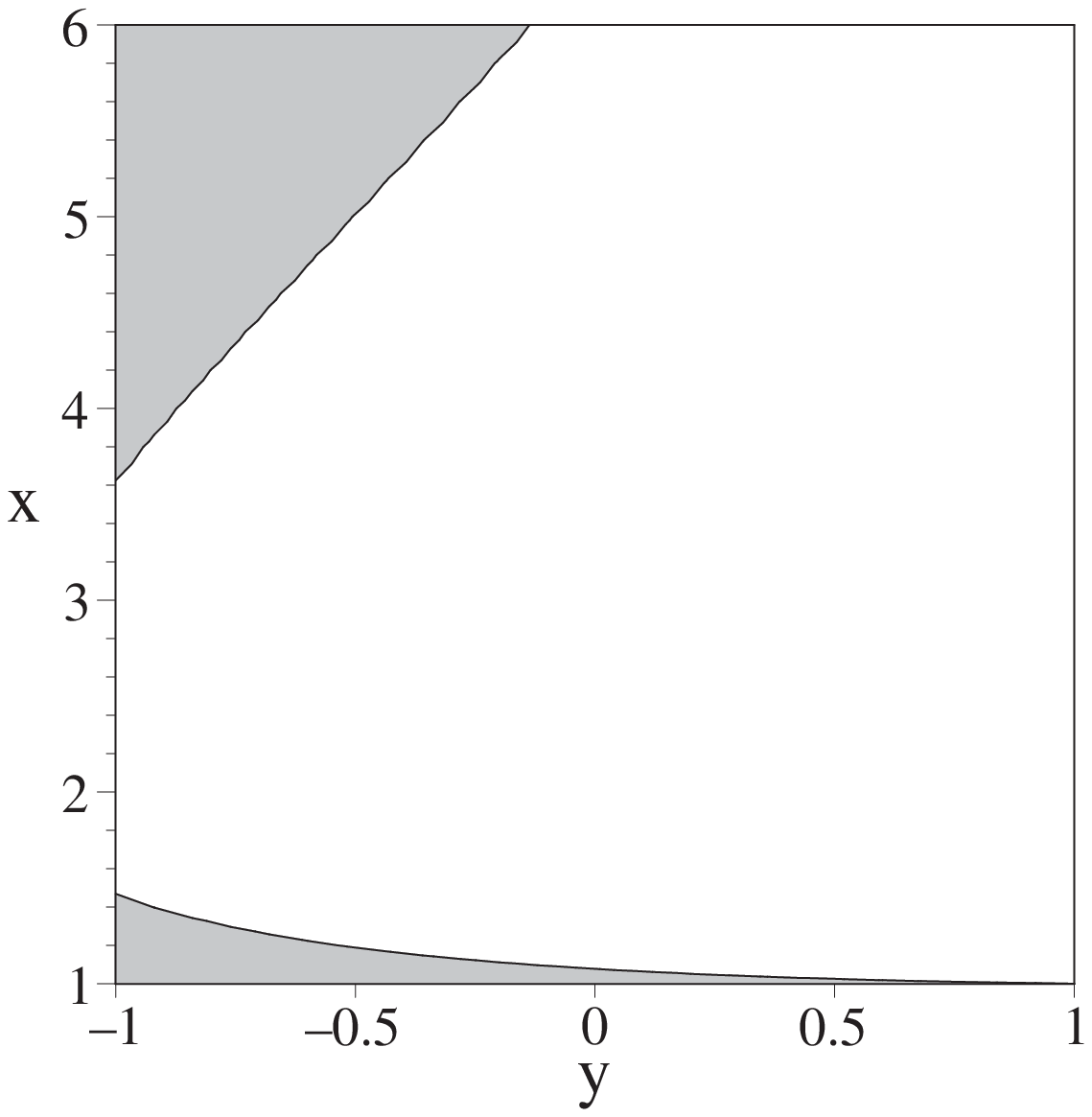} &
            \includegraphics[width=4.1cm]{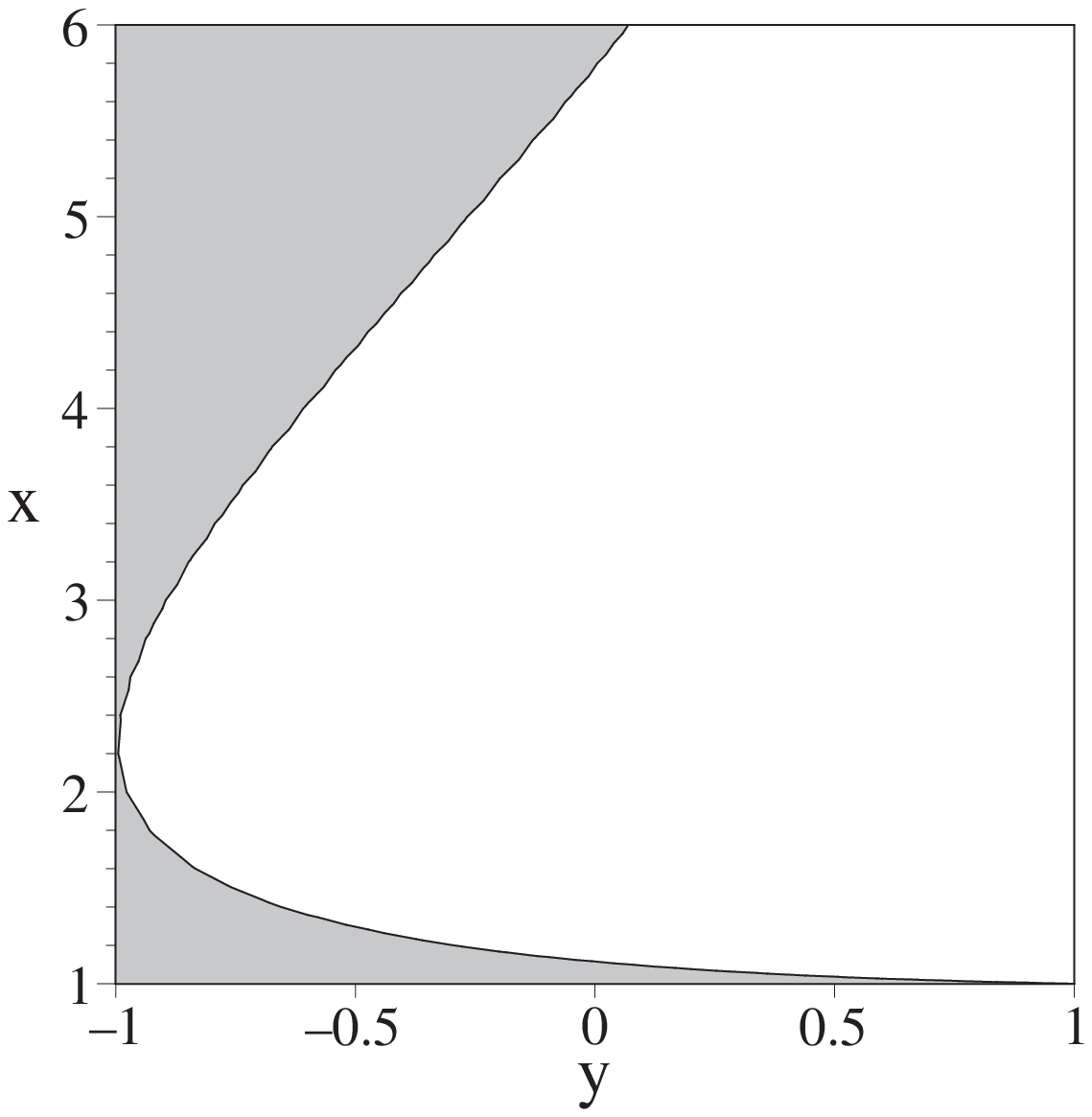} &
            \includegraphics[width=4.1 cm]{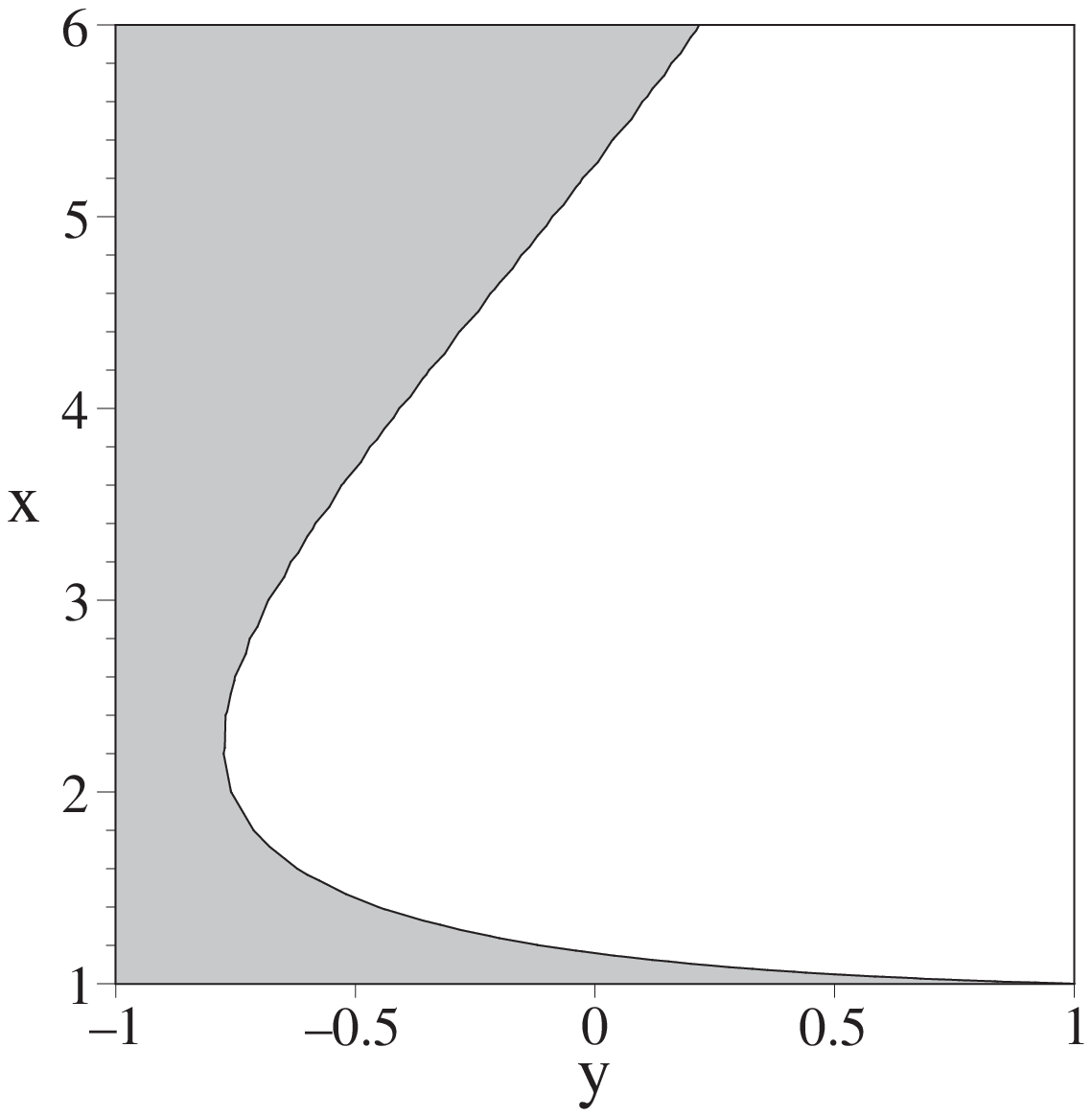} &
			\includegraphics[width=4.1cm]{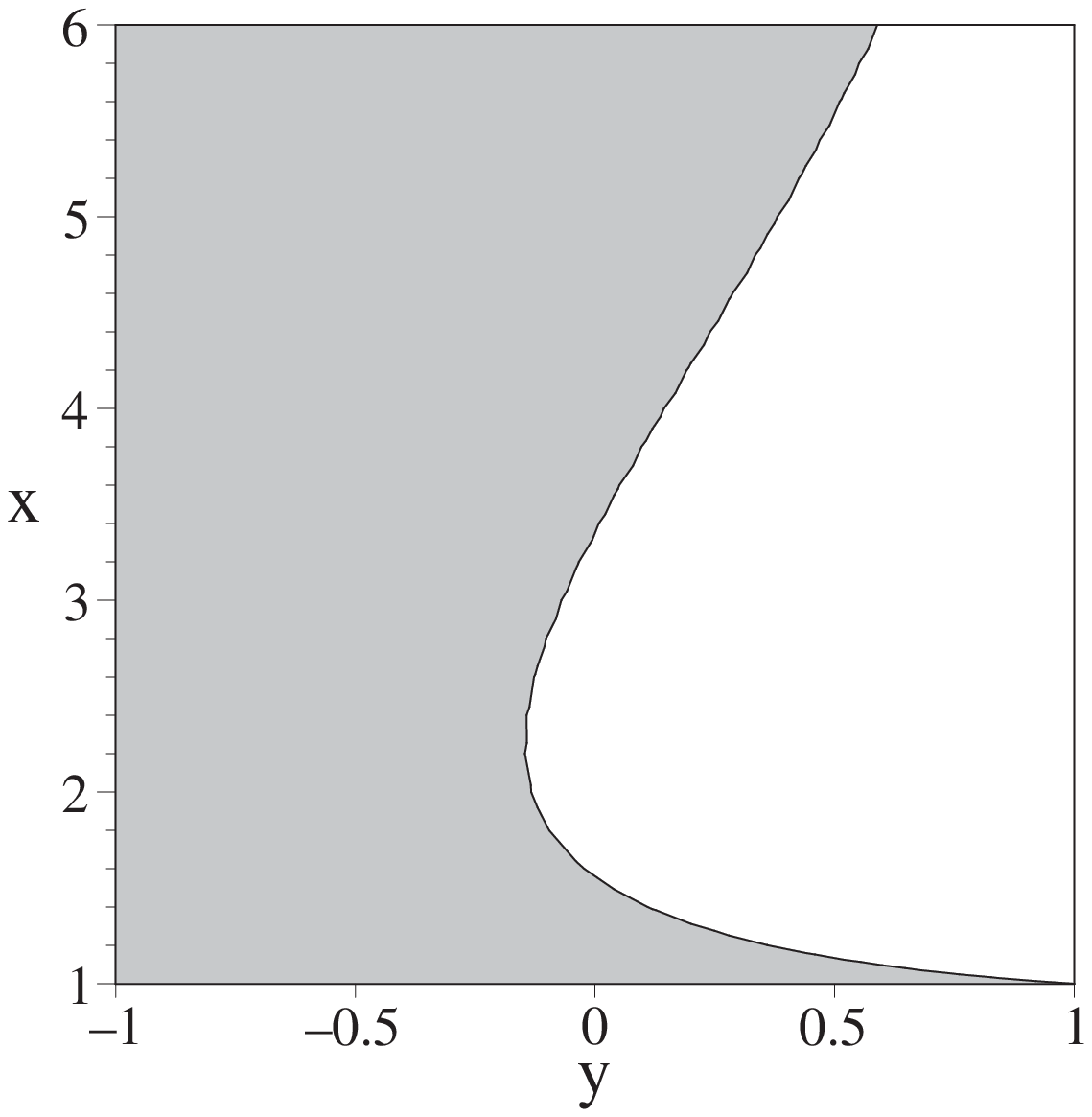} \\
			$b)\,\,\,$ $\alpha = 0.18$, $b_1 = 0.2$\  &
			$\alpha = 0.216$, $b_1 = 0.2$\  &
            $\alpha = 0.25$, $b_1 = 0.2$\  &
			$\alpha = 0.4$, $b_1 = 0.2$ \\
		\end{tabular}}
\caption{\footnotesize{Behavior of the ergoregion (grey area) for positive distortion parameter $b_1$: a) the dependence of the ergoregion on $b_1$ is investigated for fixed value of the rotation parameter $\alpha = 0.2$; b) the dependence of the ergoregion on the rotation is investigated for fixed value of $b_1=0.2$. }}
		\label{Erg_pos}
\end{figure}

If we keep the parameter $b_1$ fixed and vary the rotation parameter $\alpha$  a similar behavior is observed. Again, there exists a critical value of the rotation parameter $\alpha_{crit}$, when eq. ($\ref{Erg_int}$) has a single real root, two real roots for $0<\alpha<\alpha_{crit}$, and none for $\alpha>\alpha_{crit}$. Thus, for  $0<\alpha<\alpha_{crit}$ two separated parts of the ergoregion are observed, which touch at $\alpha=\alpha_{crit}$, and merge for $\alpha>\alpha_{crit}$ into a single ergoregion extending to infinity. The behavior of the ergoregion for positive $b_1$ is demonstrated on fig. $\ref{Erg_pos}$. On fig. $\ref{Erg_pos}$ $a)$ we investigate the ergoregion for a constant value of the rotation parameter $0<\alpha<1$, and different values of  $b_1> 0$. On fig. $\ref{Erg_pos}$ $b)$ we keep the distortion parameter $b_1$ fixed and study the influence of the rotation on the ergoregion by varying $\alpha$.

\begin{figure}[htp]
		\setlength{\tabcolsep}{ 0 pt }{\scriptsize\tt
		\begin{tabular}{ cccc }
            \includegraphics[width=4.1 cm]{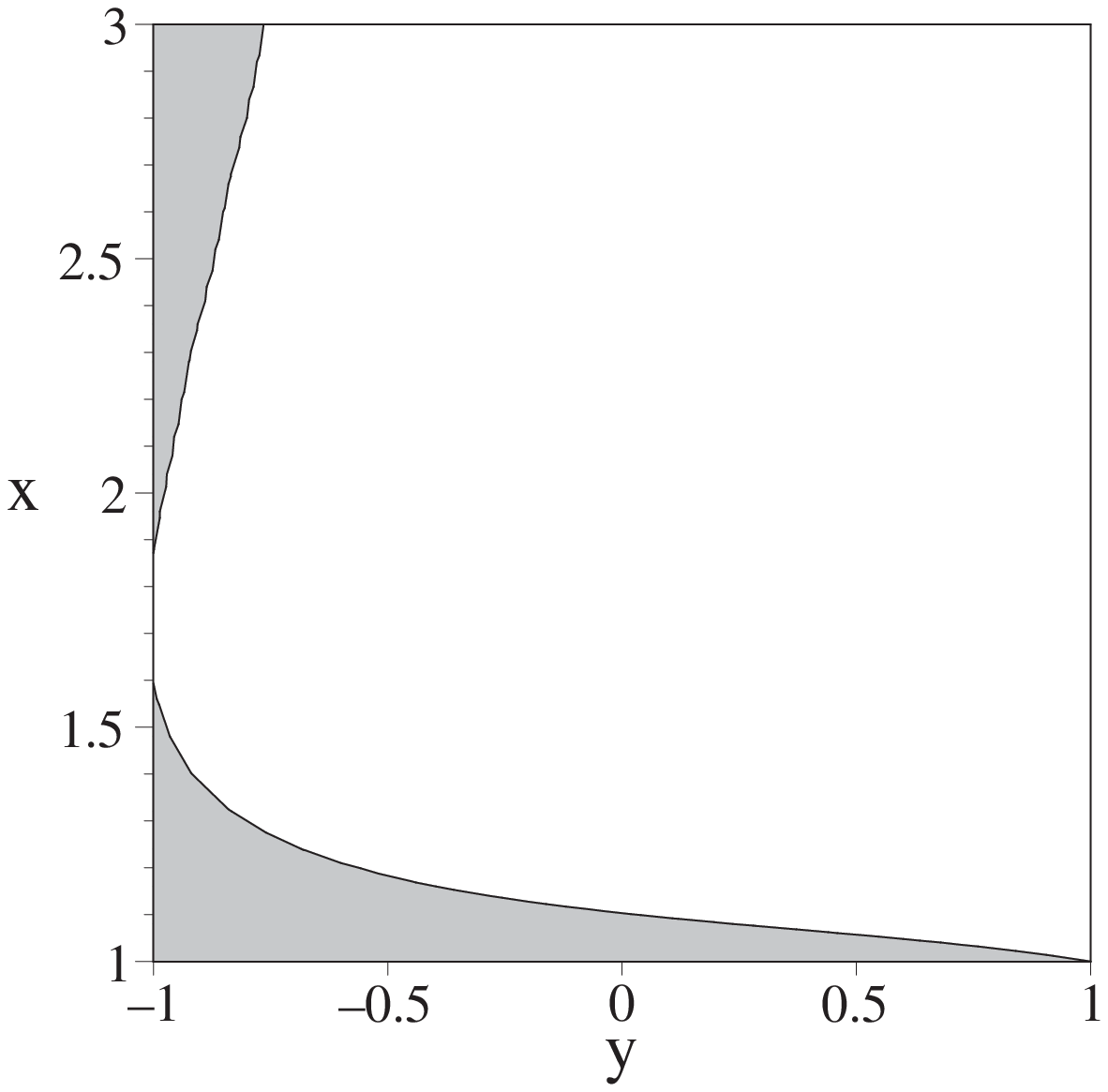} &
            \includegraphics[width=4.1 cm]{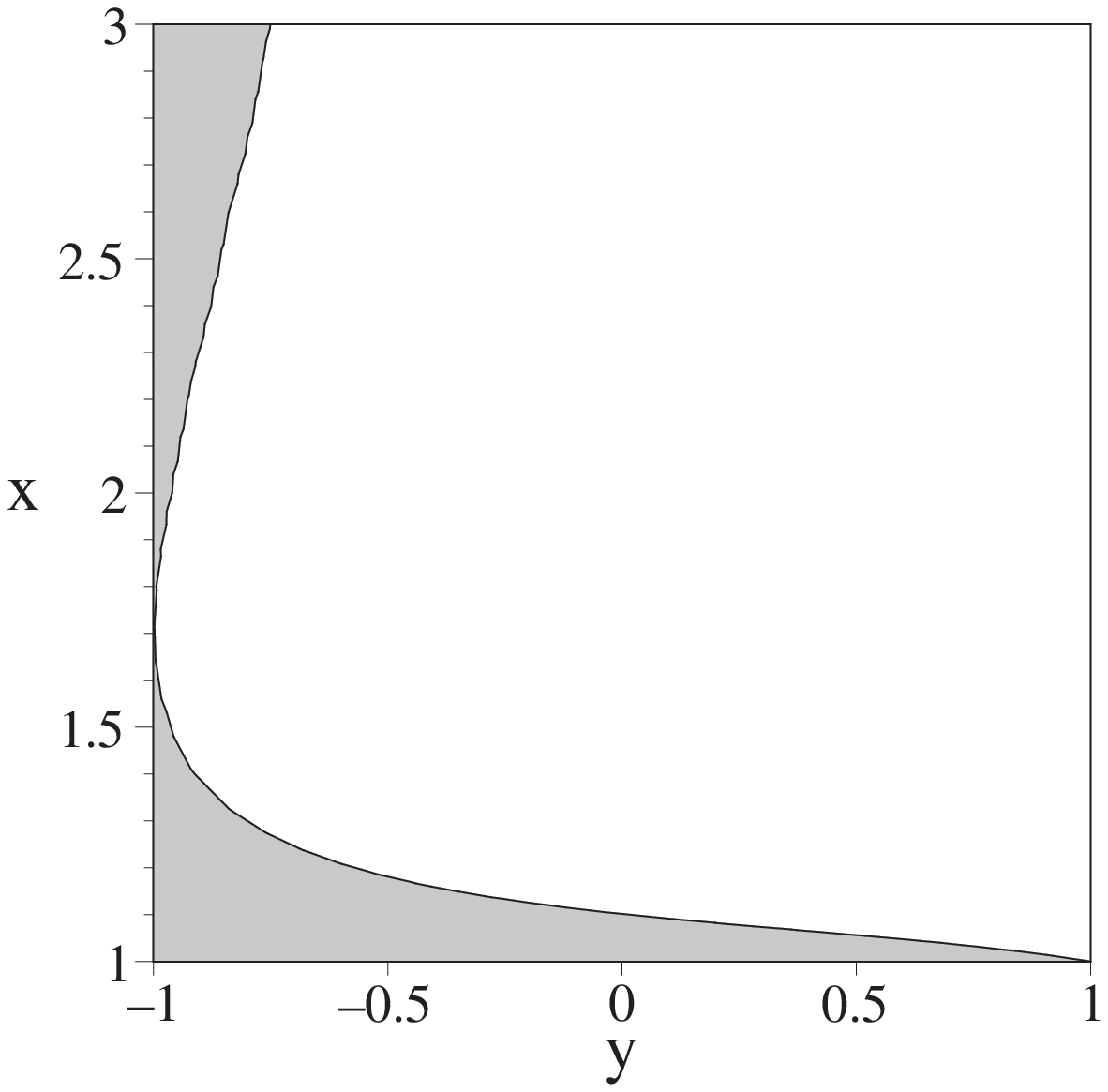} &
            \includegraphics[width=4.1 cm]{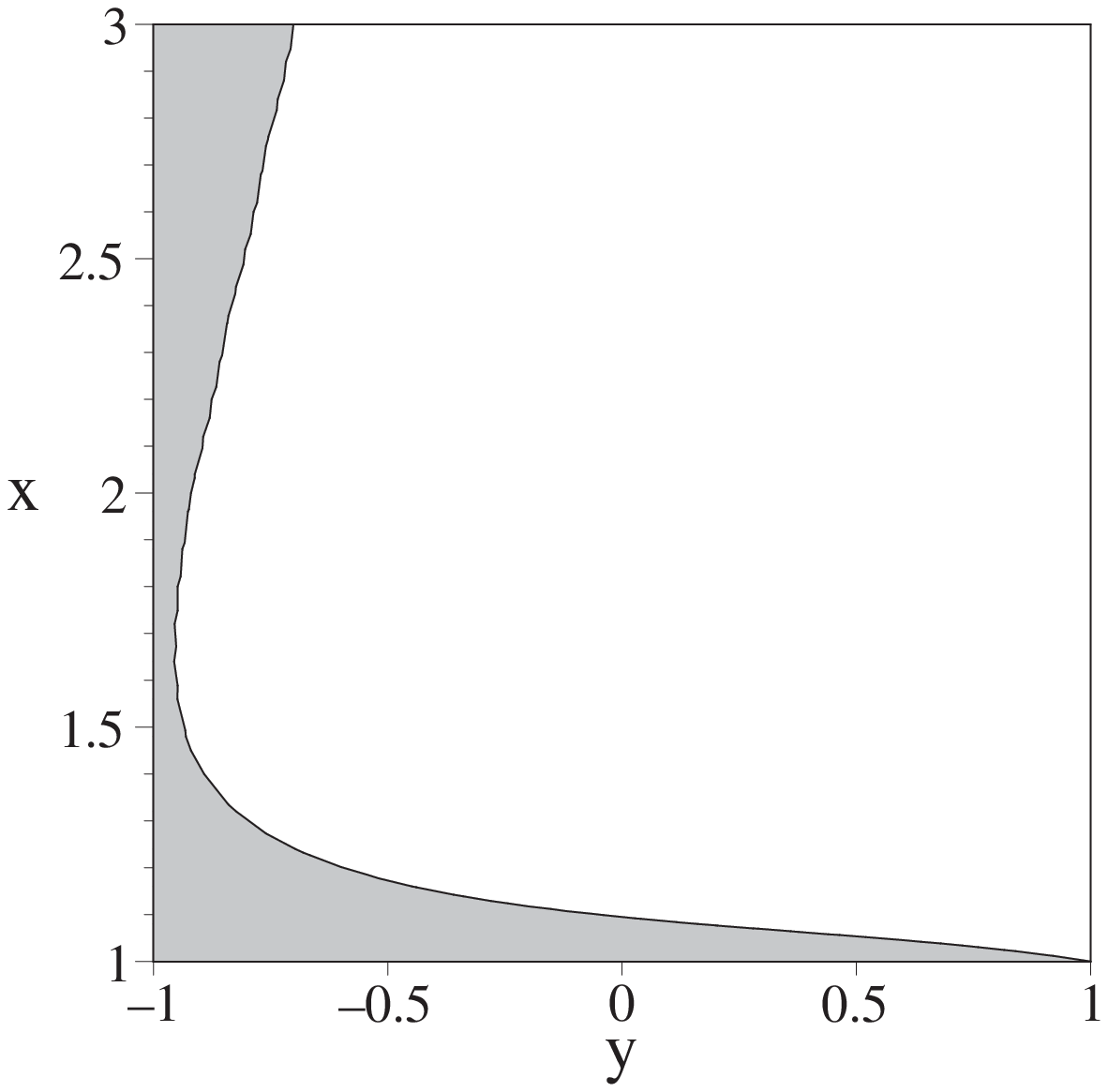} &
			\includegraphics[width=4.1 cm]{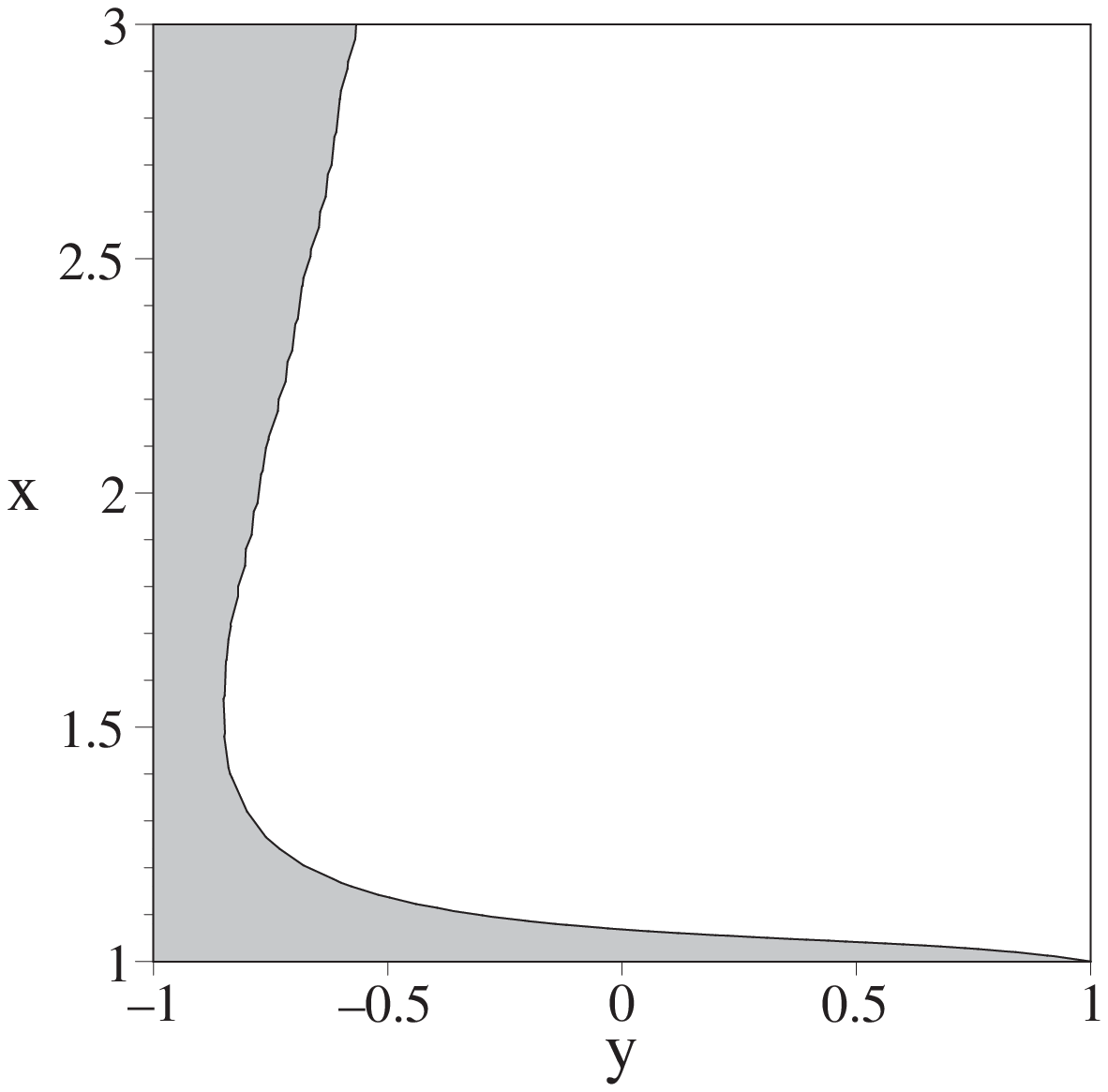} \\
            $a)\,\,\,$ $\alpha = 0.4$, $b_2 =  -0.1$\  &
			$\alpha = 0.4$, $b_2 = -0.104$\  &
            $\alpha = 0.4$, $b_2 = -0.12$\  &
			$\alpha = 0.4$, $b_2 = -0.2$ \\[2mm]
            \includegraphics[width=4.1cm]{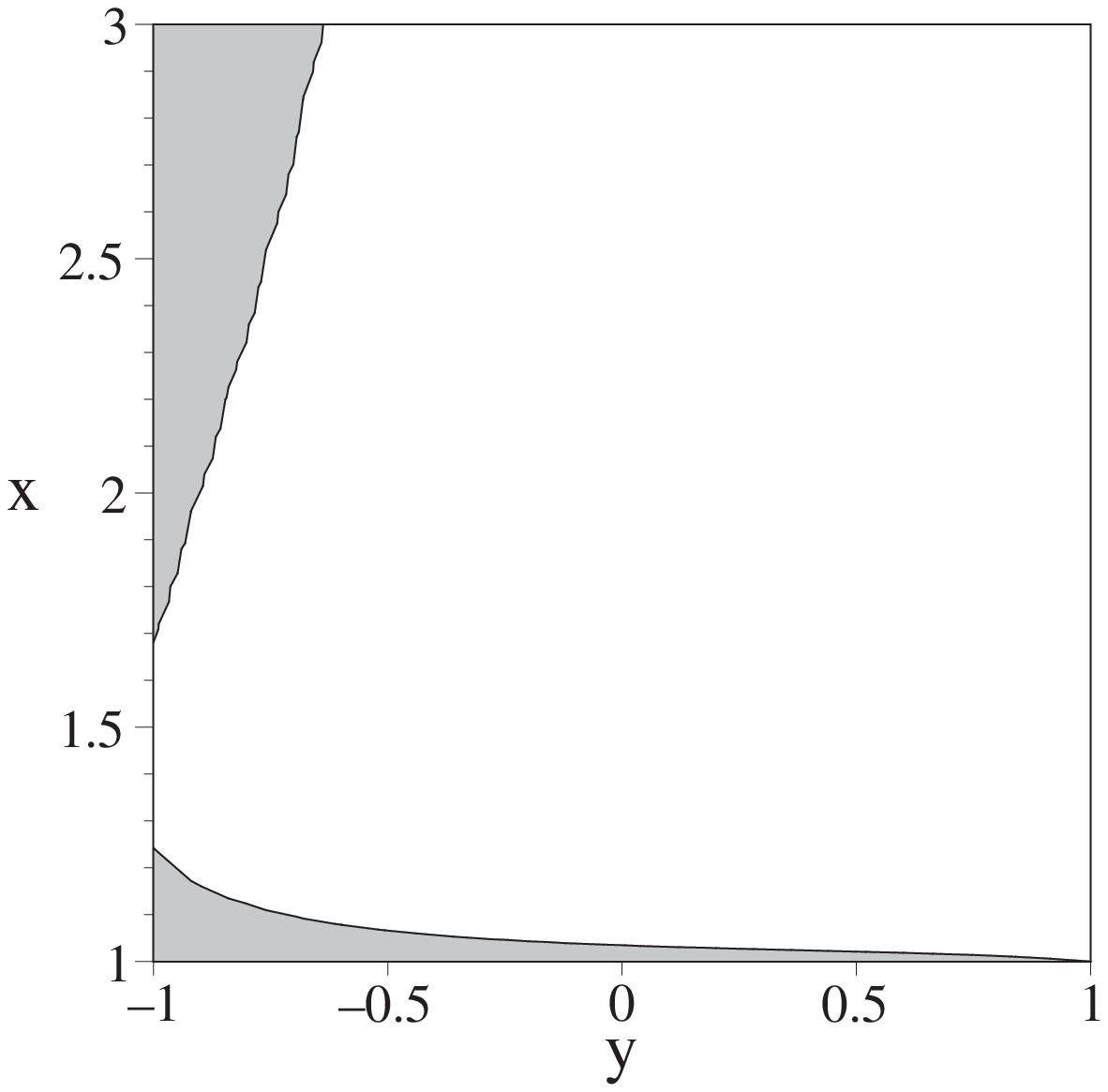} &
            \includegraphics[width=4.1cm]{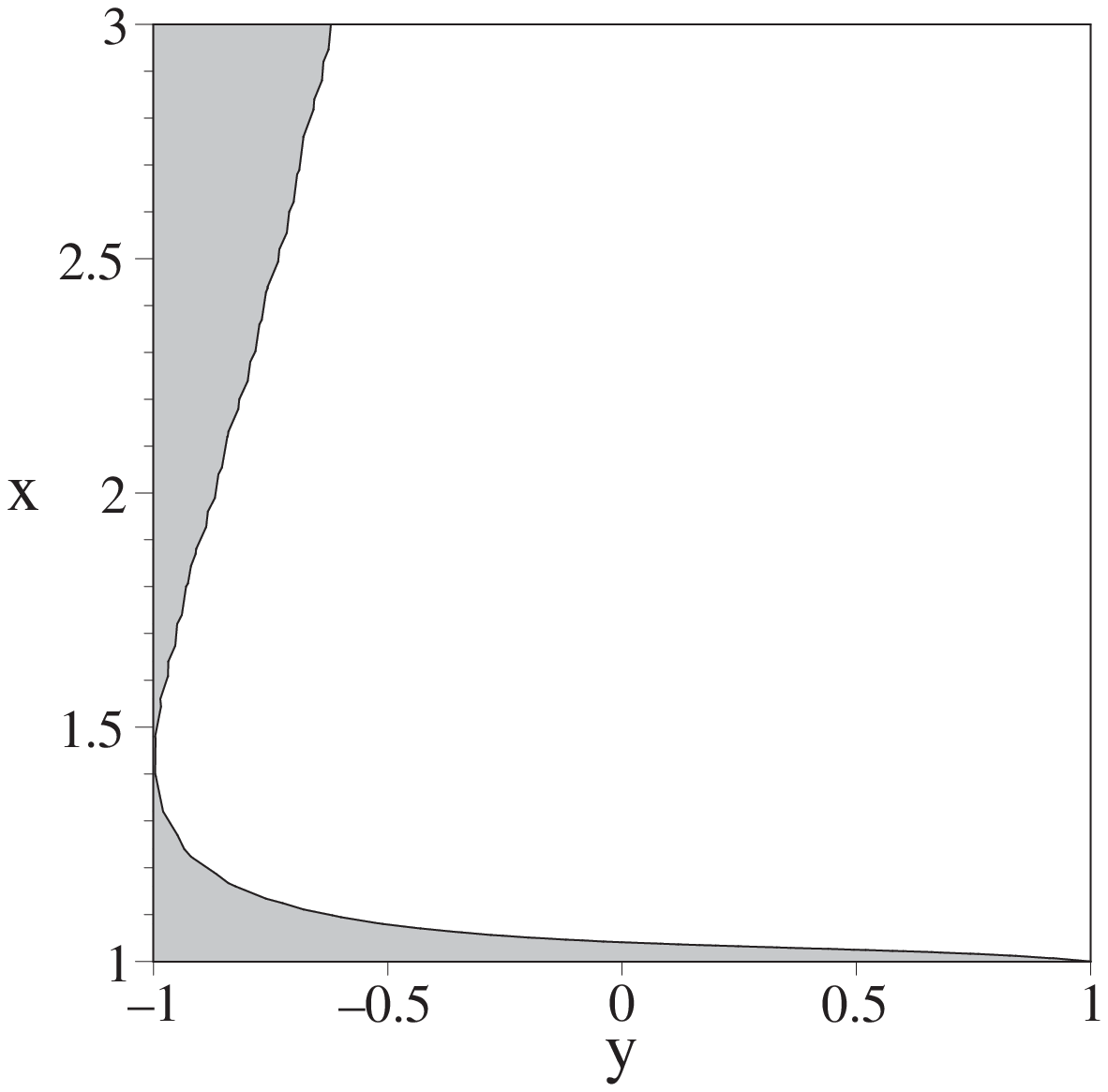} &
            \includegraphics[width=4.1 cm]{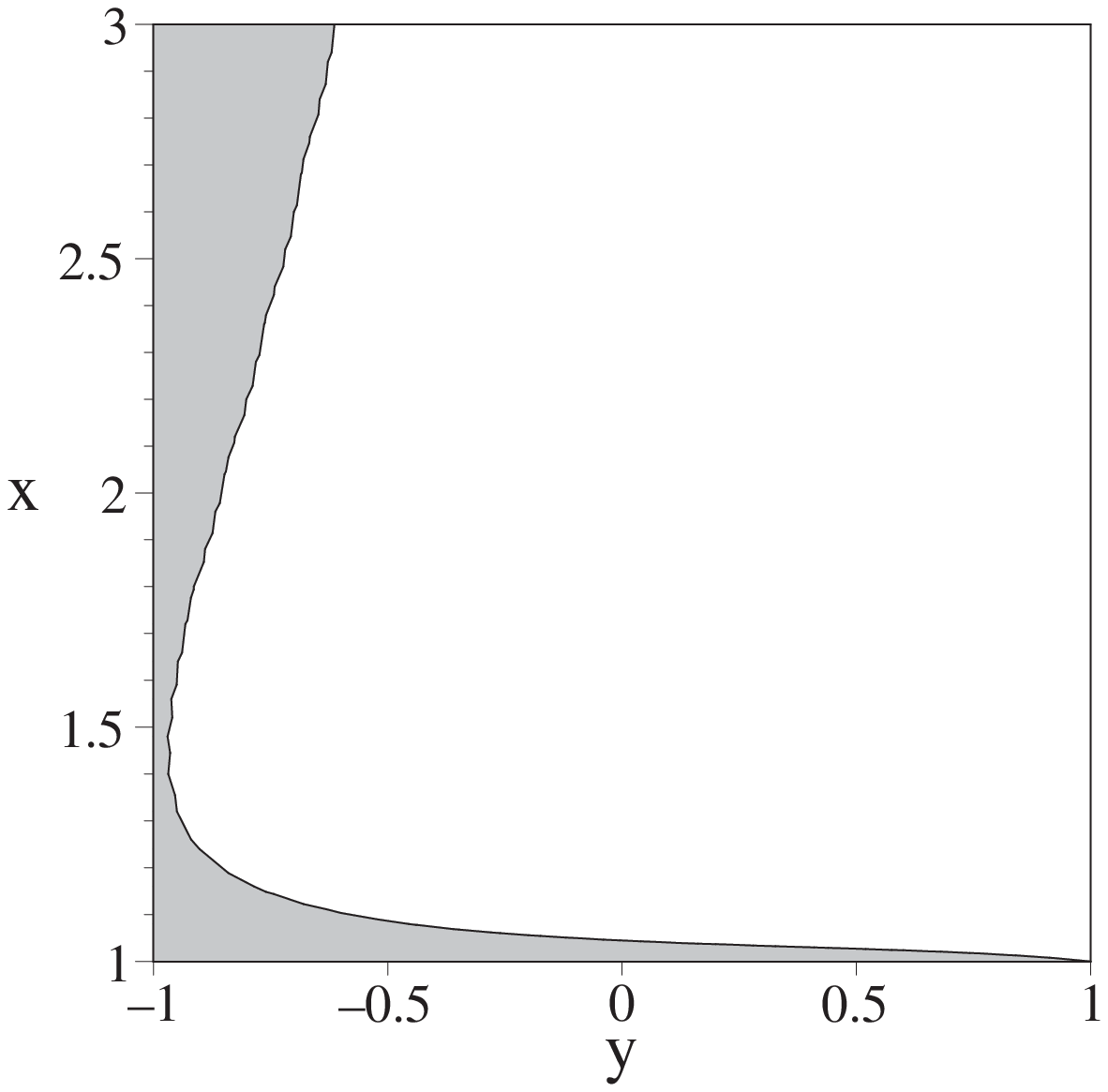} &
			\includegraphics[width=4.1cm]{ergo2-02.eps} \\
			$b)\,\,\,$ $\alpha = 0.28$, $b_2 = -0.2$\  &
			$\alpha = 0.306$, $b_2 = -0.2$\  &
            $\alpha = 0.32$, $b_2 = -0.2$\  &
			$\alpha = 0.4$, $b_2 = -0.2$ \\
		\end{tabular}}
\caption{\footnotesize{Behavior of the ergoregion (grey area) for quadrupole distortion $b_2<0$: a) the dependence of the ergoregion on $b_2$ is investigated for fixed value of the rotation parameter $\alpha=0.4$; b) the dependence of the ergoregion on the rotation is investigated for fixed value of $b_2=-0.2$. }}
		\label{Erg2_neg}
\end{figure}

\begin{figure}[htp]
		\setlength{\tabcolsep}{ 0 pt }{\scriptsize\tt
		\begin{tabular}{ cccc }
            \includegraphics[width=4.1cm]{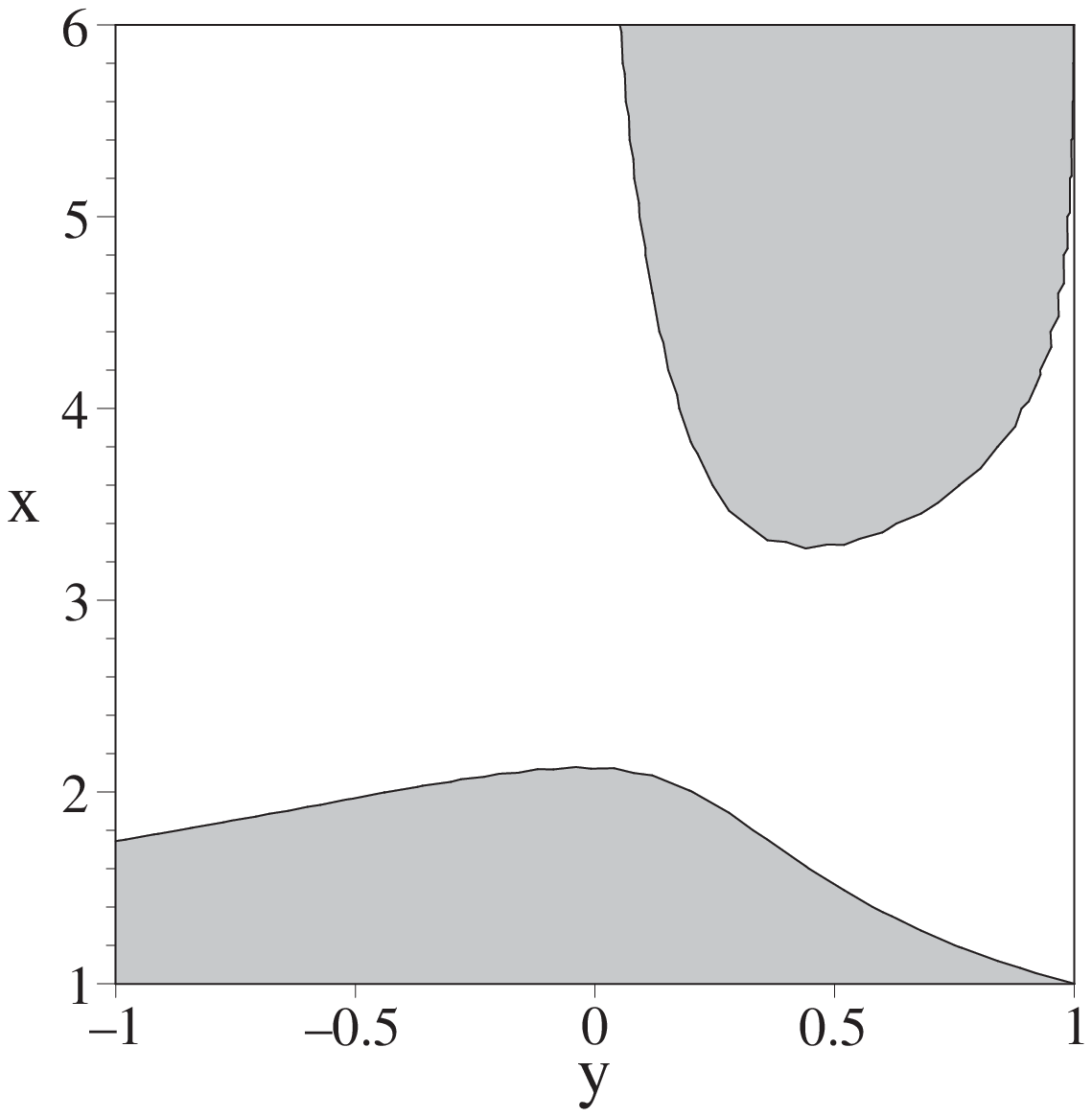} &
            \includegraphics[width=4.1cm]{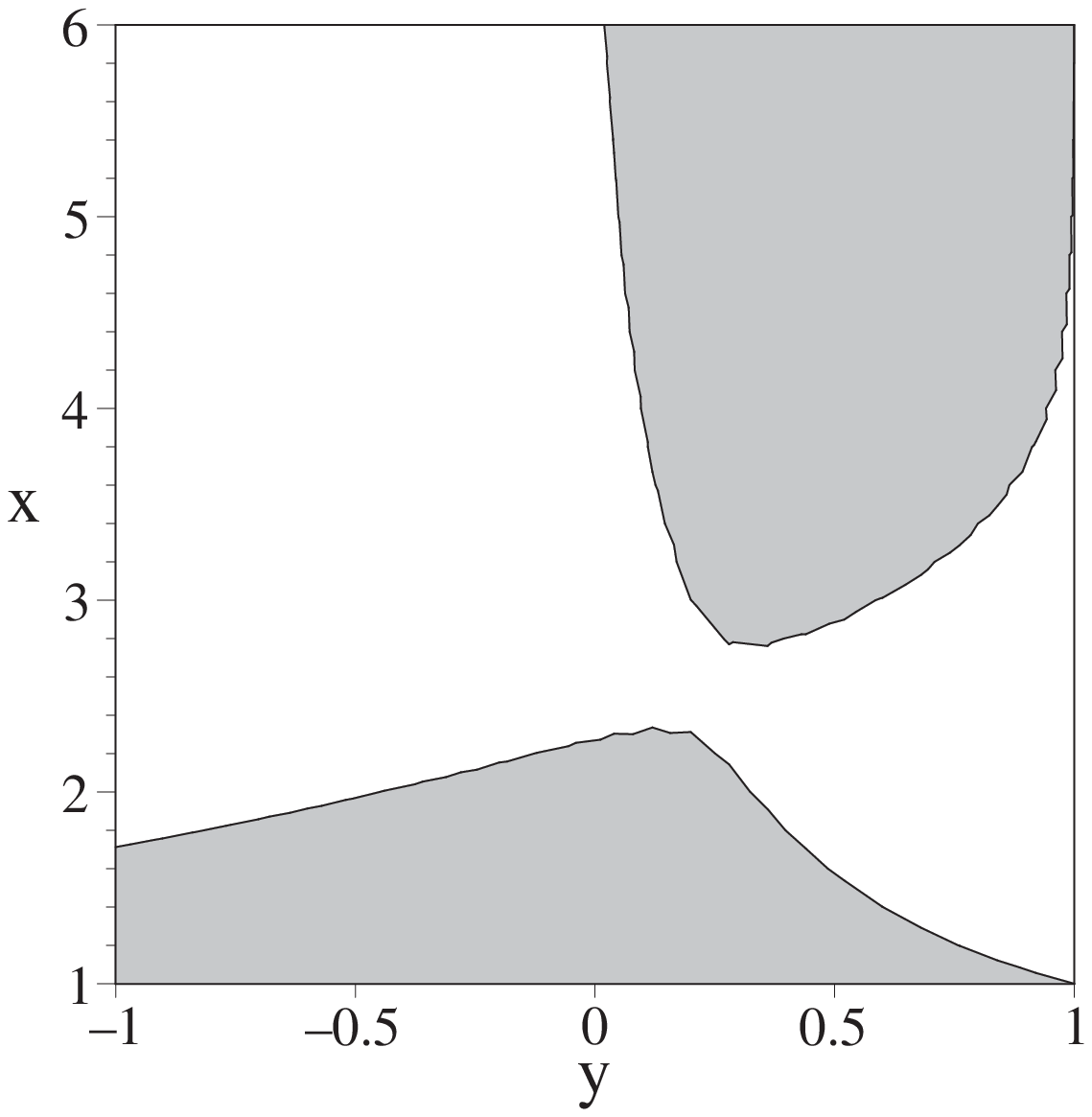} &
            \includegraphics[width=4.1 cm]{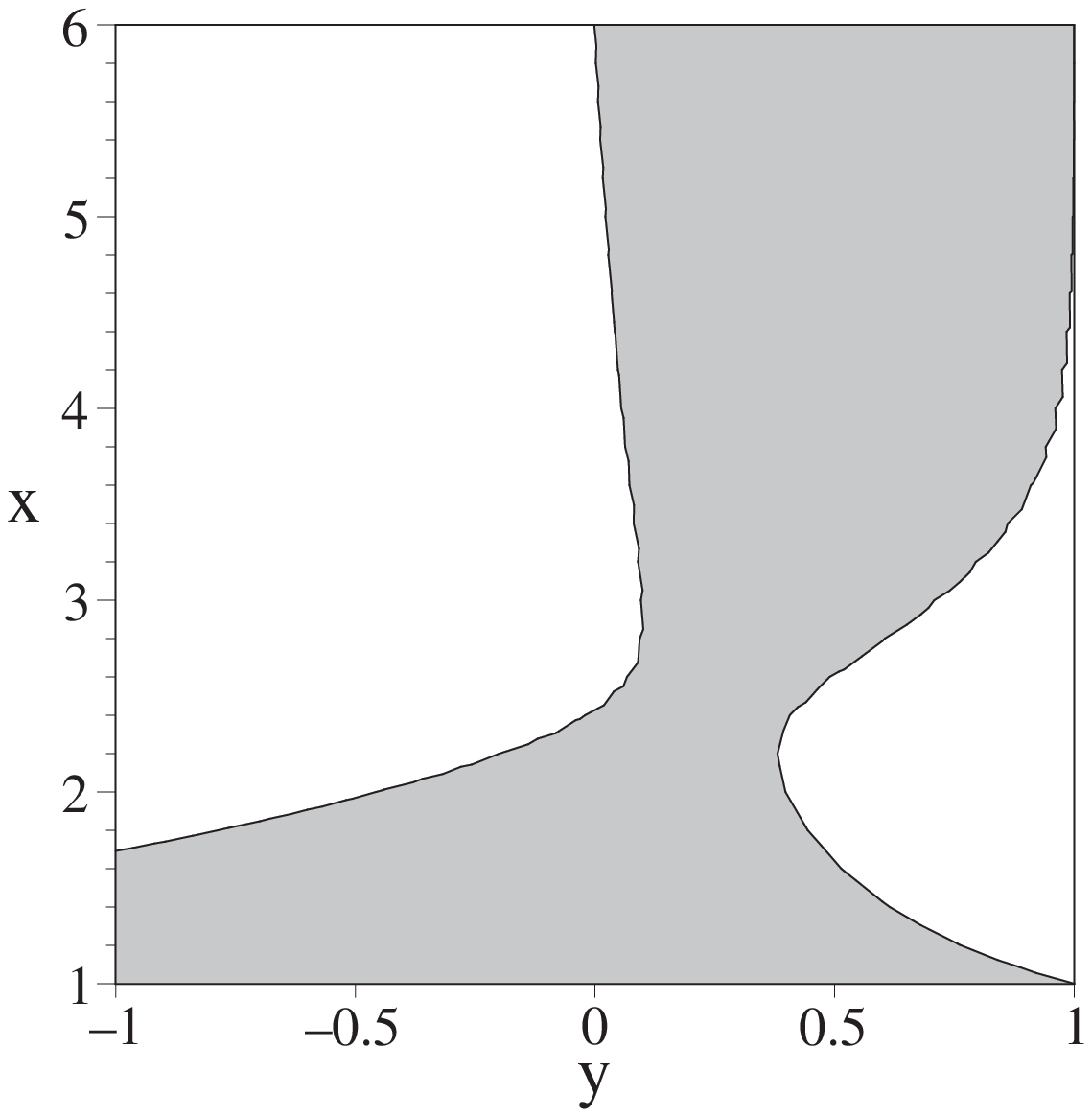} &
			\includegraphics[width=4.1cm]{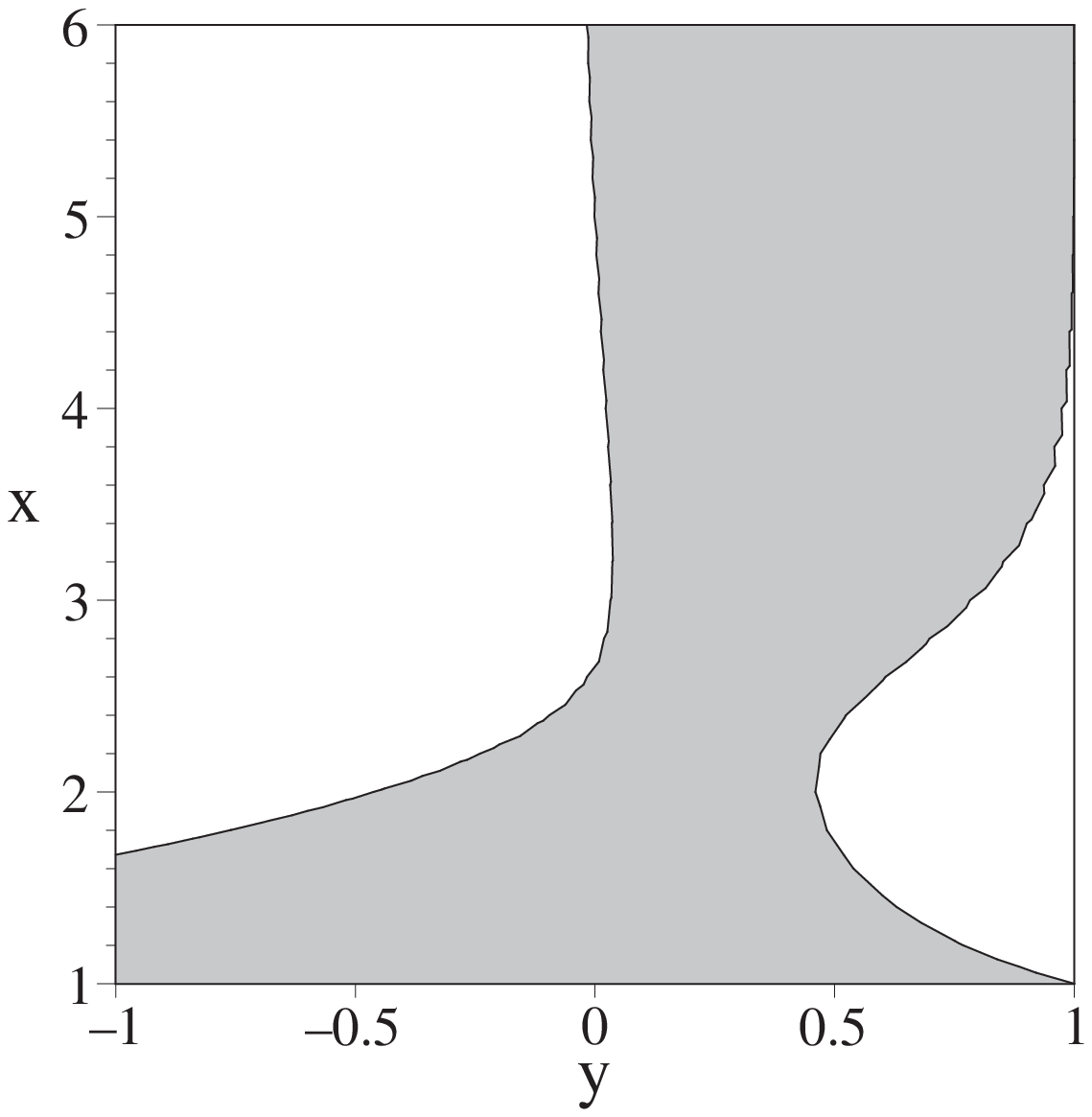} \\
			$a)\,\,\,$ $\alpha = 0.8$, $b_2 = 0.66$\  &
			$\alpha = 0.8$, $b_2 = 0.76$\  &
            $\alpha = 0.8$, $b_2 = 0.83$\  &
			$\alpha = 0.8$, $b_2 = 0.9$ \\[2mm]
            \includegraphics[width=4.1 cm]{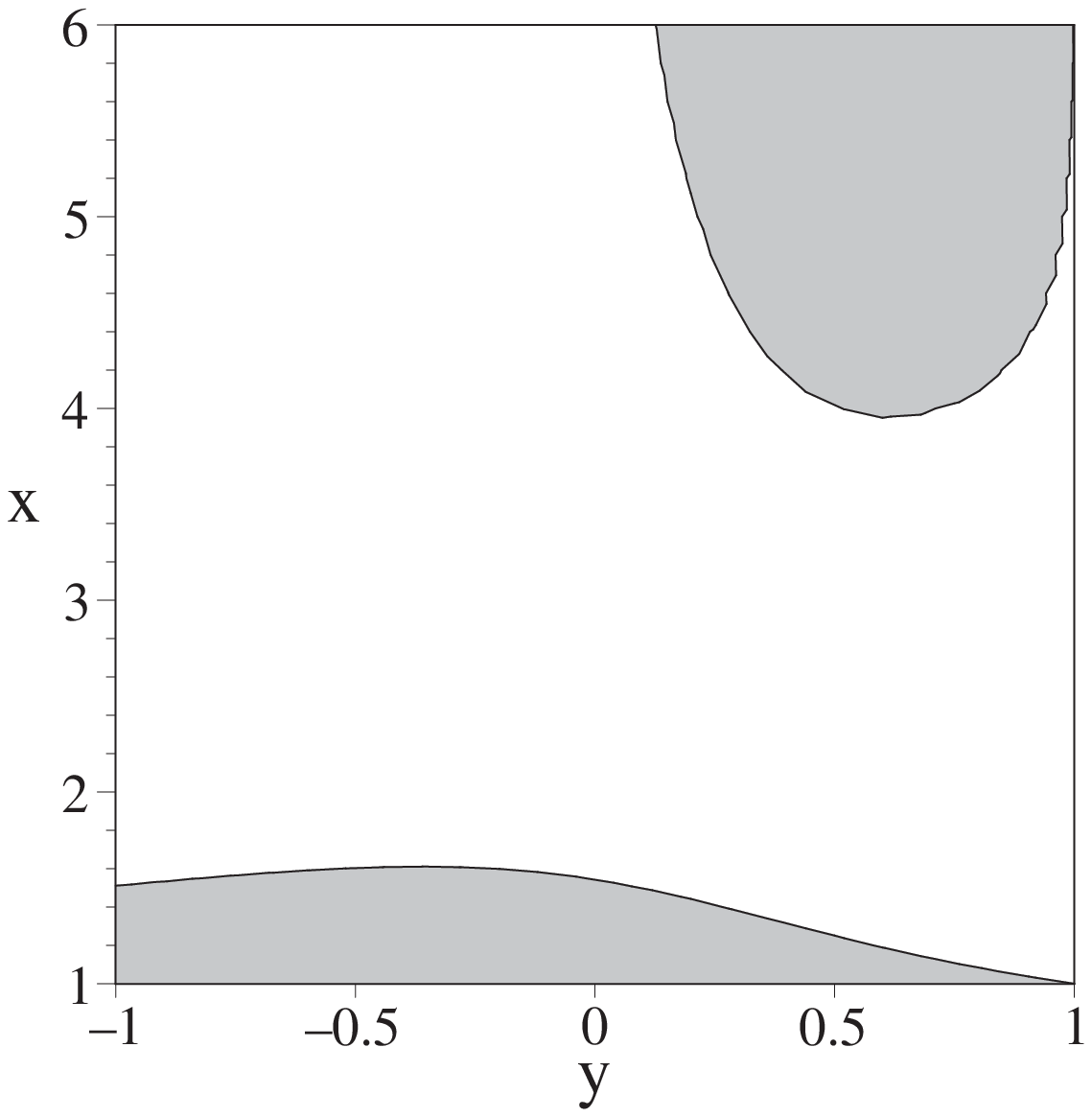} &
            \includegraphics[width=4.1 cm]{ergo2066.eps} &
            \includegraphics[width=4.1 cm]{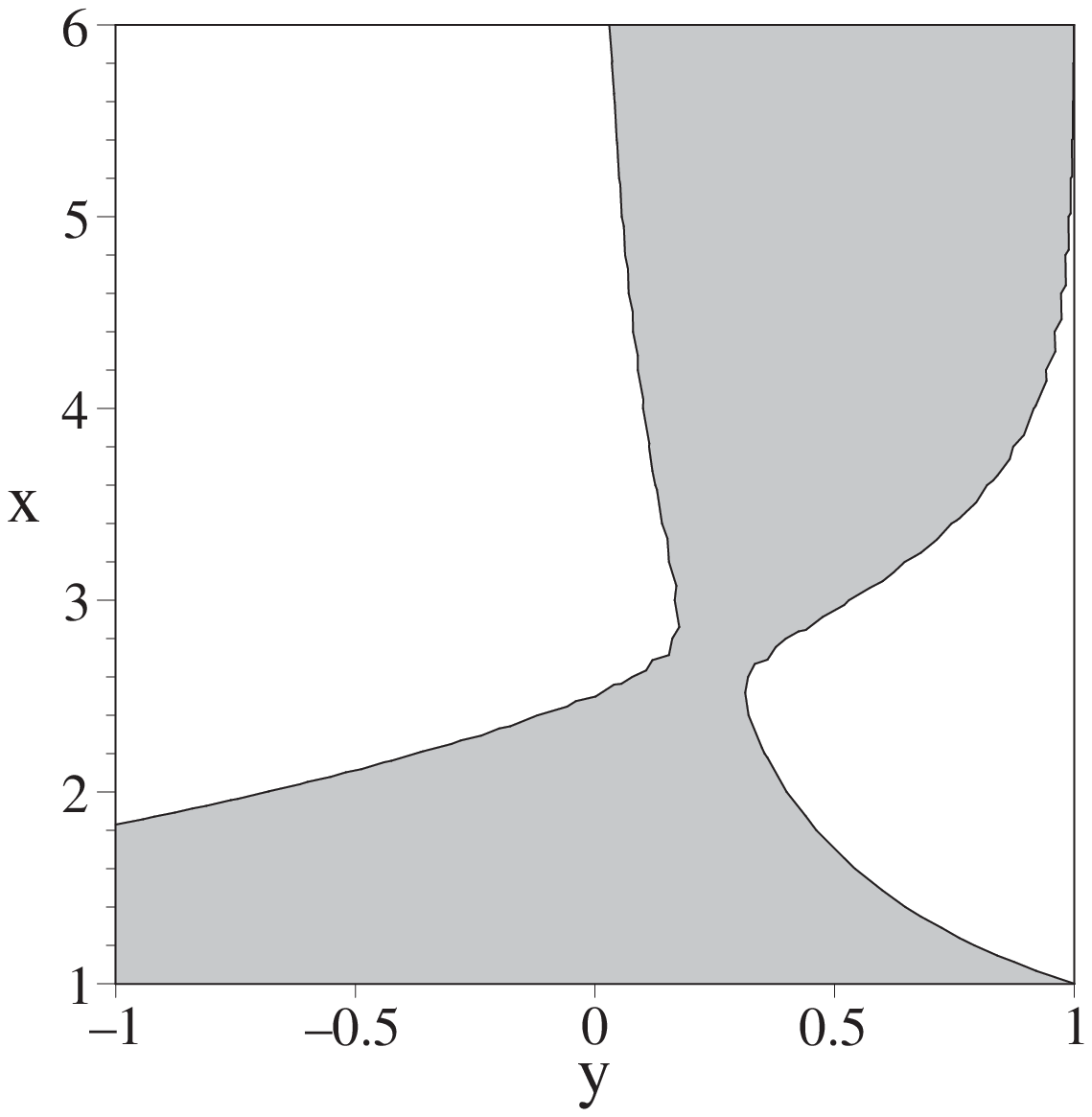} &
			\includegraphics[width=4.1 cm]{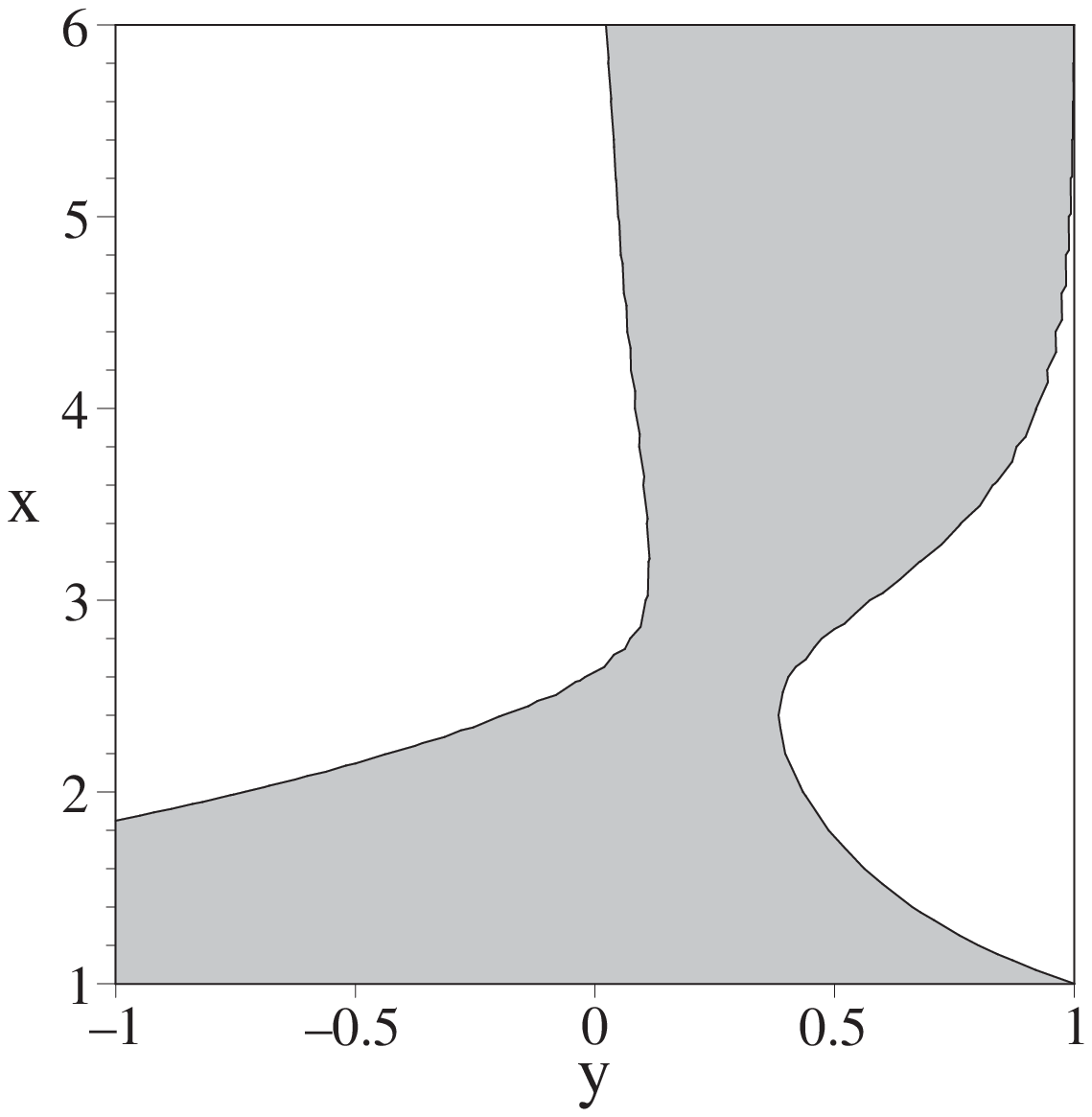} \\
            $b)\,\,\,$ $\alpha = 0.6$, $b_2 =  0.66$\  &
			$\alpha = 0.8$, $b_2 = 0.66$\  &
            $\alpha = 0.88$, $b_2 = 0.66$\  &
			$\alpha = 0.9$, $b_2 = 0.66$ \\
		\end{tabular}}
\caption{\footnotesize{Behavior of the ergoregion (grey area) for quadrupole distortion $b_2>0$: a) the dependence of the ergoregion on $b_2$ is investigated for fixed value of the rotation parameter $\alpha = 0.8$; b) the dependence of the ergoregion on the rotation is investigated for fixed value of $b_2 = 0.66$. }}
		\label{Erg2_pos}
\end{figure}

We will obtain a richer structure of the ergoregion, if we consider more complicated types of distortion by keeping other external matter parameters $a_n$ and $b_n$ different from zero. We illustrate some possible configurations in the case when the distortion parameter $b_2 \neq 0$, and the others are vanishing, which we call quadrupole distortion (see fig. $\ref{Erg2_neg}$ and $\ref{Erg2_pos}$). In contrast to the case of dipole distortion, for all values of $b_2$ the ergoregion is not compact.  For negative $b_2$ the ergoregion behaves qualitatively as in the case of positive dipole distortion $b_1>0$. For low values of $\mid b_2 \mid$ two separated parts of the ergoregion exist, intersecting the axis $y=-1$, which merge into a single one as $\mid b_2 \mid$ grows. The same structure is observed if we keep $b_2<0$ fixed and vary the rotation parameter $\alpha$ (fig. $\ref{Erg2_neg}$). For positive values of the distortion parameter $b_2$ another type of behavior of the ergoregion is demonstrated. For low values of $b_2$ the ergoregion consists of two parts - a compact one in the vicinity of the horizon, and a non-compact one separated from it in spacetime. However, the non-compact part is situated in the region $0<y<1$, and doesn't intersect the axis $y=-1$. When $b_2$ grows, the two parts merge in a single ergoregion, and the same behavior is observed if we keep $b_2>0$ fixed, and vary the rotation parameter $\alpha$ (fig. $\ref{Erg2_pos}$).

In this work we consider the distorted Myers-Perry black hole ($\ref{metric_dist}$) as a local solution, which is valid only in a certain neighbourhood of the horizon. In general, a global solution can be constructed if ($\ref{metric_dist}$) is extended to an asymptotically flat solution by some sewing technique. This can be realized by cutting the spacetime manifold in the region where the metric ($\ref{metric_dist}$) is valid and attaching to it another spacetime manifold where the solution is not vacuum anymore, but the sources of the distorting matter are also included. In the cases when the ergoregion consists of a compact region close to the horizon and another non-compact region, one can always choose to cut the manifold at such a value of the $x$ and $y$ coordinates that the non-compact region is not included. In this way one may be able to construct an extension to an asymptotically flat solution with a compact ergoregion.

\section{Conclusion}
We have obtained a new exact solution of the 5D Einstein equations in vacuum representing a Myers-Perry black hole with a single angular momentum in an external gravitational field. Locally, the solution is interpreted as a black hole distorted by a stationary $U(1)\times U(1)$ symmetric distribution of external matter. We have constructed the solution by applying a 2-fold B\"{a}cklund transformation on a 5D distorted Minkowski spacetime as a seed. The physical quantities of the solution have been calculated, and a local Smarr-like relation on the black hole horizon has been derived.
The solution satisfies the same Smarr-like relation as the asymptotically flat Myers-Perry black hole. However, in contrast to the non-distorted Myers-Perry black hole the ratio of the horizon mass and angular momentum, $J^2_H/ M^3_H$, can become arbitrarily large.
We have then considered a particular case of the general distorted Myers-Perry solution $(\ref{metric_dist})$, in which the parameters $a_n$ and $b_n$ characterizing the external matter field obey the relation $a_n + 2 b_n = 0$ for every $n$. Consequently, the angular velocity of the horizon is equal to the one of the asymptotically flat Myers-Perry case, and the ergoregion behaves also as in the asymptotically flat Myers-Perry case.

We have further considered the effect of dipole distortions and quadrupole distortions on the ergoregions. There exists a small neighbourhood of the horizon $x\rightarrow 1$, $x>1$, belonging to the ergoregion. In some cases, the ergoregion appears to be non-compact in the region of validity of the solution.  In the cases when the ergoregion consists of a compact region close to the horizon and another non-compact region, one can always choose to cut the manifold at a value of the $x$ and $y$ coordinates such that the non-compact region in not included, in order to construct an asymptotically flat solution with a compact ergoregion. However, we have not discussed the particular extension of the solution to the asymptotically flat cases.

\section*{Acknowledgment}
The authors gratefully acknowledge support by the DFG Research Training Group 1620 ``Models of Gravity''. S. A. is also grateful to the Natural Sciences and Engineering Research Council of Canada for financial support.


\begin{thebibliography}{tbds}

\bibitem{Geroch}
R. Geroch and J. B. Hartle,
``Distorted black holes,''
 J. Math. Phys. 23 (1982) 680

\bibitem{Israel}
W. Israel and K. A. Khan,
``Collinear particles and Bondi dipoles in general relativity'',
Nuovo Cimento 33 (1964) 331.

\bibitem{Doroshkevich}
A. Doroshkevich, Ya. Zel’dovich and I. Novikov,
``Gravitational Collapse of Nonsymmetric and Rotating Masses'',
Zh. Eksp. Teor. Fiz. 49 (1965) 170.

\bibitem{Erez}
G. Erez and N. Rosen,
``The gravitational field of a particle possessing a multipole moment'',
Bull. Res. Count. lsr. 8F ( 1959) 47.

\bibitem{Mysak}
 L. A. Mysak and G. Szekeres,
 ``Behavior of the Schwarzschild singularity in superimposed gravitational fields'',
  Can. J. Phys. 44 (1966) 617

\bibitem{Chandrasekhar}
 S. Chandrasekhar,
 {\it The Mathematical Theory of Black Holes, Clarendon Press}, Oxford (1983), pg. 583

\bibitem{Tomimatsu}
A. Tomimatsu,
``Distorted rotating black holes'',
Phys. Lett. A 103 (1984) 374.

\bibitem {Breton:1997}
N.~Bret\'{o}n, T.~Denisova, and V.~Manko,
``A Kerr black hole in the external gravitational field'',
Phys.\ Lett.\ A {\bf230} (1997) 7.

\bibitem{Peters}
P.~Peters,
``Toroidal black holes?'',
J. Math. Phys. 20 (1979) 1481.

\bibitem{Xanthopoulos}
B.~ Xanthopoulos,
``Local toroidal black holes that are static and axisymmetric'',
Proc. R. Soc. Lond. A 388 (1983) 117.


\bibitem{Fairhurst}
 S. Fairhurst and B. Krishnan,
 ``Distorted black holes with charge'',
 Int. J. Mod. Phys. D 10 (2001) 691 [gr-qc/0010088]

\bibitem{Yazadjiev}
S. S. Yazadjiev,
``Distorted charged dilaton black holes'',
Class. Quant. Grav. 18 (2001) 2105 [gr-qc/0012009].

\bibitem {Breton:1998}
N. Bret\'{o}n, A. A. Garc\'{\i}a, V. S. Manko, and T. E. Denisova,
``Arbitrarily deformed Kerr Newman black hole in an external gravitational field'',
 Phys. Rev. D 57 (1998) 3382.

\bibitem{Delgado}
J. Estevez-Delgado and T. Zannias,
``Distorted black holes of the Einstein-Klein-Gordon system'',
Phys.\ Rev.\ {\bf D 70} (2004) 064038.

\bibitem{Abdolrahimi:2010}
S.~Abdolrahimi,  A.~Shoom, D.~Page,
``Distorted 5-dimensional vacuum black hole'',
Phys.\ Rev.\ {\bf D 82} (2010) 124039.

\bibitem{Abdolrahimi:2013}
S.~Abdolrahimi, A.~Shoom,
``Distorted Five-dimensional Electrically Charged Black Holes'',
Phys. Rev. D 89 (2014) 024040  [arXiv:1307.440].

\bibitem{Frolov:2003}
A. V. Frolov and V. P. Frolov,
``Black holes in a compactified space-time'',
Phys. Rev. D 67 (2003) 124025 [hep-th/0302085]

\bibitem{Frolov:2006}
 V. P. Frolov and A. A. Shoom,
 ``Interior of distorted black holes'',
Phys. Rev. D. 76 (2006) 064037 [arXiv:0705.1570]

\bibitem{Frolov:2009}
A. Abdolrahimi, V. P. Frolov, and A. A. Shoom,
``Interior of a charged distorted black hole'',
Phys. Rev. D 80 (2009) 024011 [arXiv:0905.0178]

\bibitem{Papadopoulos}
D.~Papadopoulos, B.~Xanthopoulos,
``Local black holes are type D on the horizon'',
Il Nuovo Cimento {\bf B 83} (1984) 113.



\bibitem {Hennig}
J. Hennig, C. Cederbaum, and M. Ansorg,
`` A Universal Inequality for Axisymmetric and Stationary Black Holes with Surrounding Matter in the Einstein-Maxwell Theory'',
Commun. Math. Phys. 293 (2010) 449 [gr-qc/0805.4320]

\bibitem{Ansorg}
M. Ansorg, J. Hennig, C. Cederbaum,
``Universal properties of distorted Kerr-Newman black holes'',
Gen. Rel. Grav. 43 (2011) 1205,
[arXiv:1005.3128 [gr-qc]].

\bibitem{Ashtekar1}
A. Ashtekar, C. Beetle and S. Fairhurst,
``Mechanics of isolated horizons'',
Class. Quantum Grav. 17 253-298 (2000).

\bibitem{Ashtekar2}
``Isolated horizons: a generalization of black hole mechanics'',
Class. Quantum Grav. 16 L1 (1999).

\bibitem{Ashtekar3}
A. Ashtekar, S. Fairhurst and B. Krishnan,
``Isolated horizons: Hamiltonian evolution and the first law'',
Phys. Rev. D 62, 104025 (2000).

\bibitem{Booth1}
T. Pilkington, A. Melanson, J. Fitzgerald and I. Booth,
``Trapped and marginally trapped surfaces in Weyl-distorted Schwarzschild solutions'',
Class. Quant. Grav. 28 (2011) 125018,
[arXiv:1102.0999].

\bibitem{Booth2}
I.~Booth, D.~Tian,
``Some spacetimes containing non-rotating extremal isolated horizons'',
Class. Quant. Grav. 30 (2013) 145008,
[arXiv:1210.6889].

\bibitem{Ansorg1}
M.~Ansorg, D.~Petroff,
``Black holes surrounded by uniformly rotating rings'',
Phys. Rev. {\bf D 72} (2005) 024019.

\bibitem{Ansorg2}
D.~Petroff, M.~Ansorg,
``The extreme distortion of black holes due to matter'',
[gr-qc/0511102].



\bibitem{Mishima}
H.~Iguchi, T.~Mishima,
``Solitonic generation of vacuum solutions in five-dimensional general relativity,''
Phys.\ Rev.\ D {\bf74} (2006) 024029.

\bibitem{Mishima:2006}
H. Iguchi, T. Mishima,
``Solitonic generation of the five-dimensional black
ring solution,"
Phys.\ Rev.\ D {\bf 73} (2006) 121501 [arXiv:hep-th/0604050].

\bibitem{Mishima:2007a}
H. Iguchi, T. Mishima, S. Tomizawa,
``Boosted black holes on Kaluza-Klein bubbles,''
Phys.\ Rev.\ D {\bf 76} (2007) 124019; Erratum-ibid. 78 (2008) 109903 [arXiv:0705.2520[hep-th]].

\bibitem{Mishima:2008}
S. Tomizawa, H. Iguchi, T. Mishima,
``Rotating black holes on Kaluza-Klein bubbles,''
Phys.\ Rev.\ D {\bf 78} (2008) 084001 [arXiv:hep-th/0702207].

\bibitem{Mishima:2007}
H. Iguchi, T. Mishima,
``Black diring and infinite nonuniqueness,''
Phys.\ Rev.\ D {bf 75} (2007) 064018; Erratum-ibid. {\bf 78} (2008) 069903 [arXiv:hepth/
0701043].

\bibitem{Yazadjiev:2006}
S. Yazadjiev,
``Solution generating in 5D Einstein-Maxwell-dilaton gravity and derivation of dipole black ring solutions,''
JHEP 0607 (2006) 036 [arXiv:hep-th/0604140].

\bibitem{Nedkova:2010}
P. Nedkova and S. S. Yazadjiev,
 ``Rotating black ring on Kaluza-Klein bubbles,''
Phys.\ Rev.\ D {\bf 82} (2010) 044010 [arXiv:1005.5051 [hep-th]].

\bibitem{Emparan:2002}
  R.~Emparan, H.~S.~Reall,
  ``Generalized Weyl solutions,''
  Phys.\ Rev.\ D {\bf 65} (2002) 084025.


\bibitem{Myers}
R.~Myers and M.~Perry,
 ``Black holes in higher dimensional space-times,''
 \ Ann.\ Phys. {\bf172} (1986) 304.

\bibitem{Komar}
  A.~Komar,
  ``Covariant Conservation Laws in General Relativity,''
  Rhys.\ Rev.\   {\bf 113} (1959) 934.


\bibitem{Maison:1978}
D.~Maison,
``Are the stationary axially symmetric Einstein equations completely integrable?,''
\ Phys.\ Rev.\ Lett. {\bf41} (1978) 521.

\bibitem{Maison:1979a}
D.~Maison,
``On the complete integrability of the stationary axially symmetric Einstein equations,''
\ J.\ Math.\ Phys. {\bf20} (1979) 871.

\bibitem{BZ1}
V.~Belinski, V.~Zakharov,
``Integration of the Einstein equations by means of the inverse scattering problem technique and construction of exact soliton solutions,''
\ Sov.\ Phys.\ JETP {\bf48} (1978) 985.

\bibitem{Belinski:1979}
V.~Belinski, V.~Zakharov,
``Stationary gravitational solitons with axial symmetry,''
\ Sov.\ Phys.\ JETP {\bf50} (1979) 1.

\bibitem{BZ2}
V.~Belinski, E.~Verdaguer,
``Gravitational solitons,''
Cambridge University Press, Cambridge, 2001.

\bibitem {NG2}
G.~Neugebauer,
``B\"{a}cklund transformations of axially symmetric stationary gravitational fields,''
\ J.\ Phys.\ A {\bf12} (1979) L67.

\bibitem{NG}
G.~Neugebauer,
``A general integral of the axially symmetric stationary Einstein equations,''
\ J.\ Phys.\ A {\bf13}  (1980) L19.

\bibitem {Neugebauer:1980}
G.~Neugebauer,
``Recursive calculation of axially symmetric stationary Einstein fields,''
\ J.\ Phys.\ A {\bf13}  (1980) 1737.

\bibitem {Ernst}
F.~Ernst,
``New formulation of the axially symmetric gravitational field problem'',
Phys. Rev. {\bf 167}, 1175 (1968).

\bibitem{SK1}
H.~Stephani, D.~Kramer, M.~McCallum, C.~Hoenselaers, E.~Herlt,
``Exact solutions of Einstein's field equation,'' 2nd ed.
Cambridge University Press, Cambridge, 2003.

\bibitem{Yamazaki:1981}
M.~Yamazaki,
``On the Hoenselaers--Kinnersley--Xanthopoulos spinning mass fields'',
J . \ Math. \ Phys. {\bf22} (1981) 133.

\bibitem{Dietz:1982}
W. Dietz, C. Hoenselaers,
``A new class of bipolar vacuum gravitational fields'',
Proc.\ R.\ Soc. A 382 (1982) 221.


\bibitem{Amanedo}
J. Castejon-Amenedo and V.S. Manko,
``Superposition of the Kerr metric with the generalized Erez-Rosen solution'',
Phys. Rev. D {\bf 41} (1990) 2018.

\bibitem{Quevedo:1989}
H. Quevedo,
``General static axisymmetric solution of Einstein’s vacuum field equations in prolate spheroidal coordinates'',
Phys. Rev. 39 (1989) 2904.

\bibitem{Quevedo:1990}
H.~Quevedo,
``Multipole moments in general relativity. Static and stationary vacuum solutions,''
Fortschr.\ d.\ Phys. 38 (1990) 733.

\bibitem{Mars}
M. Mars, J. M. Senovilla,
``Axial symmetry and conformal Killing vectors,"
Class. Quant. Grav. 10 (1993) 1633.



\bibitem{Hollands:2007}
  S.~Hollands and S.~Yazadjiev,
  ``Uniqueness theorem for 5-dimensional black holes with two axial Killing
  fields,''
  Commum.\ Math.\ Phys.\  {\bf 283} (2008) 749
  [arXiv:0707.2775 [gr-qc]].

\end{thebibliography}
\end{document}